\documentclass[fleqn,usenatbib]{mnras}

\usepackage{newtxtext,newtxmath}

\usepackage[T1]{fontenc}

\DeclareRobustCommand{\VAN}[3]{#2}
\let\VANthebibliography\thebibliography
\def\thebibliography{\DeclareRobustCommand{\VAN}[3]{##3}\VANthebibliography}

\usepackage{graphicx}	
\graphicspath{{figures/}{../figures/}} 
\usepackage{amsmath}	
\usepackage{caption, subcaption} 
\usepackage{ulem}

\newcommand{\OmegaM}{\Omega_\mathrm{m}}
\newcommand{\fR}{f_{R_0}}

\title[Power spectrum emulator in $f(R)$CDM cosmology]{The e-MANTIS emulator: fast predictions of the non-linear matter power spectrum in $f(R)$CDM cosmology}

\author[I. Sáez-Casares et al.]{
Iñigo Sáez-Casares,$^{1}$\thanks{E-mail: inigo.saez-casares@obspm.fr (ISC)}
Yann Rasera,$^{1}$
Baojiu Li$^2$
\\
$^{1}$Laboratoire Univers et Théories, Université Paris Cité, Observatoire de Paris, Université PSL, CNRS, F-92190 Meudon, France\\
$^{2}$Institute for Computational Cosmology, Department of Physics, Durham University, South Road, Durham DH1 3LE, UK
}

\date{Accepted XXX. Received YYY; in original form ZZZ}

\pubyear{2023}

\begin{document}
\label{firstpage}
\pagerange{\pageref{firstpage}--\pageref{lastpage}}
\maketitle

\begin{abstract}
In order to probe modifications of gravity at cosmological scales, one needs accurate theoretical predictions. N-body simulations are required to explore the non-linear regime of structure formation but are very time consuming.
In this work, we release a new public emulator, dubbed \textsc{e-mantis}, that performs an accurate and fast interpolation between the predictions of $f(R)$ modified gravity cosmological simulations, run with \textsc{ecosmog}.
We sample a wide 3D parameter space given by the current background scalar field value $10^{-7}<\left|\fR\right|<10^{-4}$, matter density $0.24<\Omega_\mathrm{m}<0.39$, and primordial power spectrum normalisation $0.6<\sigma_8<1.0$, with 110 points sampled from a Latin Hypercube.
For each model we perform pairs of $f(R)$CDM and $\Lambda$CDM simulations covering an effective volume of $\left(560 \, h^{-1}\mathrm{Mpc}\right)^3$ with a mass resolution of $\sim 2 \times 10^{10} h^{-1} M_\odot$.
We build an emulator for the matter power spectrum boost $B(k)=P_{f(R)}(k)/P_{\Lambda\mathrm{CDM}}(k)$ using a Gaussian Process Regression method.
The boost is mostly independent of $h$, $n_{s}$, and $\Omega_{b}$, which reduces the dimensionality of the relevant cosmological parameter space.
Additionally, it is more robust against statistical and systematic errors than the raw power spectrum, thus strongly reducing our computational needs.
According to our dedicated study of numerical systematics, the resulting emulator has an estimated maximum error of $3\%$ across the whole cosmological parameter space, for scales $0.03 \ h\mathrm{Mpc}^{-1} < k < 7 \ h\mathrm{Mpc}^{-1}$, and redshifts $0 < z < 2$, while in most cases the accuracy is better than $1\%$.
Such an emulator could be used to constrain $f(R)$ gravity with weak lensing analyses.
\end{abstract}




\section{Introduction}%
\label{sec:intro}

The nature of the late-time accelerated expansion of the Universe remains one of the main unsolved puzzles of modern cosmology.
The unknown component responsible for such an acceleration of the expansion rate is commonly referred to as Dark Energy (DE).
In the standard cosmological model, $\Lambda$CDM, DE is modeled by a cosmological constant term $\Lambda$ in the Einstein field equations.
Other alternative scenarios are also usually considered, such as the quintessence models, where DE is made of a dynamical scalar field~\citep[see for example][for a review of different DE models]{amendola_tsujikawa_2010}.
Another possibility is that DE comes from a modification of General Relativity (GR) at cosmological scales, such as in several Modified Gravity (MG) theories~\citep[see][for an extensive review of MG theories in a cosmological context]{Clifton_2012_mg_review}.

Cosmic structure formation is the result of the interplay between gravitational collapse and the expansion of the Universe.
Therefore, the Large Scale Structure (LSS) of the Universe offers a unique window to probe simultaneously the nature of DE and the laws of gravity at cosmological scales.
This is precisely one of the main objectives of current and future next generation photometric and spectroscopic galaxy surveys, such as DESI~\citep{DESI}, LSST~\citep{LSST}, and \textit{Euclid}~\citep{EUCLID}.

Accurate theoretical predictions will be required to fully exploit the potential of those experiments.
One of the basic theoretical tools to characterise the growth of cosmic structures is the matter power spectrum.
Even if not directly observable, it serves as a building block for other observables.
However, getting accurate predictions for the power spectrum in the non-linear regime of structure formation requires the use of computationally expensive N-body simulations.
Constraining theoretical models with observations requires searching a large parameter space formed by cosmological and DE/MG parameters.
Therefore, predictions from N-body simulations cannot directly be used in an observational analysis, which would require to run simulations for hundreds of thousands of cosmological models.

In order to overcome this limitation, the concept of cosmic emulators was introduced in the pioneering work of~\citet{habib_emulator}.
An emulator is calibrated on a reduced number of simulations.
It is then able to produce accurate and fast predictions for any cosmological model within the parameter space covered by the training simulations.
In the past years, several emulators have been developed to predict the matter power spectrum in $\Lambda$CDM and some extended DE models~\citep[e.g.][]{Coyote_Lawrence_2010,agarwal_emu,Lawrence_2017,euclid_emu_1,euclid_emulator,bacco_emulator}.

When it comes to modified gravity theories, cosmological simulations require significantly more computing resources than in standard gravity.
Indeed, modified gravity simulations require to solve for the dynamics of at least one extra degree of freedom, which often exhibits a non-linear equation of motion~\citep[see][for a recent review on MG simulations]{llinares_review_MG}.

In this work, we focus on a particular MG model, the $f(R)$ gravity theory~\cite[see][for other MG models]{Clifton_2012_mg_review}.
It is part of the scalar-tensor family of MG theories, where the gravitational dynamics are described by the usual metric tensor and an extra scalar field.
More precisely, we work with a particular form of $f(R)$ gravity introduced by~\citet{Hu_Sawicki}.
More details are given in Section~\ref{sec:theory}.
The Hu \& Sawicki $f(R)$ gravity model is of cosmological interest since it is able to produce cosmic acceleration without an explicit cosmological constant.
Additionally, it exhibits the so-called chameleon screening mechanism~\citep{Khoury2004}, which hides the deviations from GR in dense environments such as the solar system.
Such a mechanism is required to pass laboratory and solar system scale precision tests of gravity.

Several parallel and adaptive N-body codes have been developed over the past years to run cosmological simulations in $f(R)$ gravity, such as \textsc{ecosmog}~\citep{Li2012}, \textsc{mg-gadget}~\citep{mg-gadget}, \textsc{isis}~\citep{isis}, a modified version of \textsc{arepo}~\citep{arnold_arepo,Hernandez-Aguayo:2020kgq} and \textsc{mg-glam}~\citep{Ruan:2021wup,Hernandez-Aguayo:2021kuh}.
A first emulator for the matter power spectrum in $f(R)$ gravity was introduced by~\cite{ramachandra_emulator}, named \textsc{mgemu}.
However, their emulator is based on COmoving Lagrangian Acceleration (COLA) simulations, which are less accurate than N-body methods in the non-linear regime.
{Very recently, generic pipelines to build emulators for the matter power spectrum in $f(R)$ gravity and othe MG theories using COLA simulations have developped~\citep{Mauland_2023, Fiorini_2023}.
To date, the only $f(R)$ emulator based on N-body simulations is the \textsc{forge} emulator~\citep{Arnold2021}, which is built on top of cosmological simulations run with \textsc{arepo}.

The purpose of this paper is to present a new emulator, dubbed as \textsc{e-mantis}\footnote{\url{https://doi.org/10.5281/zenodo.7738362}} (Emulator for Multiple observable ANalysis in extended cosmological TheorIeS), for the matter power spectrum boost in $f(R)$ gravity, defined as $B(k) = P_{f(R)}(k)/P_{\Lambda\mathrm{CDM}}(k)$.
The boost is less affected by statistical and systematic errors and has a smoother cosmological dependence than the power spectrum.
Therefore, by emulating the boost instead of the raw power spectrum, we significantly reduce our computational needs.
The simulations used to calibrate \textsc{e-mantis} are run with the modified gravity code \textsc{ecosmog}.
Our emulation strategy and simulation code are different than the ones used in \textsc{forge}.
We therefore expect both emulators to be complementary and we will investigate the differences within this article.
We also make a detailed and careful study of numerical systematic which is generally not present in publications dedicated to modified gravity emulators.

The paper is structured as follows.
Section~\ref{sec:theory} presents some theoretical aspects about $f(R)$ gravity and the Hu \& Sawicki model in the context of cosmological perturbations.
In Section~\ref{sec:methods}, we introduce the methodology and numerical codes used to run the N-body simulations and build the emulator.
Section~\ref{sec:conv_stud} estimates the accuracy of the emulator against statistical and systematic errors in the training data as well as pure emulation errors.
Finally, in Section~\ref{sec:results}, we compare the predictions from \textsc{e-mantis} to other prescriptions in the literature and give an example of usage.

\section{f(R) gravity}
\label{sec:theory}

In this section, we briefly introduce the basic equations of $f(R)$ gravity applied to cosmological perturbations and their phenomenology.
For more details, we refer the reader to reviews on the subject~\citep[e.g.][]{sotiriou_review, Clifton_2012_mg_review}.
In $f(R)$ gravity the Einstein-Hilbert part of the action is modified by the addition of a new arbitrary function $f$ of the Ricci scalar $R$
\begin{equation}\label{eq:modified_action}
  S = \frac{1}{2\kappa^{2}} \int \mathrm{d}^{4}x \sqrt{-g}\left[R + f(R) \right] + S_{m},
\end{equation}
where $\kappa^{2}=8\pi G/c^4$ with $G$ being Newton's gravitational constant and $c$ the speed of light in vacuum, $g$ is the determinant of the spacetime metric $g_{\mu\nu}$ and $S_{m}$ is the action describing the matter content of the Universe.
By extremising the total action with respect to the metric, we obtain the modified field equations
\begin{equation}\label{eq:modified_field_eqs}
  G_{\mu\nu} +f_{R}R_{\mu\nu} - \left(\frac{f}{2}-\Box{f_{R}}\right)g_{\mu\nu}-\nabla_{\mu}\nabla_{\nu}f_{R} = \kappa^{2} T_{\mu\nu},
\end{equation}
where $R_{\mu\nu}$ is the Ricci curvature tensor, $G_{\mu\nu}$ is the Einstein tensor, $T_{\mu\nu}$ is the matter stress-energy tensor, $\nabla_{\mu}$ is the covariant derivative compatible with the metric and $\Box \equiv g^{\mu\nu}\nabla_{\mu}\nabla_{\nu}$.
We have introduced the derivative of the $f$ function with respect to the Ricci scalar
\begin{equation}\label{eq:scalaron_def}
  f_{R} \equiv \frac{\mathrm{d}f\left(R\right)}{\mathrm{d}R}.
\end{equation}
This quantity is a new dynamical scalar field introduced by the $f(R)$ term in the modified gravitational action.
It is sometimes referred to as the scalaron.
The equation of motion for this scalar field can be obtained by taking the trace of equation~\eqref{eq:modified_field_eqs}
\begin{equation}\label{eq:scalaron_eom}
  \Box{f_{R}} = \frac{1}{3}\left(R-f_{R}R+2f+\kappa^{2}T\right),
\end{equation}
where $T=T^{\mu}_{\mu}$ is the trace of the stress-energy tensor.
We are interested in modeling the late-time evolution of the Universe.
Therefore, we make the usual assumption that the Universe is filled with a pressureless non-relativistic fluid called Cold Dark Matter (CDM).
In such a case, as long as we focus on sufficiently small scales, we can write $T=-\rho c^2$, where $\rho$ is the mass density of CDM~\citep{Chisari2011_New_vs_GR_sim}.

\subsection{Cosmological perturbations}%
\label{subsec:cosmo_pert}

We want to solve the field equations for cosmological scalar perturbations around an homogeneous Friedmann-Lemaître-Robertson-Walker (FLRW) background.
In order to do so, we use the Newtonian gauge where the line element is given by
\begin{equation}\label{eq:metric}
  \mathrm{d}s^{2} = -\left(1+2\psi/c^2\right)c^2\mathrm{d}t^{2} + a^{2}(t)\left(1-2\phi/c^2\right)\delta_{ij}\mathrm{d}x^{i}\mathrm{d}x^{j},
\end{equation}
where $\psi$ and $\phi$ are the two gauge invariant Bardeen potentials~\citep{Bardeen_1980} and $a$ is the scale factor.
We work in the weak field limit, i.e. $|\psi/c^2|, |\phi/c^2|, |f_{R}| \ll 1$.
We also adopt the quasi-static approximation, which means that we neglect all time derivatives of the three scalar fields with respect to their spatial ones.
The validity of this approximation for $f(R)$ gravity cosmological simulations has been confirmed by~\citet{Bose_2015}.
Under these approximations and by subtracting the background component from Eq.~\eqref{eq:scalaron_eom}, we obtain
\begin{equation}\label{eq:scalaron_eom_perturbed}
 \frac{1}{a^{2}}\vec{\nabla}^{2} f_{R} = \frac{1}{3}\left[\delta R\left(f_R\right) - \frac{8\pi G}{c^2}\delta\rho\right],
\end{equation}
where $\delta R \equiv R - \bar{R}$ and $\delta \rho \equiv \rho - \bar{\rho}$, the over-bars denoting background quantities, and $\vec{\nabla}$ is the gradient operator in euclidean space with respect to the comoving coordinate $\vec{x}$.
In order to obtain Eq.~\eqref{eq:scalaron_eom_perturbed}, we have used the fact that $\delta f\sim f_R \delta R \ll \delta R$.
From the $00$ component of Eq.~\eqref{eq:modified_field_eqs}, we derive the equation for the lapse function $\phi$,
\begin{equation}\label{eq:phi_eom}
\frac{1}{a^2}\vec{\nabla}^{2}\phi = \frac{8\pi G}{3}\delta\rho + \frac{c^2}{6}\delta R\left(f_R\right),
\end{equation}
where we have neglected terms of order $H^2\phi/c^2 \ll \partial_i^2\phi/a^2$.
Such an approximation remains valid as long as we focus on small scales such that $k \gg aH/c$, where $k$ is the comoving wavenumber.
In the same way, from the trace of the $ij$ component of Eq.~\eqref{eq:modified_field_eqs} we obtain
\begin{equation}\label{eq:psi-phi_eom}
    \frac{2}{a^2}\vec{\nabla}^{2}\left(\psi-\phi\right) = 8\pi G\delta\rho - c^2\delta R\left(f_R\right) + \frac{c^2}{a^2}\vec{\nabla}^{2}f_R.
\end{equation}
We can now combine the last three equations, which gives
\begin{equation}\label{eq:psi_eom}
    \frac{1}{a^2}\vec{\nabla}^{2}\psi = \frac{16\pi G}{3}\delta\rho - \frac{c^2}{6}\delta R\left(f_R\right).
\end{equation}
Finally, by combining Eqs.~\eqref{eq:phi_eom} and \eqref{eq:psi_eom}, one can obtain
\begin{equation}\label{eq:psi+psi_eom}
    \frac{1}{a^2}\vec{\nabla}^{2}\left(\psi+\phi\right) = 8\pi G\delta\rho.
\end{equation}
This equation shows that the relation between the potential in lensing studies and the matter distribution is the same as in GR. More details about the derivations of these equations can be found for example in~\citet{Hu_Sawicki, Oyaizu2008, Bose_2015}.

\subsection{The Hu \& Sawicki model}%
\label{subsec:hu_sawicki}

From now on we focus on a particular $f(R)$ model, proposed by~\citet{Hu_Sawicki}, which uses the functional form
\begin{equation}\label{eq:hu_sawicki}
  f(R) = -m^{2}\frac{c_{1}\left(R/m^{2}\right)^{n}}{c_{2}\left(R/m^{2}\right)^{n}+1},
\end{equation}
where $c_{1}, c_{2}$ and $n$ are dimensionless parameters and $m$ is a curvature scale.
This scale is taken to be
\begin{equation}\label{eq:mass_scale}
m^{2} \equiv \frac{\Omega_{m}H_{0}^{2}}{c^2},
\end{equation}
where $\Omega_{m}$ is the current fractional matter density and $H_{0}$ is the Hubble parameter today.

It is interesting to perform an expansion of the $f(R)$ function at high curvature with respect to $m^{2}$
\begin{equation}\label{eq:high_R_expansion}
 \lim_{R\gg m^{2}} f(R) \simeq -\frac{c_{1}}{c_{2}}m^{2} + \frac{c_{1}}{c_{2}^{2}}m^{2}\left(\frac{m^{2}}{R}\right)^{n}.
\end{equation}
First, we see that the Hu \& Sawicki model is equivalent to a cosmological constant in the limit where $c_{1}/c_{2}^{2}\to 0$ and $c_{1}/c_{2}$ is kept constant.
Therefore, this model is able to produce cosmic acceleration.
In order to closely match the $\Lambda$CDM expansion history one can fix
\begin{equation}\label{eq:c1/c2}
  \frac{c_{1}}{c_{2}} = 6\frac{\Omega_{\Lambda}}{\Omega_{m}},
\end{equation}
where $\Omega_{\Lambda}$ is the current fractional density of dark energy.

Secondly, at finite $c_{1}/c_{2}^{2}$ and fixed $n$, the deviation from a pure cosmological constant vanishes as the density increases.
This is an example of a screening mechanism~\citep[see for example][for a recent review]{Brax2022_review}, and in particular the so-called chameleon mechanism~\citep{Khoury2004}.
Such a mechanism allows the Hu \& Sawicki model to exhibit significant deviations from $\Lambda$CDM at cosmological scales while recovering GR in the Solar System, where the laws of gravity are more tightly constrained.
The remaining free parameters $n$ and $c_{1}/c_{2}^{2}$ control the efficiency of the chameleon screening.

The scalar field $f_{R}$ can be written in terms of the scalar curvature $R$ as
\begin{equation}\label{eq:scalaron_from_R}
  f_{R} = -n \frac{c_{1}\left(\frac{R}{m^{2}}\right)^{n-1}}{\left[c_{2}\left(\frac{R}{m^2}\right)^{n}+1\right]^{2}} \simeq -n \frac{c_{1}}{c_{2}^{2}}\left(\frac{m^{2}}{R}\right)^{n+1}.
\end{equation}
The second equality is an approximation valid in the high curvature regime $R\gg m^{2}$.
Such an approximation remains valid up to $z=0$ for the choice of $m^{2}$ given by equation~\eqref{eq:mass_scale} as explained in~\citet{Hu_Sawicki} and~\citet{Oyaizu2008}.
In practice, it is useful to replace the parameter $c_{1}/c_{2}^{2}$ by the background value of the scalaron field today $\fR$.
Indeed, by combining Eqs.~\eqref{eq:high_R_expansion}, \eqref{eq:c1/c2} and \eqref{eq:scalaron_from_R}, one can write in the high curvature limit,
\begin{equation}\label{eq:high_R_expansion_fR0}
    f(R) \simeq -2\Lambda + \frac{\fR}{n}\frac{R_0^{n+1}}{R^{n}},
\end{equation}
where $R_0$ is the current background value of the Ricci scalar and $\Lambda$ is the cosmological constant.

At the background level, the deviation from an equation of state $w=-1$ is of order $\fR$~\citep{Hu_Sawicki}.
In this work, the strongest modification of gravity that we consider is $\fR = -10^{-4}$.
As a consequence, for all practical purposes, the background expansion history is identical to $\Lambda$CDM.
Additionally, for such values of $\fR$, the effect of $f(R)$ gravity on the evolution of the matter density field at high redshift is negligible.
The differences between $f(R)$ and $\Lambda$CDM will arise at the level of the late-time non-linear matter clustering.

We restrict our study to the case $n=1$, which is the most common case in the literature.
Performing cosmological simulations with $n\neq1$ is also much more cpu-time consuming (see Section~\ref{subsec:codes}).
We refer the reader to other papers such as~\citet{ramachandra_emulator,Ruan:2021wup} for the case $n\neq1$.

\section{Methods}%
\label{sec:methods}

\subsection{Codes}%
\label{subsec:codes}

The suite of simulation was performed with the automated pipeline introduced in \citet{Blot2015} and further refined in \citet{Blot2021}.
In this study, we have extended it to $f(R)$ modified gravity.
Note that as stated above, the effect of modified gravity on the structure formation at high redshift (initial condition), and its impact on the background evolution are both negligible.
The main changes are therefore limited to the dynamical solver.

The linear matter power spectrum was computed using the \textsc{camb} Boltzmann solver \citep{Lewis2000}. Initial conditions were generated using \textsc{mpgrafic}~\citep{Prunet2008} which assumes gaussian random field and second-order Lagrangian perturbation theory (2LPT) \citep{Crocce2006}. The initial redshift was chosen as low as possible while avoiding particle crossing (in practice $z_i\sim 50$).
With this choice, numerical errors remain limited \citep{Michaux2020}.

Particles are evolved in a periodic cubic box using the $f(R)$ version of the \textsc{ramses} N-body code \citep{Teyssier2002} called \textsc{ecosmog} \citep{Li2012}.
\textsc{ramses} implements a particle-mesh (PM) method with adaptive mesh refinements (AMR) in overdense regions.
A mesh is refined (i.e. divided into $8$ sub-cells), whenever the number of particles inside it exceeds a given threshold $m_{\mathrm{ref}}$, up to a maximum refinement level $l_{\mathrm{max}}$.
The (linear) modified Poisson equation for the newtonian potential (Eq.~\eqref{eq:psi_eom}) is solved through a standard multigrid approach \citep{Guillet2011}.
On the other hand, the non-linear equation of motion for the extra scalar field (Eq.~\eqref{eq:scalaron_eom_perturbed}) was solved using a Newton-Gauss-Seidel method embedded in the multigrid approach \citep{Li2012}.
The solver has been further optimised in \citet{Bose2017} thanks to the existence of analytical solutions for the particular case $n=1$.
We have therefore used this optimised \textsc{ecosmog} version.
An important parameter is the required precision for the non-linear scalar field solver: we have set it to $\epsilon_{\mathrm{sf}}=10^{-7}$, and we have checked that with this choice, the errors on the power spectrum boost are much smaller than $1\%$ in the studied range of wavenumbers.

The matter power spectrum was estimated using \textsc{powergrid} \citep{Prunet2008} with a Cloud-In-Cell assignment (CIC).
We use a mesh grid with $64$ times more cells than the \textsc{ramses} coarse grid.
We have dropped the matter power spectrum values for wavenumbers beyond half the Nyquist frequency of the \textsc{powergrid} grid to avoid aliasing.

\subsection{Emulation}%
\label{subsec:emulation}

\subsubsection{The matter power spectrum boost}
\label{subsubsec:boost}
Instead of emulating the matter power spectrum itself, we decide to focus on the matter power spectrum boost defined as
\begin{equation}\label{eq:boost_def}
  B(k) = P_{f(R)}(k)/P_{\Lambda\mathrm{CDM}}(k).
\end{equation}
This quantity encodes the enhancement of the matter power spectrum due to $f(R)$ gravity with respect to a $\Lambda$CDM scenario.
It brings several benefits that simplify the emulation procedure.
First, the dynamical range of the power spectrum boost is significantly smaller than the one of the full power spectrum.
We go from emulating a quantity spanning several orders of magnitude to one that is of order $\mathcal{O}(1)$ at all scales, which is a simpler task to perform~\citep{bacco_emulator}.

Second, we have verified that the power spectrum boost is mostly independent of the baryon density parameter $\Omega_{b}$, the reduced Hubble parameter $h$\footnote{Defined as $H_{0}= 100h$ km/s/Mpc, where $H_{0}$ is the Hubble constant today.} and the primordial spectral index $n_{s}$, even at deeply non-linear scales.
This is in agreement with previous studies~\citep{winther_emulator}.
In order to build an emulator, we only need to consider the $3$D parameter space given by the background value of the scalar field today $\fR$, the total matter density parameter $\OmegaM$, and the primordial power spectrum normalisation\footnote{
  We follow the convention of previous works in the field, and by $\sigma_{8}$ we refer to the value obtained assuming a linear $\Lambda$CDM evolution, even for $f(R)$CDM cosmologies.
  The parameter $\sigma_{8}$ is used as an indirect normalisation of the primordial power spectrum, which is therefore the same in both $\Lambda$CDM and $f(R)$CDM for a given set of cosmological parameters.
  }$\sigma_{8}$.
The dependence of the power spectrum boost on the cosmological parameters is discussed in more details in Section~\ref{subsubsec:parameter_range}.

Third, by taking the ratio of the power spectrum obtained in $f(R)$ gravity and $\Lambda$CDM from the same exact initial conditions, we obtain a significant cancellation of cosmic variance and mass resolution errors, which dominate the error budget at large and small scales respectively.
With this method, we can build an accurate emulator without having to perform large volume and very high resolution simulations, as it would be necessary for the raw matter power spectrum.
More details are given in Section~\ref{subsec:sys_sym_err}.

An important caveat of emulating the boost is that in order to get the full power spectrum in $f(R)$ models, it is necessary to combine the predictions of \textsc{e-mantis} with an independent $\Lambda$CDM emulator.
However, current state-of-the-art $\Lambda$CDM emulators usually achieve per cent level accuracy~\citep{bacco_emulator, euclid_emulator_2, CosmicEmu_IV}.

\subsubsection{Defining the parameter range}%
\label{subsubsec:parameter_range}

\begin{table}
  \caption{Cosmological models used in the \textit{cosmo} simulation suite.
    The first line (in italics) is our reference model F$5(\star)$.
    Models $4-13$ vary one of the $5$ $\Lambda$CDM parameters by $\pm 25\%$.
    Models $2$ (F$4$) and $3$ (F$6$) vary the $\fR$ parameter.
    The parameters in bold are the ones that are being varied by a specific model.
    The values of $\Omega_{\Lambda}$ are not explicitly given, but they are fixed by the flatness condition $\Omega_{\Lambda} = 1 - \Omega_{m}-\Omega_r$.
    The radiation density parameter is fixed by the CMB temperature to $\Omega_{r}\sim10^{-4}$.
    For each model in this table we run the corresponding $\Lambda$CDM simulations in addition to the $f(R)$ ones.
    }
  \label{tab:cosmo_suite}
  \begin{tabular}{|c|c|c|c|c|c|c|}
    \hline\hline
    Model & $|\fR$| & $\Omega_{m}$ & $\sigma_{8}$ & $h$ & $\Omega_{b}$ & $n_{s}$ \\
    \hline\hline
    $\mathit{F5}(\star)$ & $\mathit{10^{-5}}$ & $\mathit{0.3153}$ & $\mathit{0.8111}$ & $\mathit{0.6736}$ & $\mathit{0.049302}$ & $\mathit{0.9649}$ \\
    \hline
    F$4$ & $\mathbf{10^{-4}}$ & $0.3153$ & $0.8111$ & $0.6736$ & $0.049302$ & $0.9649$ \\
    \hline
    F$6$ & $\mathbf{10^{-6}}$ & $0.3153$ & $0.8111$ & $0.6736$ & $0.049302$ & $0.9649$ \\
    \hline
    $\Omega_{m}^{+}$ & $10^{-5}$ & $\mathbf{0.3941}$ & $0.8111$ & $0.6736$ & $0.049302$ & $0.9649$ \\
    \hline
    $\Omega_{m}^{-}$ & $10^{-5}$ & $\mathbf{0.2365}$ & $0.8111$ & $0.6736$ & $0.049302$ & $0.9649$ \\
    \hline
    $\sigma_{8}^{+}$ & $10^{-5}$ & $0.3153$ & $\mathbf{1.0140}$ & $0.6736$ & $0.049302$ & $0.9649$ \\
    \hline
    $\sigma_{8}^{-}$ & $10^{-5}$ & $0.3153$ & $\mathbf{0.6083}$ & $0.6736$ & $0.049302$ & $0.9649$ \\
    \hline
    $h^{+}$ & $10^{-5}$ & $0.3153$ & $0.8111$ & $\mathbf{0.8420}$ & $0.049302$ & $0.9649$ \\
    \hline
    $h^{-}$ & $10^{-5}$ & $0.3153$ & $0.8111$ & $\mathbf{0.5052}$ & $0.049302$ & $0.9649$ \\
    \hline
    $\Omega_{b}^{+}$ & $10^{-5}$ & $0.3153$ & $0.8111$ & $0.6736$ & $\mathbf{0.061620}$ & $0.9649$ \\
    \hline
    $\Omega_{b}^{-}$ & $10^{-5}$ & $0.3153$ & $0.8111$ & $0.6736$ & $\mathbf{0.036980}$ & $0.9649$ \\
    \hline
    $n_{s}^{+}$ & $10^{-5}$ & $0.3153$ & $0.8111$ & $0.6736$ & $0.049302$ & $\mathbf{1.2060}$ \\
    \hline
    $n_{s}^{-}$ & $10^{-5}$ & $0.3153$ & $0.8111$ & $0.6736$ & $0.049302$ & $\mathbf{0.7237}$ \\
    \hline
  \end{tabular}
\end{table}

\begin{figure*}
  \centering
  \begin{subfigure}{0.33\linewidth}
    \includegraphics[width=\linewidth]{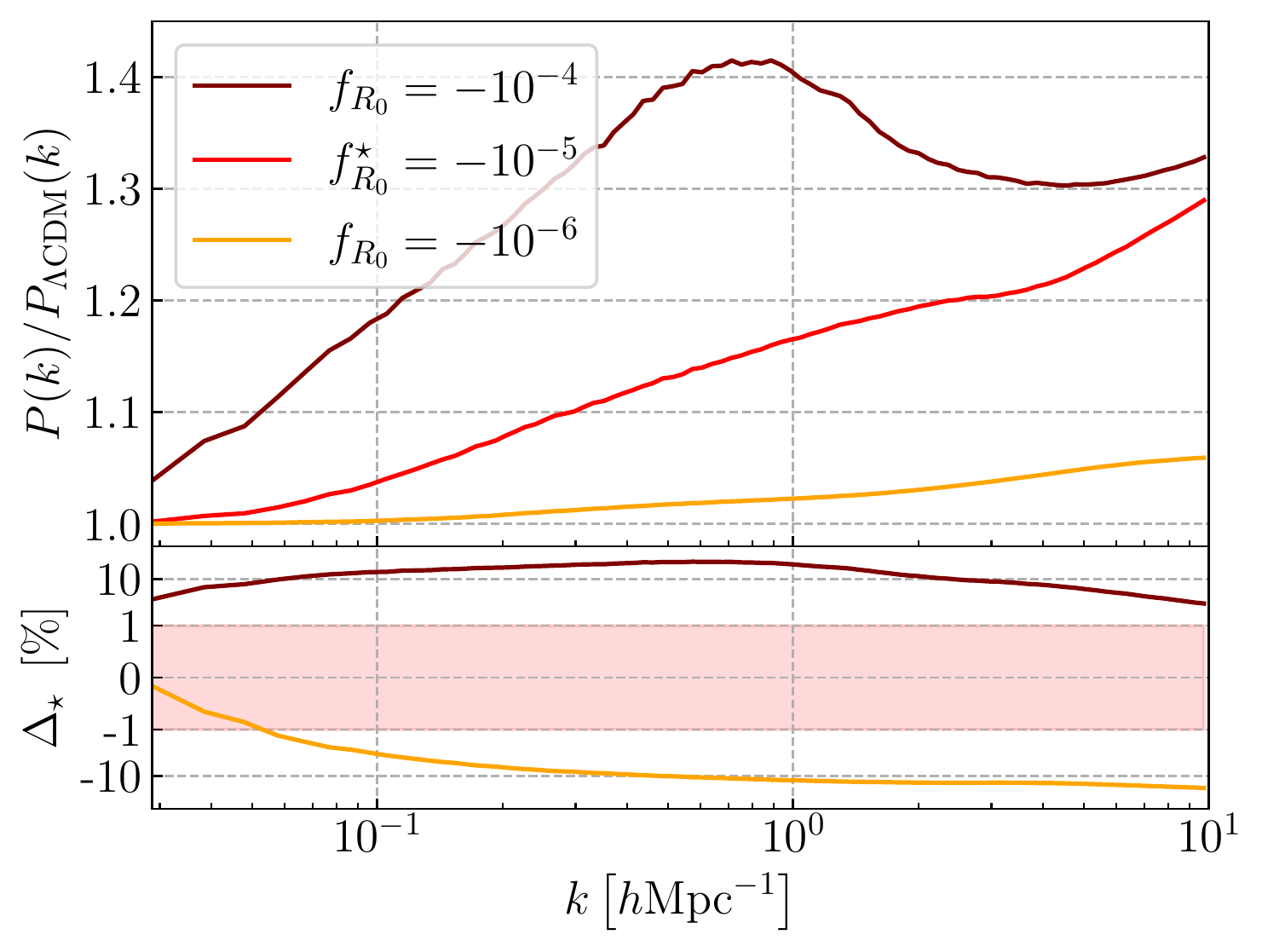}
  \end{subfigure}
  \begin{subfigure}{0.33\linewidth}
    \includegraphics[width=\linewidth]{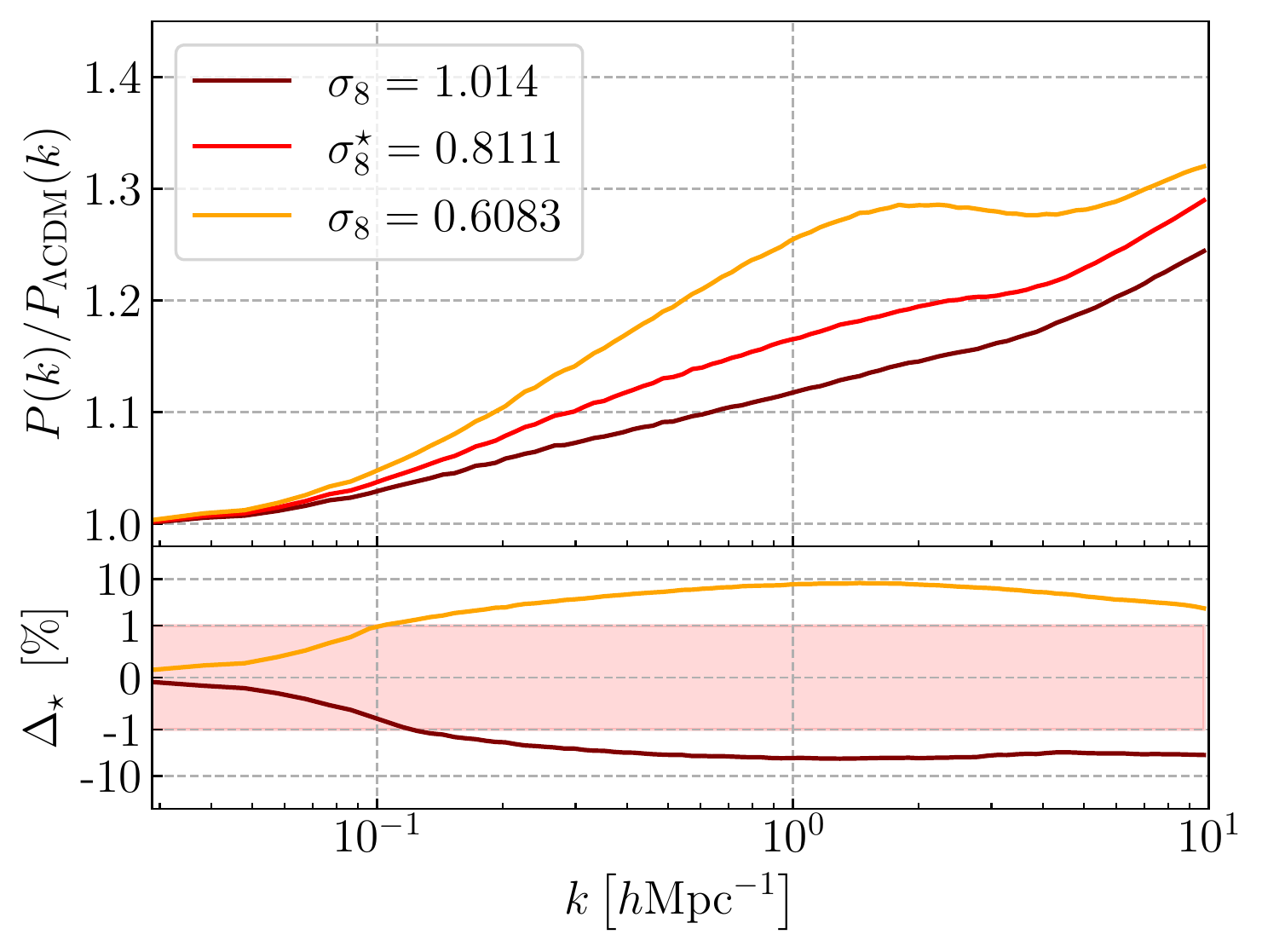}
  \end{subfigure}
  \begin{subfigure}{0.33\linewidth}
    \includegraphics[width=\linewidth]{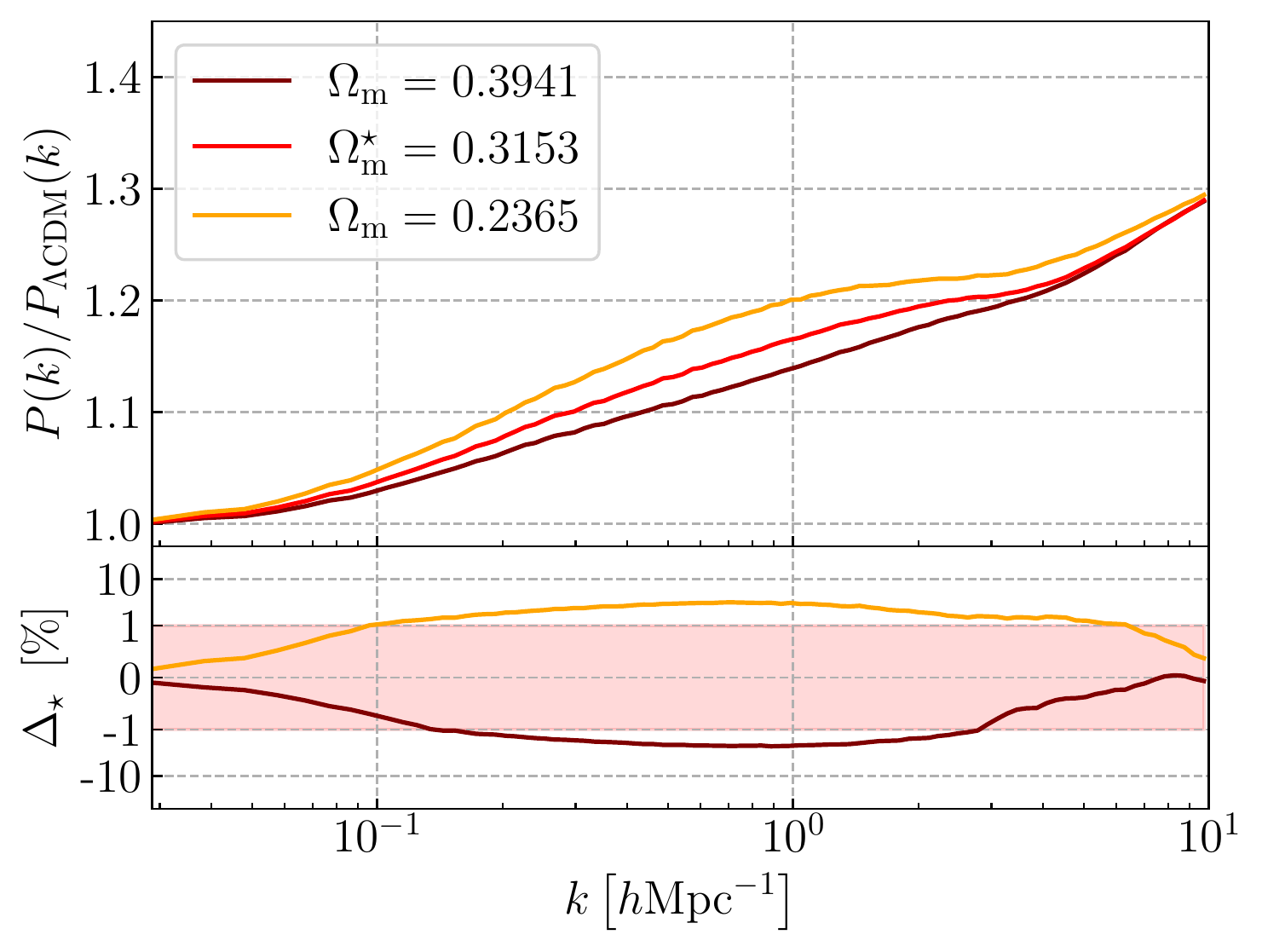}
  \end{subfigure}

  \begin{subfigure}{0.33\linewidth}
    \includegraphics[width=\linewidth]{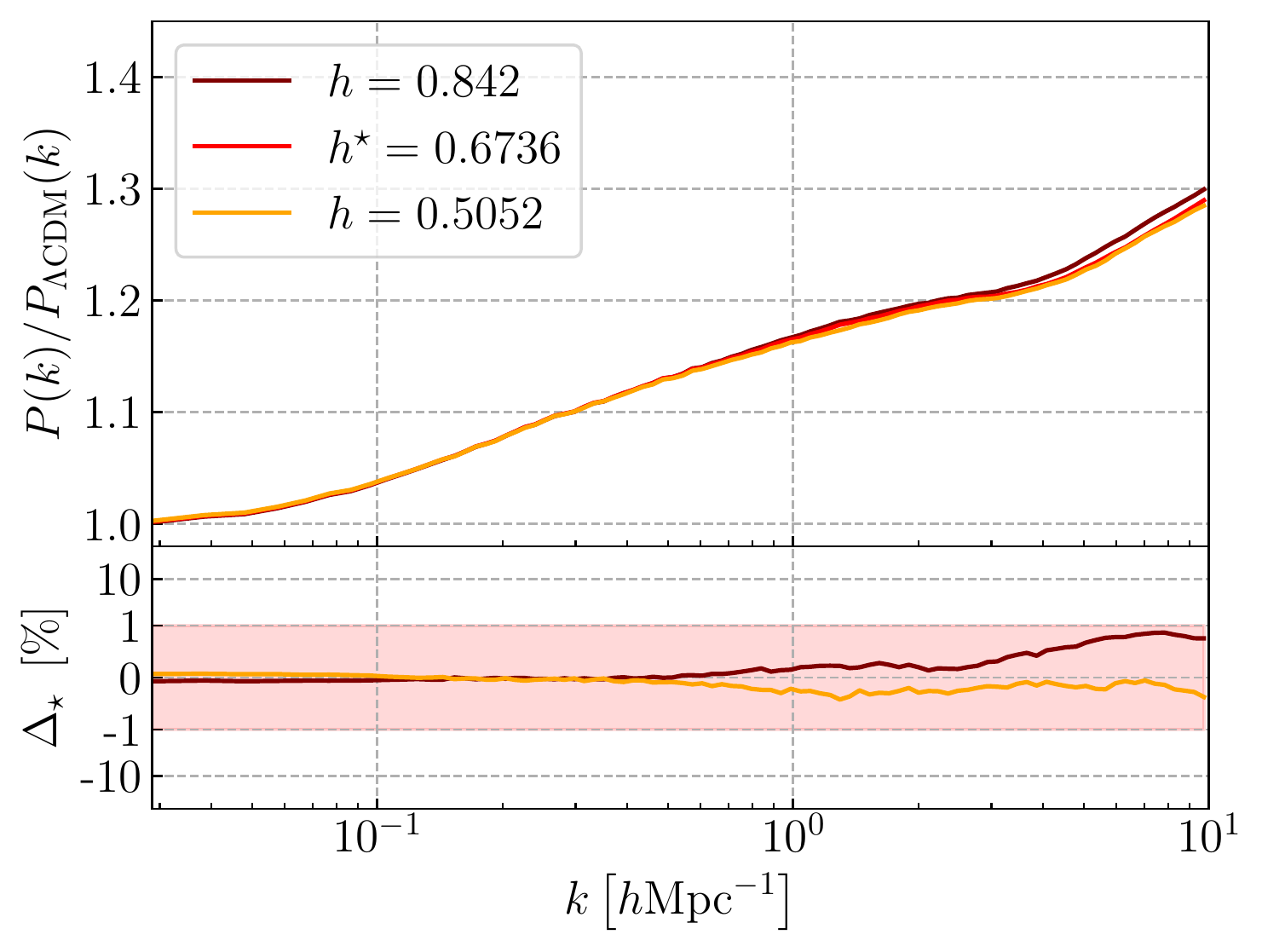}
  \end{subfigure}
  \begin{subfigure}{0.33\linewidth}
    \includegraphics[width=\linewidth]{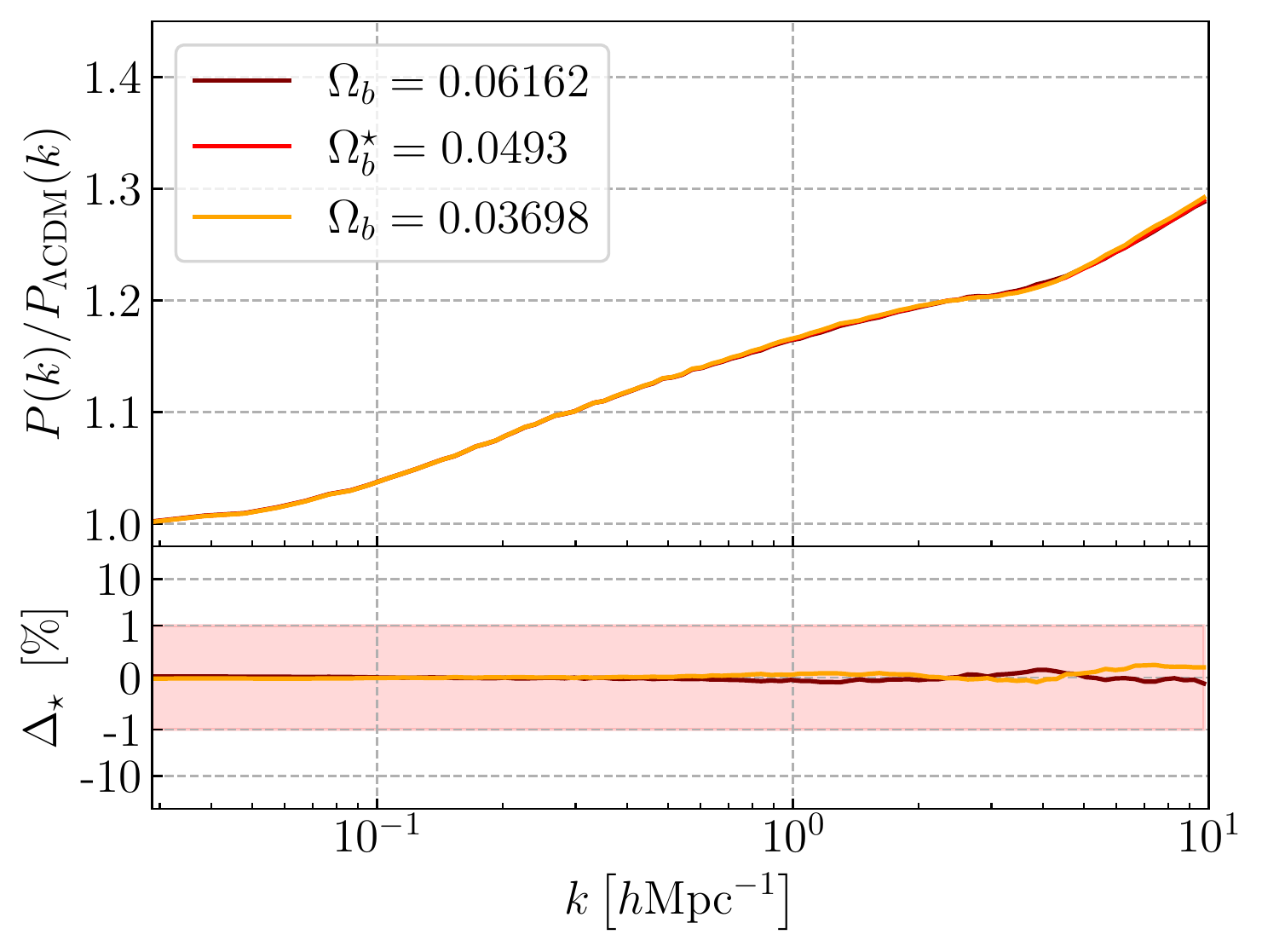}
  \end{subfigure}
  \begin{subfigure}{0.33\linewidth}
    \includegraphics[width=\linewidth]{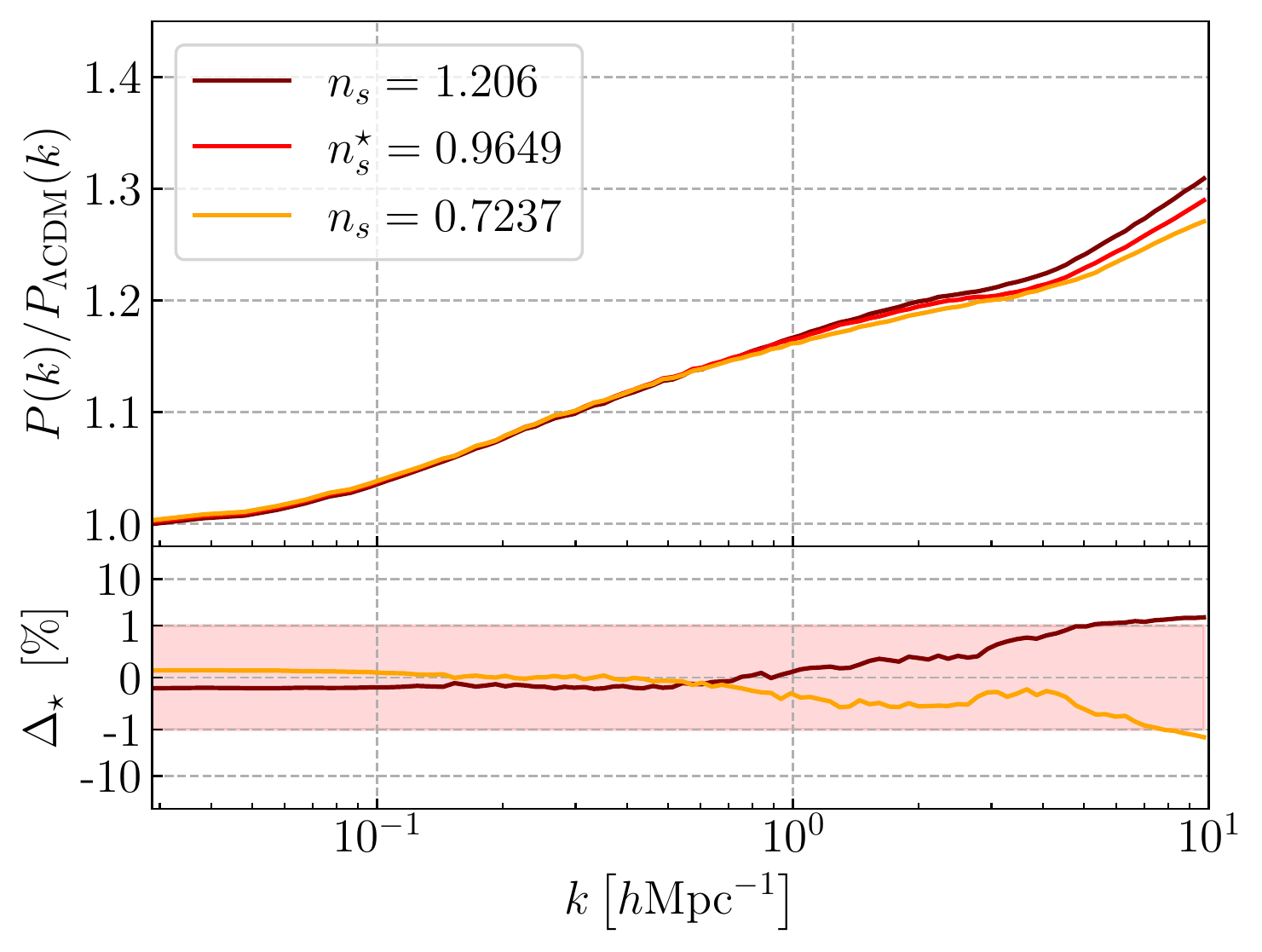}
  \end{subfigure}
  \caption{Cosmological dependence of the matter power spectrum boost at $z=0$.
    In each subplot, one of the $6$ cosmological parameters (as indicated in the inset) is varied by $\pm25\%$ around the reference model F$5(\star)$ from Table~\ref{tab:cosmo_suite}.
    The red line corresponds to F$5(\star)$, the brown line to the $+25\%$ variation and the orange line to the $-25\%$ variation.
    The lower panels give the relative difference with respect to the reference.
    The light red bands mark the $\pm1\%$ variation limit.
    Inside it the $y$ axis scaling is linear and outside it is logarithmic.
    Only the parameters $\fR$, $\sigma_{8}$ and $\Omega_{m}$ have an impact greater than $1\%$ on the matter power spectrum boost.
    We can therefore neglect the effect of the other parameters when building the emulator and keep them fixed to their reference values.
  }
  \label{fig:cosmo_dependence}
\end{figure*}

Before fixing the emulation parameter space, we explore the dependence of the matter power spectrum boost with the $6$ cosmological parameters $\{\fR, \Omega_{m}, \sigma_{8}, h, \Omega_{b}, n_{s}\}$.
Our aim is to show that in order to emulate the matter power spectrum boost, we only need to consider a subset of these parameters.

We define two reference cosmological models (one with usual gravity and the other one with modified gravity): the $\Lambda$CDM parameters take the best fit values from~\citet{Planck18}, while the corresponding $f(R)$CDM parameters takes the same values but with $\fR = -10^{-5}$.
From this reference we build pairs of models where each $\Lambda$CDM parameter is varied by $\pm 25\%$ at a time.
We also study models with $\fR = -10^{-4}$ (F4) and $\fR = -10^{-6}$ (F6) to investigate the effect of $\fR$ (corresponding to a variation of $\pm 20\%$ for $\log{|\fR|}$ while matching standard values found in the literature for comparison).
We always consider a flat Universe such that $\Omega_{\Lambda}=1-\Omega_{m}-\Omega_{r}$ and take a constant radiation density parameter fixed by the CMB temperature to $\Omega_{r}\sim10^{-4}$.
The parameter values of these models are summarized in Table~\ref{tab:cosmo_suite}.

For each model, we evolve $N_{\mathrm{part}}=512^{3}$ DM particles in a periodic cubic box of length size $L_{\mathrm{box}} = 328.125 \, h^{-1}\mathrm{Mpc}$ in a $\Lambda$CDM and a $f(R)$ simulation.
Both simulations are initialised using the exact same initial conditions and same resolution.
We perform such pairs of simulations for $5$ independent initial conditions, which are shared across the different models.
In total, we realise a set of $65$ pairs of $\Lambda$CDM \& $f(R)$ simulations to study the cosmological dependence of the matter power spectrum boost.
We refer to this ensemble of simulations as the \textit{cosmo} simulation suite.
From this simulation suite, we measure the matter power spectrum boost for each cosmology and average it over the $5$ realisations.
The numerical parameters of these simulations are the same as the \textit{emulator} simulation suite that is introduced in Section~\ref{subsec:simu_set_emu} (see Table~\ref{tab:emulation_suite}).

Figure~\ref{fig:cosmo_dependence} shows the dependence of the matter power spectrum boost with each cosmological parameter.
Each subplot varies one of the $6$ parameters independently from the others.
The lower panels show the relative difference with respect to the reference model F$5(\star)$.
These results show that the variation of the boost with the parameters $h$, $\Omega_{b}$, and $n_{s}$ is very weak up to small scales of order $k\sim 10$ $h \mathrm{Mpc}^{-1}$.
For $\Omega_{b}$ and $h$, the relative variation of the boost remains smaller than $0.2\%$ and $1\%$ respectively at all scales considered, i.e. for $0.03 \ h\mathrm{Mpc}^{-1} < k < 10 \ h\mathrm{Mpc}^{-1}$.
In the case of $n_{s}$, the relative variation is smaller than $1\%$ except at the smallest scales, where it starts to cross the $1\%$ line.
We stress here the fact that we have performed very strong variations of the cosmological parameters.
If we had considered only variations within the current observational constraints~\citep{Planck18}, the effect on the boost would have been completely negligible.
The weak dependence on $h$ shown in Fig.~\ref{fig:cosmo_dependence} explains why \citet{Harnois-Deraps:2022bie} find, in their forecast, that the weak lensing convergence power spectrum places very weak constraint on this parameter.

As expected the parameter with the strongest impact is $\fR$, which for the F$4$ model, can lead to an enhancement of the matter power spectrum of up to $40\%$ with respect to the $\Lambda$CDM scenario.
The parameters $\sigma_{8}$ and $\Omega_{m}$ lead to variations of up to $8\%$ and $3\%$ respectively.
These results are in qualitative agreement with previous works such as~\citet{winther_emulator}.

One can interpret why $\sigma_{8}$ and $\Omega_{m}$ play an important role in terms of the screening mechanism.
For instance, if the amount of matter in the Universe increases, the overall density does the same, and because of the chameleon screening mechanism, the effect of $f(R)$ gravity gets weaker.
Indeed, we can see in Figure~\ref{fig:cosmo_dependence} that the matter power spectrum boost is a decreasing function of $\Omega_{m}$, the other parameters being fixed.
The same goes for $\sigma_{8}$.
With a higher $\sigma_{8}$ the amount of clustering in the Universe increases and overdensities are globally denser.
The screening is therefore stronger and the effect of $f(R)$ gravity is weaker.

\begin{figure*}
  \begin{subfigure}{0.49\linewidth}
    \centering
    \includegraphics[width=\linewidth]{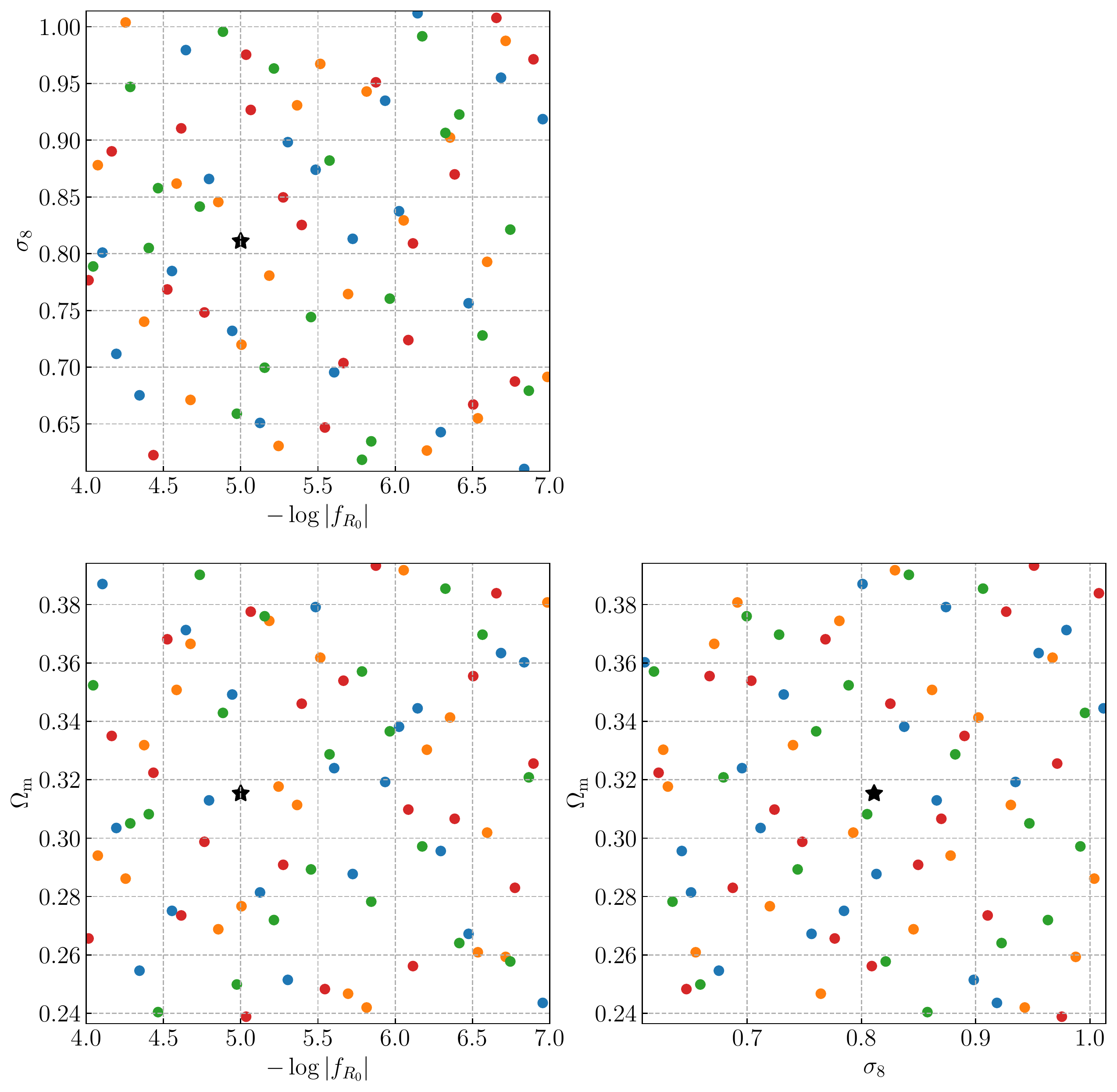}
  \end{subfigure}
  \begin{subfigure}{0.49\linewidth}
    \centering
    \includegraphics[width=\linewidth]{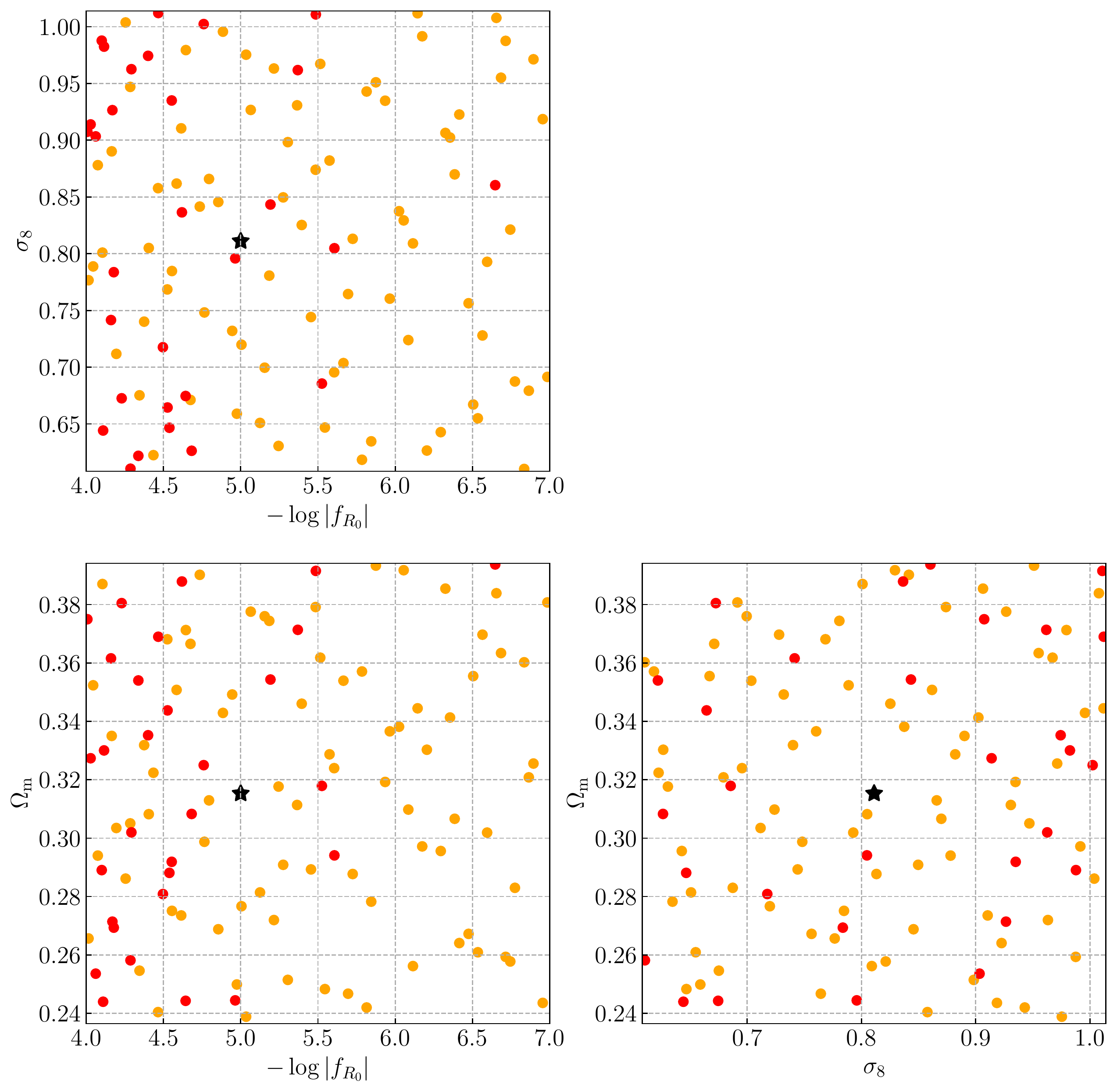}
  \end{subfigure}
  \caption{\label{fig:SLHD} $2$D projections of the model points used to train and test the emulator.
  \textit{Left:} each color represents one of the first four slices of the primary Sliced Latin Hypercube Design (SLHD).
  Each slice is composed of $20$ points, forming a LHD on the 3D parameter space $\{\OmegaM, \sigma_{8}, \fR\}$.
  The black star corresponds to our reference F5 cosmology.
  \textit{Right:} training points of the final emulator.
  The points in orange correspond to the four slices of the primary SLHD.
  The red ones are the additional training models drown from the refinement SLHD by the iterative procedure described at the end of Section~\ref{subsubsec:sampling}.
  They are mostly distributed towards the high values of $\fR$, while staying uniform in $\OmegaM$ and $\sigma_8$.
  This is the region where the emulation errors are larger.
  }
\end{figure*}

We have established the impact of each individual cosmological parameter on the matter power spectrum boost around the reference F$5(\star)$model.
In particular, the parameters $n_s$, $h$ and $\Omega_b$ have an effect smaller than $1\%$, even for very large variations of those parameters.
We could therefore ignore them in the construction of the emulator.
However, the simultaneous variation of multiple parameters might produce a non-negligible effect.
In Appendix~\ref{app:cosmo_cross_dependence}, we investigate the evolution of errors in the most sensitive parts of the parameter space.
We find that the cost of ignoring those parameters on the accuracy of the power spectrum boost determination is still smaller than $1\%$ in most cases.
In the worst cases, mostly around the edges of our wide cosmological parameter space, the induced error can reach the $2-3\%$ level.
As we show in Section~\ref{subsubsec:res_effects}, this is less than the systematic errors in our training data.

From this study, we can conclude that in order to build an emulator, with a few percent level accuracy, for the matter power spectrum boost, it is enough to focus on the three parameters $\{\fR, \Omega_{m}, \sigma_{8}\}$.
Considering a $3$D parameter space, instead of the full $6$-dimensional, reduces the number of simulations we need to perform in order to achieve a given emulation accuracy.
Indeed, performing an interpolation in $3$ dimensions requires less samples than in $6$ dimensions.
Given the computational cost of carrying out N-body simulations in $f(R)$ gravity, this is an important aspect.

Ideally, we would build an emulator that covers all possible cosmological models within this 3-dimensional parameter space.
However, we can only emulate a finite volume of it.
This volume needs to be large enough to accommodate all reasonable cosmological models as well as some margin around them, so that the emulator can be properly used in any theoretical, numerical or observational analysis.
We decide to consider variations of $\Omega_{m}$ and $\sigma_{8}$ of up to $25\%$ around the current best fit values in $\Lambda$CDM from~\citet{Planck18}.
This range is enough to explore very conservative deviations of order $\sim5\sigma$ from the constraints obtained in $f(R)$ gravity with the galaxy clustering ratio~\citep{Bel2015_clustering_ratio} for $\Omega_{m}$ and with galaxy clusters~\citep{Cataneo2015_galaxy_clusters} for $\sigma_{8}$.
Both studies exclude models with $-\fR \geq 10^{-4}$ at the $2\sigma$ level.
We will use this value as the upper bound for $\fR$.
For the lower bound we take $\fR = -10^{-7}$, which corresponds to an almost vanishing modification of the matter power spectrum with respect to $\Lambda$CDM.
In the end the emulator covers the parameter range given by
\begin{align}
  -7  & \leq \log{|\fR|}  \leq  -4, \\
  0.2365 & \leq \Omega_{m} \leq  0.3941, \\
  0.6083 & \leq \sigma_{8} \leq  1.0140.
\end{align}
Such large range of variation of the parameters is also important to be able to measure their impact on the boost.
This is key to accurately estimate the derivatives of the power spectrum boost with respect to the cosmological parameters, so as to make an accurate interpolation.
In practice \textsc{e-mantis} works with $-\log{\left|\fR\right|}$ instead of $\fR$, since it improves its numerical accuracy.
We fix the other parameters to their best fit values from~\cite{Planck18}: $h=0.6736$, $\Omega_b=0.049302$ and $n_s=0.9649$.

We conclude this Section by noting that, as mentioned in Section~\ref{subsec:codes}, we limit our study to the case where the modified gravity parameter $n=1$. 
This is mainly due to computational limitations, since performing N-body simulations for other values of this parameter is much more time consuming~\citep{Bose2017}.
However, \citet{ramachandra_emulator} shows that a variation of $\pm 25\%$ around the value $n=1$ has an effect smaller than $1\%$ on the matter power spectrum boost.
Therefore, our emulator would also be valid in that regime, at least for scales $k\lesssim 1 \ h\mathrm{Mpc}^{-1}$, which is the range of validity of their predictions.
Moreoever this work suggests that the most sensitive part of the boost with respect to a variation of $n$ is the one at intermediate scales near $k\simeq 0.5\ h\mathrm{Mpc}^{-1}$. 

\subsubsection{Sampling the parameter space}
\label{subsubsec:sampling}

\begin{table*}
  \caption{
    Table summarising the main characteristics of the \textsc{emulator} simulation suite.
    $L_{\mathrm{box}}$ is the side length of the simulation box, $N_{\mathrm{part}}$ the number of DM particles, $m_{\mathrm{part}}$ the mass resolution (which varies depending on the particular cosmology, here are given the two most extreme values), $l_{\mathrm{min}}$ the coarse refinement level (corresponding to a coarse grid of $512^{3}$), $l_{\mathrm{max}}$ the maximum refinement level, $m_{\mathrm{ref}}$ the refinement threshold, $\Delta x$ the spatial resolution of the maximum refinement level, $\epsilon_\mathrm{p}$ and $\epsilon_\mathrm{sf}$ are the convergence criteria for the modified Poisson equation (Eq~\eqref{eq:psi_eom}) and the extra scalar field equation of motion (Eq.~\eqref{eq:scalaron_eom_perturbed}) respectively, $N_{\mathrm{real}}$ the number of realisations per model, $N_{\mathrm{cosmo}}$ the number of cosmological models (the $\times2$ factor accounts for the corresponding $\Lambda$CDM simulations performed for each $f(R)$ model) and $z_{i}$ the approximate initial redshift (which is also cosmology dependent).
  }
  \label{tab:emulation_suite}
  \begin{tabular}{|c|c|c|c|c|c|c|c|c|c|c|}
    \hline\hline
    $L_{\mathrm{box}} [h^{-1}\mathrm{Mpc}]$ & $N_{\mathrm{part}}$ & $m_{\mathrm{part}} [h^{-1}M_{\odot}]$ & $l_{\mathrm{min}}$ & $l_{\mathrm{max}}$ & $m_{\mathrm{ref}}$ & $\Delta x [h^{-1}\mathrm{kpc}]$ & $\epsilon_\mathrm{p}(\epsilon_\mathrm{sf})$ & $N_{\mathrm{real}}$ & $N_{\mathrm{cosmo}}$ & $z_{i}$ \\
    \hline\hline
    $328.125$ & $512^{3}$ & $\left(1.73-2.88\right)\times10^{10}$ & $9$ & $15$ & $14$ & $10$ & $10^{-4}(10^{-7})$ & $5$ & $110 (\times 2)$ & $\sim50$ \\
    \hline
  \end{tabular}
\end{table*}

We now detail the strategy employed to sample the model parameter space with training points.
This ensemble of training models is usually referred to as the Experimental Design (ED).
Since N-body simulations are highly time consuming, we are limited by the number that can be performed.
Usually, cosmological emulators use a number of training models of order $\mathcal{O}(10-100)$~\citep{Nishimichi_2019, euclid_emulator_2}.
In order to efficiently sample the parameter space with a reduced number of points, it is common to use a Latin Hypercube Sampling (LHS) method.
It has been shown to give good performances when combined with a Gaussian Process (GP) emulation~\citep{habib_emulator}.

A Latin Hypercube Design (LHD) is generated by first dividing the D-dimensional parameter space into a $N^{D}$ regular grid.
Then $N$ grid cells are randomly selected with the condition that each cell cannot have any common coordinate with the other ones.
Finally, within each selected grid cell a point is randomly sampled~\citep{LHS_review}.
Usually, additional criteria are imposed.
For instance, one can generate a set of $k$ independent LHD and then select the one that maximises the minimal distance between sample points (maximin-distance condition).
This kind of method ensures that the ED optimally covers the whole parameter space in a homogeneous way.

One possible modification of the LHS method is the maximin-distance Sliced LHD (SLHD) introduced by~\citet{LHS_sliced}.
A LHD is divided into sub-samples with an equal number of points each.
The particularity is that each sub-sample, also referred to as a \textit{slice}, verifies the LHD condition on its own.
A version of the maximin-distance condition is also applied, where the maximised distance is a mixture of the distances between points in each slice and in the whole sample.
We refer the reader to the original paper for more details about the precise implementation.
The SLHD sampling technique was used in~\citet{Nishimichi_2019} in order to separate the ED points into a training and a validation sample.
In this way, the validation points are homogeneously distributed over the parameter space, while at the same time being as far as possible from the training points.
In this work, we use the R package SLHD\footnote{\url{https://CRAN.R-project.org/package=SLHD}} to generate a SLHD with $5$ slices of $20$ points each (hereafter named \textit{primary} SLHD).
We have performed cosmological simulations only for the first four slices (i.e. $80$ cosmological models).
Indeed, we found that it was more beneficial to sample additional points by targeting specific regions of the parameter space, instead of adding an additional set of homogeneously distributed models (see next paragraph).
The left panel of Figure~\ref{fig:SLHD} shows the distribution of models of the first four slices from the \textit{primary} SLHD.
Each slice is represented by a different colour.
We have sampled $\OmegaM$ and $\sigma_{8}$ with an homogeneous prior.
For $\fR$, we sample homogeneously in $\log\left|\fR\right|$ instead.

Sampling training points with a LHS method is a good strategy when we have no a priori knowledge of how the signal that we want to emulate varies over the whole parameter space.
However, if the gradient (or its derivatives) of the signal is stronger in a particular region, it will require a higher density of training points to get an accurate emulation.
In this work, we use a strategy similar to the one followed by~\citet{bacco_emulator} in order to add training points in the regions where emulation is more difficult.
We start by building a preliminary emulator using $3$ slices (i.e. a total of $60$ training models) from the primary SLHD.
We can use the remaining slice (i.e. $20$ validation models) to assess the accuracy of this emulator and decide whether or not it is necessary to add new training points.

Since additional training points are needed, we generate a second SLHD (hereafter the \textit{refinement} SLHD) with $100$ models per slice.
In order to sample new training points from these slices, we follow an iterative procedure.
First, we randomly select $10$ points from the first slice, with a selection probability proportional to the emulation error.
We use the interpolation error directly given by the GP itself.
Even though this estimation is not exact, it gives a good order of magnitude for our purpose here.
We then re-train the emulator with the newly selected training models.
Finally, we assess once again the accuracy of the emulator with the validation models.
We repeat this procedure until the desired accuracy is obtained.

In the end, we add a total of $30$ training models using this method.
With the obtained gain in emulation accuracy, the emulation errors are now smaller than $1\%$ even for the most extreme models (see Appendix~\ref{app:emu_err_nreal_refinements} for more details).
The final version of the emulator is trained using all available simulated models, i.e. the $80$ models from the first four slices of the primary SLHD and the additional $30$ refinement models.
The final distribution of the $110$ training points is shown in the right panel of Figure~\ref{fig:SLHD}.

\subsubsection{Smoothing the power spectrum boost}
\label{subsubsec:smoothing_boost}

For each cosmological node, we measure, from our simulations, the binned matter power spectrum in $f(R)$ gravity and the corresponding $\Lambda$CDM cosmology.
These raw power spectra need some treatment before they can be used to train the emulator.
Indeed, they are noisy due to the fine binning in $k$ and the finite volume of the simulations.
Such noise reduces the accuracy of the emulator.

In order to smooth out the training data it is common to use filtering techniques~\citep[e.g.][]{ramachandra_emulator, Arnold2021}.
In this work, we adopt a different strategy.
We start by rebinning the power spectra into larger $k$ bins.
We adopt a constant linear binning with $\Delta k \simeq 0.009 h \mathrm{Mpc}^{-1}$ from $k=0.024 h\mathrm{Mpc}^{-1}$ to $k=0.18 h\mathrm{Mpc}^{-1}$ (in terms of bin edges).
For the remaining scales we use a constant logarithmic binning up to $k=10 h\mathrm{Mpc}^{-1}$.
In order to avoid a discontinuity in the bin widths, the first logarithmic bin has the same size as the linear bins.
The rebinned power spectra have a number of bins $N_{k}=90$ instead of the initial $N_{\mathrm{raw}}=1040$ bins.
We then compute the boosts by taking the ratio of the power spectra in $f(R)$CDM with $\Lambda$CDM.
Additionally, we simulate a number $N_{\mathrm{real}}=5$ of independent realisations per cosmological model and average the boosts over them.
In the end, we obtain a signal that is smooth enough to train the emulator and with enough bins in $k$.
Appendix~\ref{app:emu_err_nreal_refinements} shows how the number of realisations impacts the emulation accuracy.

\subsubsection{Emulation strategy}
\label{subsubsec:emulation_strategy}

We detail in this section the procedure adopted to interpolate between the training cosmological models.
We focus initially on a single redshift.
The strategy to extend the emulator to arbitrary redshifts is explained at the end.

Our training data are the smoothed power spectrum boosts from the cosmological nodes.
They each have $N_{k}=90$ bins in $k$.
Emulating every bin independently is a quite inefficient strategy.
Indeed, the different $k$-bins are highly correlated with each other with respect to the variation of the cosmological parameters.
We can therefore use a Principal Component Analysis (PCA) to reduce the dimensionality of the training data.
In a PCA, the training data is decomposed in a basis of orthogonal functions
\begin{equation}\label{eq:PCA}
B(\theta; k) = \sum_{i=1}^{N_{\mathrm{PCA}}}\alpha_{i} (\theta)\phi_{i}(k) +\epsilon,
\end{equation}
where $\theta$ represents the vector of cosmological parameters, the $\phi_{i}$ are a set of orthogonal basis functions and $N_{\mathrm{PCA}}$ is the number of basis functions used to describe the signal.
Such a decomposition is obtained by diagonalising the covariance matrix of the training data.
The quantity $\epsilon$ represents the part of $B$ that is not captured by the truncated decomposition.
By setting $N_{\mathrm{PCA}}=N_{k}$, we would have $\epsilon=0$ identically.
In practice, only a reduced number of PCA coefficients will be required to describe most of our signal.
The base functions $\phi_{i}$ are independent of the cosmological parameters and all the cosmological dependence is carried by the $\alpha_{i}$ coefficients.
Furthermore, these coefficients are now fully independent from each other, which was not the case of the $k$ bins.
We find that with $N_{\mathrm{PCA}}=5$, we are able to describe more than $99.99\%$ of the total variance contained in the power spectrum boosts.
By increasing the number of PCA coefficients, we would mostly capture the noise in the training signals.
Additionally, we have verified that, given our strategy of binning in $k$, and the number of PCA coefficients that we keep, the absolute error introduced by this decomposition is on average smaller than $0.1\%$.

\begin{table*}
  \centering
  \caption{Table summarising the different simulation boxes used to study the large scale and mass resolution errors of the matter power spectrum boost.
  $L_\mathrm{box}$ is the side length of the simulation box and $m_\mathrm{part}$ the mass of an N-body particle.
  The number of DM particles in each simulation is $512^3$.
  The other numerical parameters are identical to those desribed in Table~\ref{tab:emulation_suite}.
  The models F4, F5 and F6 correspond to the cosmological parameters given in Table~\ref{tab:cosmo_suite}.
  Inside the parentheses are given the number of realisations performed per $f(R)$ model.
  For each $f(R)$ model and realisation, we also perform the corresponding $\Lambda$CDM simulation.
  }
  \label{tab:res_suite}
  \begin{tabular}{|c|c|c|c|}
    \hline\hline
    Name  & $L_{\mathrm{box}} \left[h^{-1}\mathrm{Mpc}\right]$ & $m_{\mathrm{part}}\left[h^{-1}M_{\odot}\right]$ & Models \\
    \hline\hline
    High Res. (HR) & $164.0625$ & $2.9\times10^{9}$ & F4, F5, F6 ($\times1$)\\
    \hline
    Standard Res. (SR)  & $328.125$ & $2.3\times10^{10}$ & F4, F6 ($\times1$), F5 ($\times30$)\\
    \hline
    Low Res. (LR)  & $656.25$ & $1.8\times10^{11}$ & F4, F5, F6 ($\times1$)\\
    \hline
    Very Low Res. (VLR) & $1312.5$ & $1.5\times10^{12}$ & F4, F5, F6 ($\times1$)\\
    \hline
  \end{tabular}
\end{table*}

We now build an independent emulator for each of the $N_{\mathrm{PCA}}$ coefficients.
We use a Gaussian Process (GP) regression to interpolate the PCA coefficients between training nodes.
Gaussian Processes are commonly used in cosmological emulators~\citep[e.g.][]{habib_emulator, euclid_emulator, bacco_emulator, ramachandra_emulator, Arnold2021}.
One of the most used kernels is the Radial-basis function (RBF) kernel.
This kernel states that the covariance of the signal decreases as a squared exponential with the distance in the parameter space.
It is characterised by a length scale in the isotropic case, or a length scale per parameter space dimension in the anisotropic case.
The RBF kernel is most efficient at emulating smooth quantities, since it is infinitely differentiable.
For our emulator we use a generalisation of the RBF kernel to less smooth functions, the Matern kernel.
We find that an anisotropic Matern-5/2 kernel gives the best emulation accuracy given our training data.
We refer the reader to~\citet{GPforML} for more in depth details about GPs and kernels.
It is worth mentioning that before performing the PCA, we standardise the training data and model points.
This means that we remove the mean and rescale them in order to obtain a unit variance distribution.
This simple step simplifies the usage of the PCA and GP emulation and improves their performance.
We use the Python package \textsc{scikit-learn}\footnote{\url{https://scikit-learn.org/stable/index.html}}\citep{scikit-learn} to perform the standardisation of the training data, the PCA and the GP regression.

We have just described the procedure to build the emulator at a given redshift.
We have a total of $19$ redshift nodes between $0 < z < 2$ at our disposal (see Section~\ref{subsec:simu_set_emu}).
In practice, we build an independent emulator for each redshift node.
Then, for any arbitrary redshift, we linearly interpolate the predicted power spectrum boost between the two closest redshift nodes.
We have checked that this method is sufficient to obtain interpolation errors smaller than $1\%$ at all scales and for all cosmological models.
More details are given in Appendix~\ref{app:emu_err_z}.
Such linear interpolation in redshift is also used in other modified gravity emulators~\citep{ramachandra_emulator, Arnold2021}.

\subsection{Emulator simulation set}%
\label{subsec:simu_set_emu}

We perform N-body simulations for each of the $N_{\mathrm{cosmo}}=110$ cosmological models sampled in Section~\ref{subsubsec:sampling}.
We use the numerical simulation chain described in Section~\ref{subsec:codes}.
Each simulation evolves $512^{3}$ DM particles in a cubic periodic box of side length $L_{\mathrm{box}}=328.125 h^{-1}\mathrm{Mpc}$, which gives a mass resolution of order $\sim 2\times10^{10}h^{-1}M_{\odot}$ (depending on the particular cosmology), from an initial redshift of $z_{i}\sim50$ to $z=0$.
For each of the $N_{\mathrm{cosmo}}$ models, we simulate $N_{\mathrm{real}}=5$ independent random realisations, covering a total effective volume of $\left(560 h^{-1}\mathrm{Mpc}\right)^{3}$.
For each $f(R)$ cosmological model and realisation, we perform the equivalent $\Lambda$CDM simulation.
In the end, this gives a total of $1100$ cosmological simulations.
We refer to this large set of simulations as the emulator simulation set.
The CPU-time usage for this simulation suite was of $\sim3.5$ million core-hours using the AMD Irene ROME partition of the Joliot-Curie supercomputer hosted at the Très Grand Centre de Calcul (TGCC).
The main characteristics of our emulator simulation set are summarised in Table~\ref{tab:emulation_suite}.
We save the matter power spectra at $19$ redshift outputs per simulation: $z=0$, 0.05, 0.1, 0.15, 0.25, 0.3, 0.42, 0.5, 0.6, 0.7, 0.8, 0.9, 1, 1.1, 1.25, 1.4, 1.5, 1.74, 2.

\begin{figure*}
  \centering
  \begin{subfigure}{0.48\linewidth}
    \includegraphics[width=\linewidth]{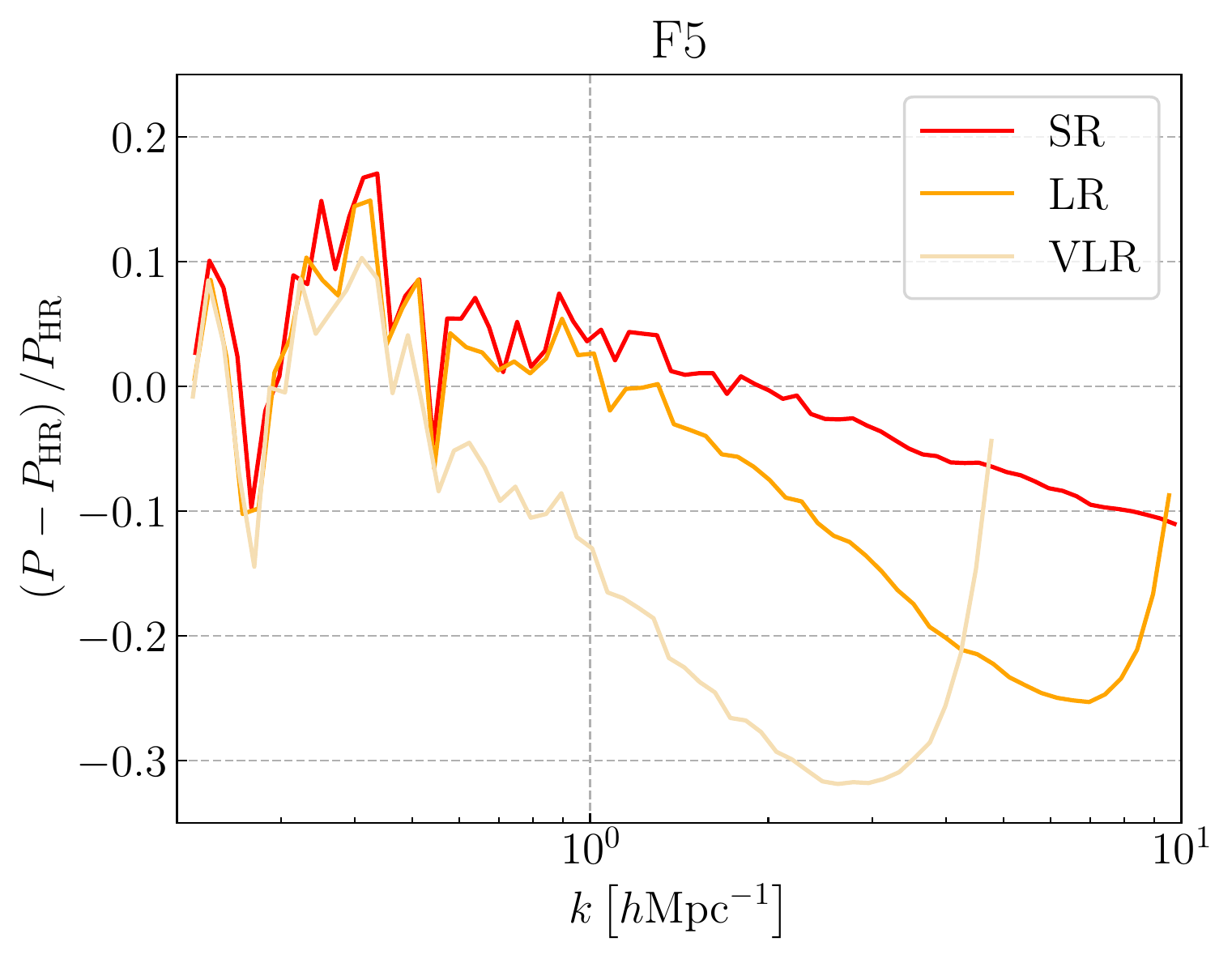}
  \end{subfigure}
  \begin{subfigure}{0.48\linewidth}
	\includegraphics[width=\linewidth]{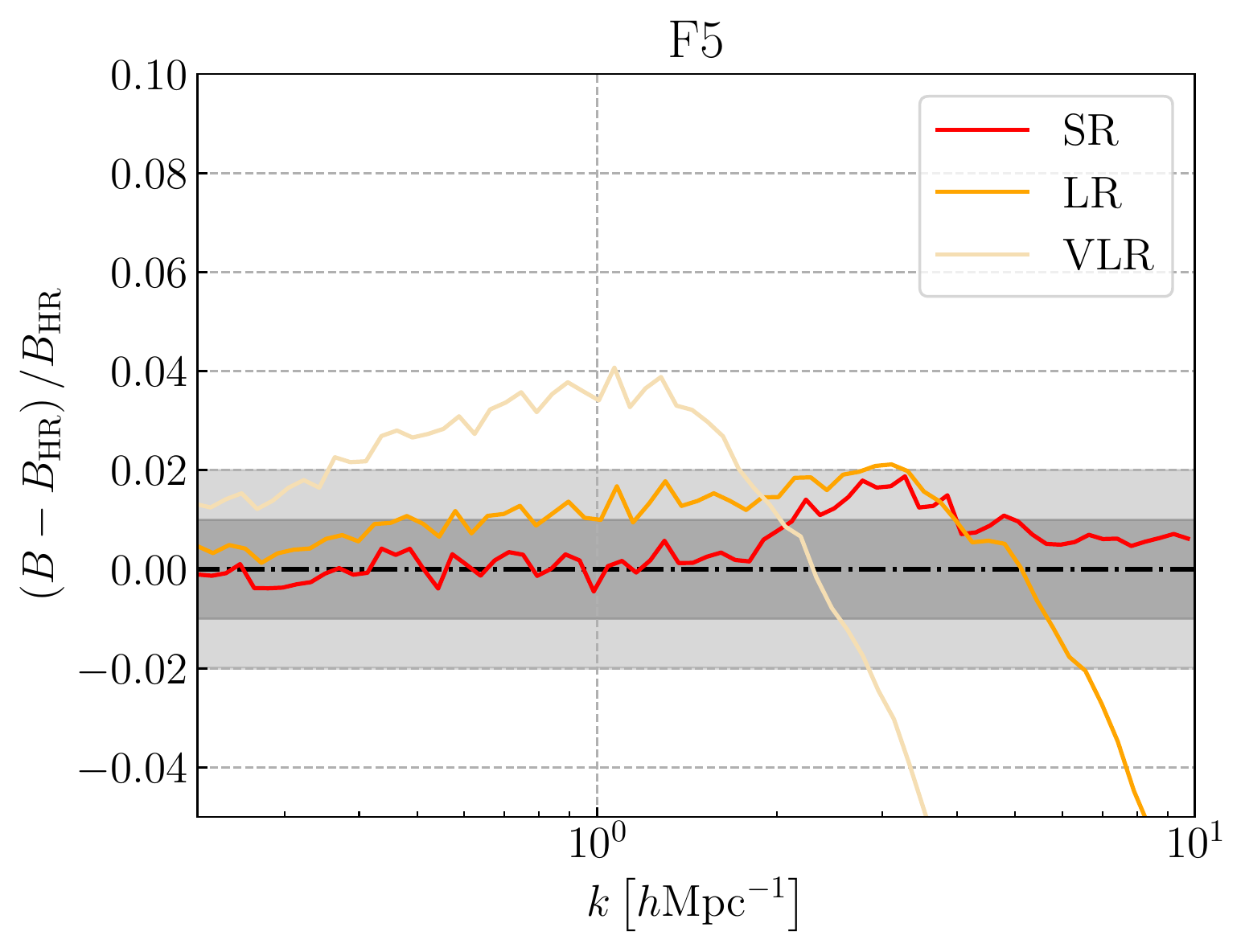}
  \end{subfigure}
  \caption{\label{fig:boxsize_small_scales_z0} Results of the mass resolution convergence test at $z=0$ for the F5 model.
    The HR box is used as a reference and the measurements from the SR (red), LR (orange) and VLR (wheat) boxes are compared to it.
    The left panel shows the relative difference of the total matter power spectrum.
    The systematic drop of the SR power spectrum starts around $k\simeq2h\mathrm{Mpc}^{-1}$ and reaches the $10\%$ level at $k=10h\mathrm{Mpc}^{-1}$.
    The right panel shows the same comparison for the matter power spectrum boost.
    This time the resolution error in the SR box remains under the $2\%$ level at all considered scales.
    The mass resolution errors are up to an order of magnitude smaller in the power spectrum boost than in the full power spectrum.}
\end{figure*}

\subsection{Simulation set for the evaluation of systematic and statistical errors}%
\label{subsec:simu_set_err}

The finite volume and mass resolution of the N-body simulations produce non negligible errors on the measured power spectrum boosts.
A proper estimation of such effects would require to perform a dedicated set of simulations with fixed mass resolution and varying volume to assess the finite volume and sample variance errors.
Another set with fixed volume and varying number of DM particles would be needed to estimate the resolution errors.
However, it would require a significant amount of computing resources.
This is particularly true in $f(R)$ gravity, since, as we will show in the next section, the resolution errors are highly dependent on the value of $\fR$.
It is therefore important to assess the accuracy of our simulations for different $f(R)$ models, which again increases the computational cost.

We resort instead to a more approximate method.
We perform a set of simulations where we vary the simulation box-size while keeping the number of DM particles fixed.
The simulations used to build our emulator evolve $N_{\mathrm{part}}=512^{3}$ DM particles in a cubic periodic box of side $L_{\mathrm{box}} = 328.125$ $h^{-1}\mathrm{Mpc}$.
We refer to this box-size and resolution as the \textit{Standard Resolution} (SR).
In order to assess the systematic errors at small scales due to the finite mass resolution we consider a second class of simulations with $L_{\mathrm{box}} = 164.0625$ $h^{-1}\mathrm{Mpc}$.
We label them as the \textit{High Resolution} (HR) simulations.
The HR box has a volume $8$ times smaller than the SR box and the same number of DM particles, which corresponds to a mass resolution $8$ times higher.
The HR box-size is relatively small and will suffer from an important cosmic variance.
As long as it only affects the largest scales, we can still use it to test the convergence of the SR simulations at small scales.
We also run simulations with larger box-sizes in order to assess the impact of the simulation box length.
More precisely we consider box-sizes of $L_{\mathrm{box}} = 656.25$ and $1312.5$ $h^{-1}\mathrm{Mpc}$, which we refer to as \textit{Low Resolution} (LR) and \textit{Very Low Resolution} (VLR) respectively.
These simulation boxes have a volume $8$ and $64$ times bigger than the SR respectively and the same number of DM particles.
We use them to estimate the accuracy of our measurements at large scales.
More precisely, with this method we are probing a combination of sample variance and finite volume effects.

For each box-size, we run a single realisation of the models F4, F5 and F6 as well as the corresponding $\Lambda$CDM cosmology.
In order to study the impact of the number of realisations, we also simulate an ensemble of $30$ independent realisations with the SR box for the F$5$ model (and the equivalent $\Lambda$CDM cosmology).
The volumes and mass resolutions of this suite of simulations is given in Table~\ref{tab:res_suite}.
The other numerical parameters are identical to the ones used in the \textit{emulator} suite (see Table~\ref{tab:emulation_suite}).

\section{Numerical convergence studies}%
\label{sec:conv_stud}

The final accuracy of the emulator is affected by both the errors on the training data and the pure emulation errors of the GP regression.
In this section, we estimate the amplitude of both sources of errors.

\subsection{Systematic and statistical simulation errors}%
\label{subsec:sys_sym_err}

In this section, we study the impact of the finite simulation volume and mass resolution on the matter power spectrum boost.
In particular, we show that by computing the matter power spectrum boost we get more accurate results, both at large and small scales, than with the full power spectrum.
We use the simulation suite described in Section~\ref{subsec:simu_set_err}.

\subsubsection{Small scale convergence: impact of the mass resolution}
\label{subsubsec:res_effects}

\begin{figure*}
  \centering
  \begin{subfigure}{0.48\linewidth}
    \includegraphics[width=\linewidth]{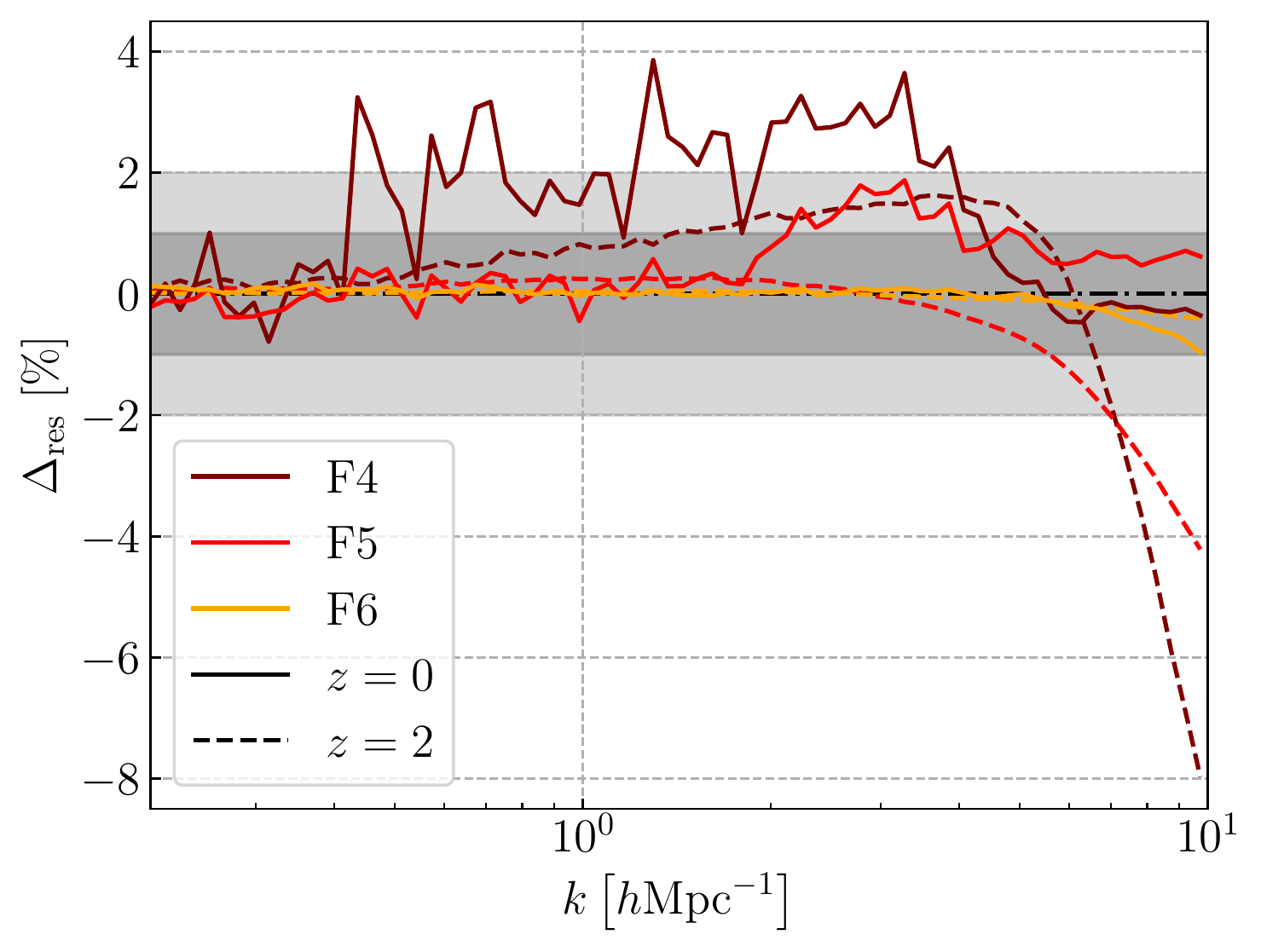}
  \end{subfigure}
  \begin{subfigure}{0.48\linewidth}
    \includegraphics[width=\linewidth]{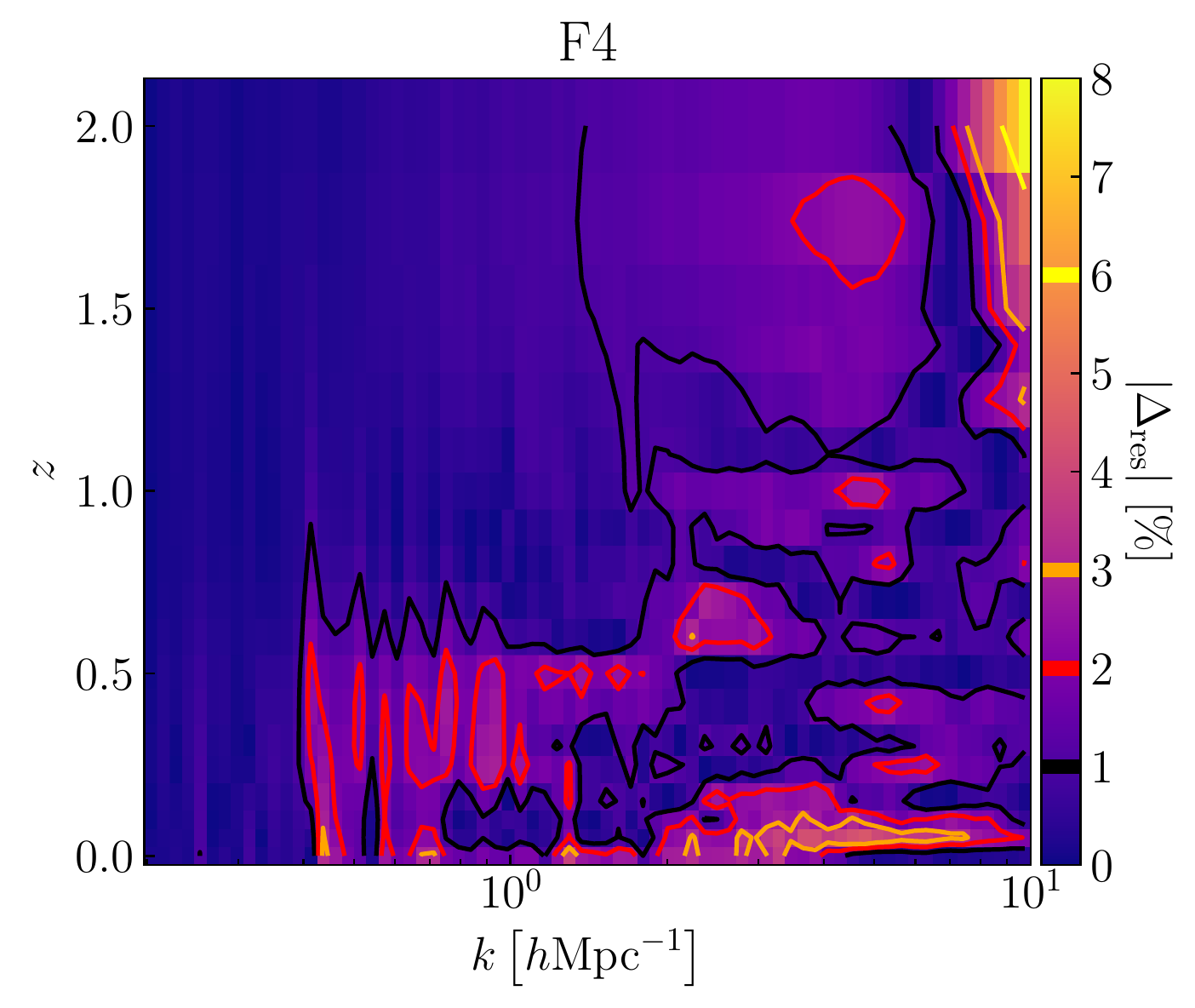}
  \end{subfigure}

  \begin{subfigure}{0.48\linewidth}
    \includegraphics[width=\linewidth]{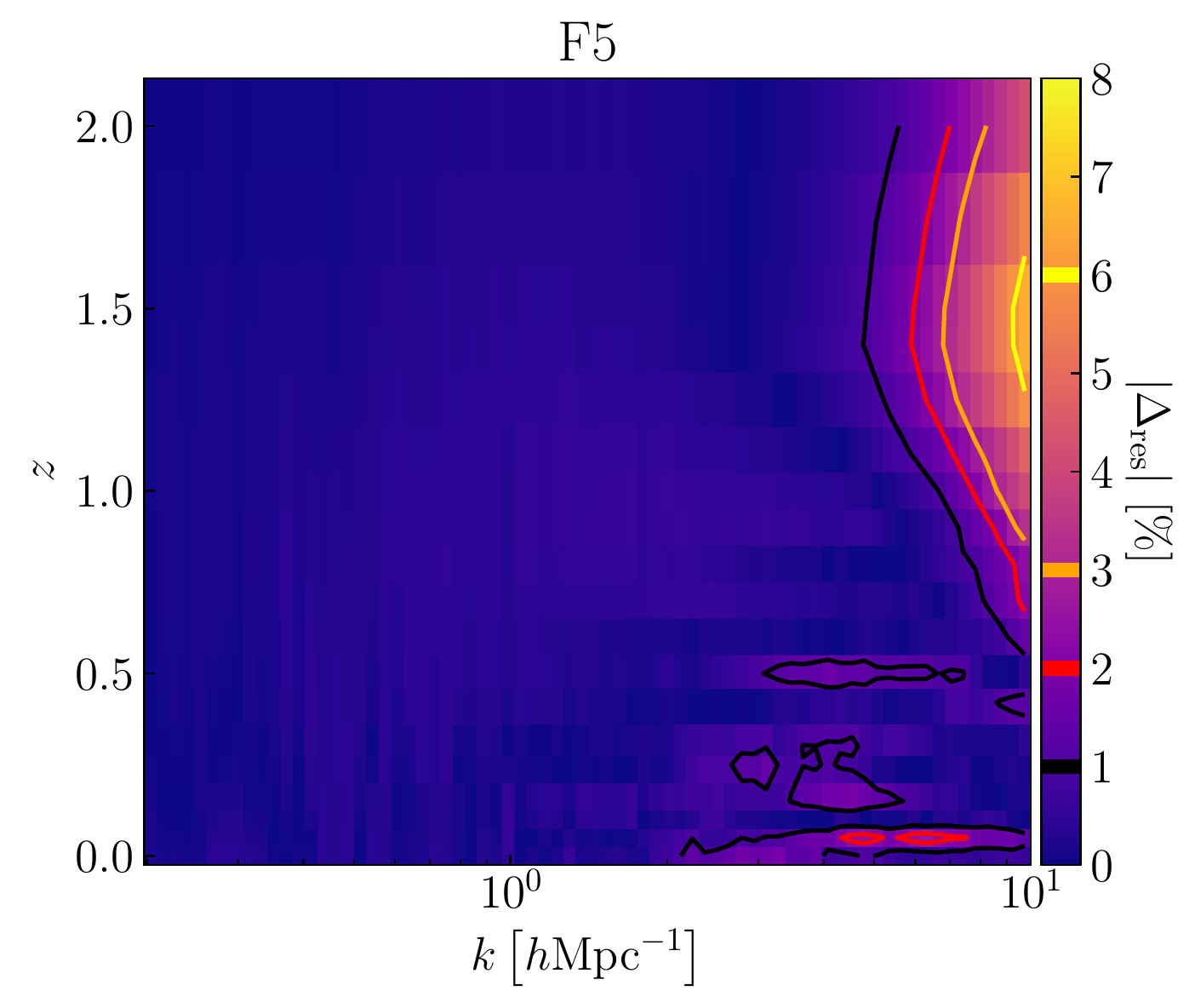}
  \end{subfigure}
  \begin{subfigure}{0.48\linewidth}
    \includegraphics[width=\linewidth]{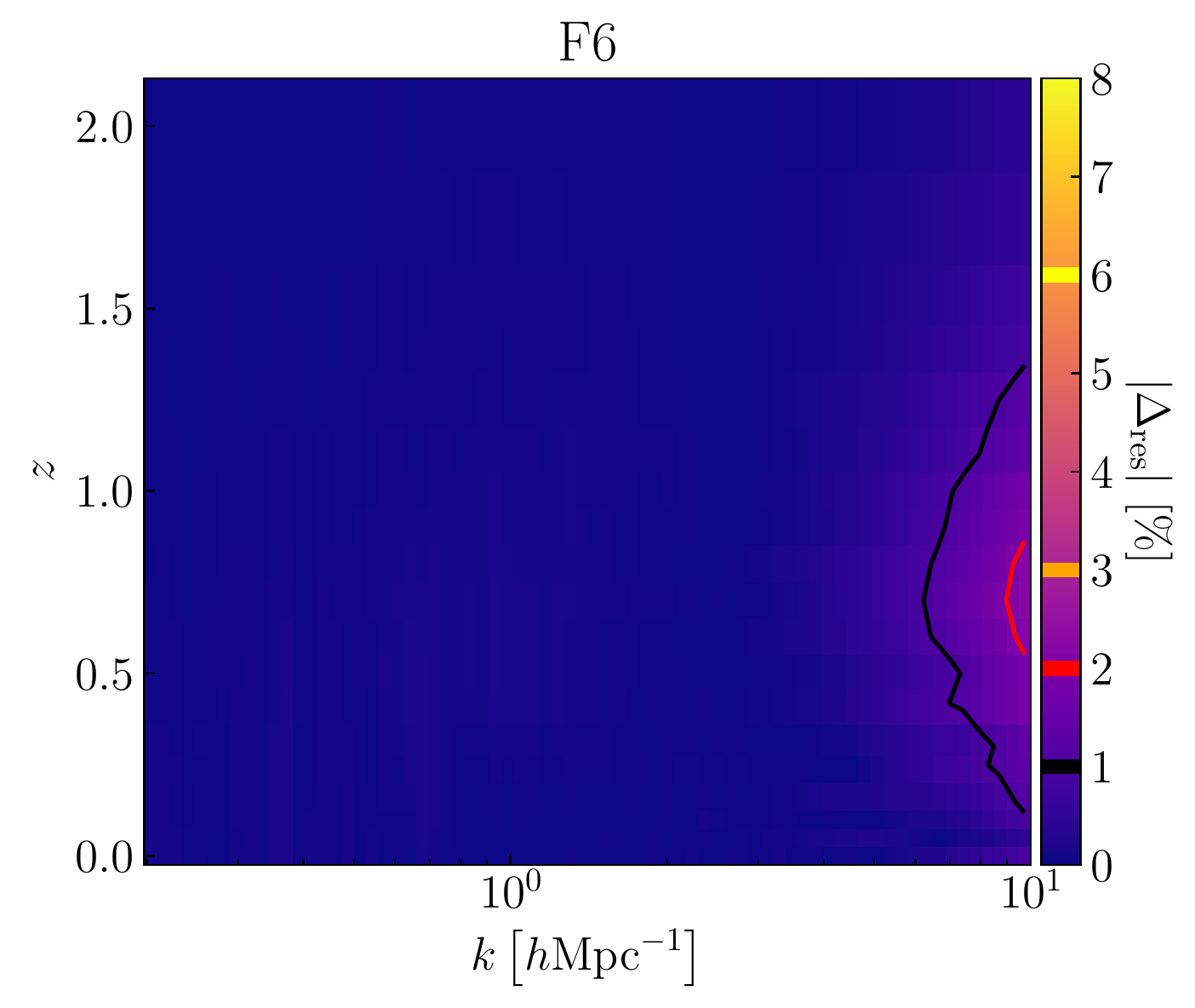}
  \end{subfigure}
  \caption{\label{fig:boxsize_small_scales_contour} Results of the mass resolution convergence test for the matter power spectrum boost in the SR simulation box for different scales, redshifts and $f(R)$ models.
  The top left panels shows the resolution error $\Delta_{\mathrm{res}}$ as a function of the wavenumber $k$ for three different $f(R)$ models (F$4$ in maroon, F$5$ in red and F$6$ in orange) and two redshifts ($z=0$ with solid lines and $z=2$ with dashed lines).
  The remaining panels give $\left|\Delta_{\mathrm{res}}\right|$ as a function of redshift and scale for F$4$ (top right), F$5$ (bottom left) and F$6$ (bottom right).
  Overall, the systematic errors of the power spectrum boost are smaller than $3\%$ for all models in the range $0<z<2$ and $k\lesssim7$ $h\mathrm{Mpc}^{-1}$.
  }
\end{figure*}

We start by comparing the measured matter power spectra and boosts from the different boxes at small scales.
In this test we use the HR simulations as a reference.
For the SR, LR and VLR boxes we compute the quantity
\begin{equation}
  \Delta_{\mathrm{res}}(z,k) = \frac{Y-Y_{\mathrm{HR}}}{Y_{\mathrm{HR}}},
\end{equation}
where $Y$ is a placeholder for the power spectrum $P$ or the boost $B$.
The results at $z=0$ for the F$5$ model are shown in Figure~\ref{fig:boxsize_small_scales_z0}.
The left panel shows $\Delta_{\mathrm{res}}$ for the total matter power spectrum.
The resolution drop in the SR box starts around $k\simeq2 h\mathrm{Mpc}^{-1}$.
At smaller scales the SR power spectrum is systematically under estimated, reaching an error of $\sim10\%$ at $k=10$ $h\mathrm{Mpc}^{-1}$.
Such finite resolution effects are well known.
They are in agreement with previous $\Lambda$CDM simulations run with \textsc{ramses} and the same mass resolution~\citep{RayGal}.
The fluctuations at larger scales are mainly due to the large cosmic variance of the HR box and the difference in realisations from one box-size to another.

We now turn to the power spectrum boost.
The corresponding resolution error $\Delta_{\mathrm{res}}$ is shown in the right panel of Figure~\ref{fig:boxsize_small_scales_z0}.
The lower resolution runs tend to over-estimate the boost at intermediate to small scales.
In the case of the SR box, this effect induces an error smaller than $2\%$.
As explained in~\citet{Li2012_boost}, a low resolution simulation is not able to properly resolve the screening mechanism in high density environments.
As a consequence, the clustering in $f(R)$ gravity, and therefore the matter power spectrum, is over-estimated in a scale-dependent manner.
In addition to this over-estimation at intermediate to small scales, the LR and VLR boosts show a severe drop at the very small scales.
However, this drop is not visible in the SR run, where $\left|\Delta_{\mathrm{res}}\right|$ is smaller than $1\%$ at $k=10h\mathrm{Mpc}^{-1}$.
In any case, this comparison shows that the mass resolution error is reduced by almost an order of magnitude when going from the total power spectrum in $f(R)$ gravity to its boost with respect to $\Lambda$CDM.

In addition to the wavenumber $k$, the resolution errors also depend on redshift and the parameter $\fR$.
This is shown in Figure~\ref{fig:boxsize_small_scales_contour} for the SR simulation box.
The top left panel gives $\Delta_{\mathrm{res}}$ for the three models F$4$, F$5$ and F$6$ and at two different redshifts $z=0,2$.
The over-estimation of the boost at intermediate scales is more important for the F$4$ model and vanishes for F$6$.
Indeed, the screening mechanism enters into play in denser environments for higher values of $\left|\fR\right|$.
At $z=2$, the power spectrum boost suffers from a resolution drop at very small scales, which also depends on the value of $\fR$.
Indeed, at $z=2$, $f(R)$ gravity has just started having an effect on matter clustering.
This effect being small, the lower resolution simulations are not able to resolve it, while the higher resolution ones predict it more accurately.
On the other hand, the over-estimation at intermediate scales is smaller.

The remaining panels of Figure~\ref{fig:boxsize_small_scales_contour} give $\left|\Delta_{\mathrm{res}}\right|$ at all redshifts ($0<z<2$) and scales of interest for the models F$4$, F$5$ and F$6$.
For F$6$ the resolution error on the boost is smaller than $1\%$ most of the time and smaller than $2\%$ virtually for all redshifts ($0<z<2$) and scales considered ($k<10 h\mathrm{Mpc}^{-1}$).
In the case of F$5$, the accuracy is better than $3\%$ in most of the $z-k$ plane.
However, the resolution error in the $z\gnsim1$ and $k\gnsim7 h\mathrm{Mpc}^{-1}$ region is of order $3-6\%$.
Finally, for the F$4$ model the accuracy is of order $1-2\%$ in most of the $z-k$ plane and only reaches the $3\%$ level at very low redshifts and small scales as well as at high redshifts and very small scales.
Overall, we estimate that the systematic resolution errors in our power spectrum boost measurements are always smaller than $3\%$ in the range $0<z<2$ and $k\lesssim 7$ $h\mathrm{Mpc}^{-1}$.

\subsubsection{Large scale convergence: impact of the simulation box length}
\label{subsubsec:vol_effects}

\begin{figure*}
  \centering
  \begin{subfigure}{0.48\linewidth}
	\includegraphics[width=\linewidth]{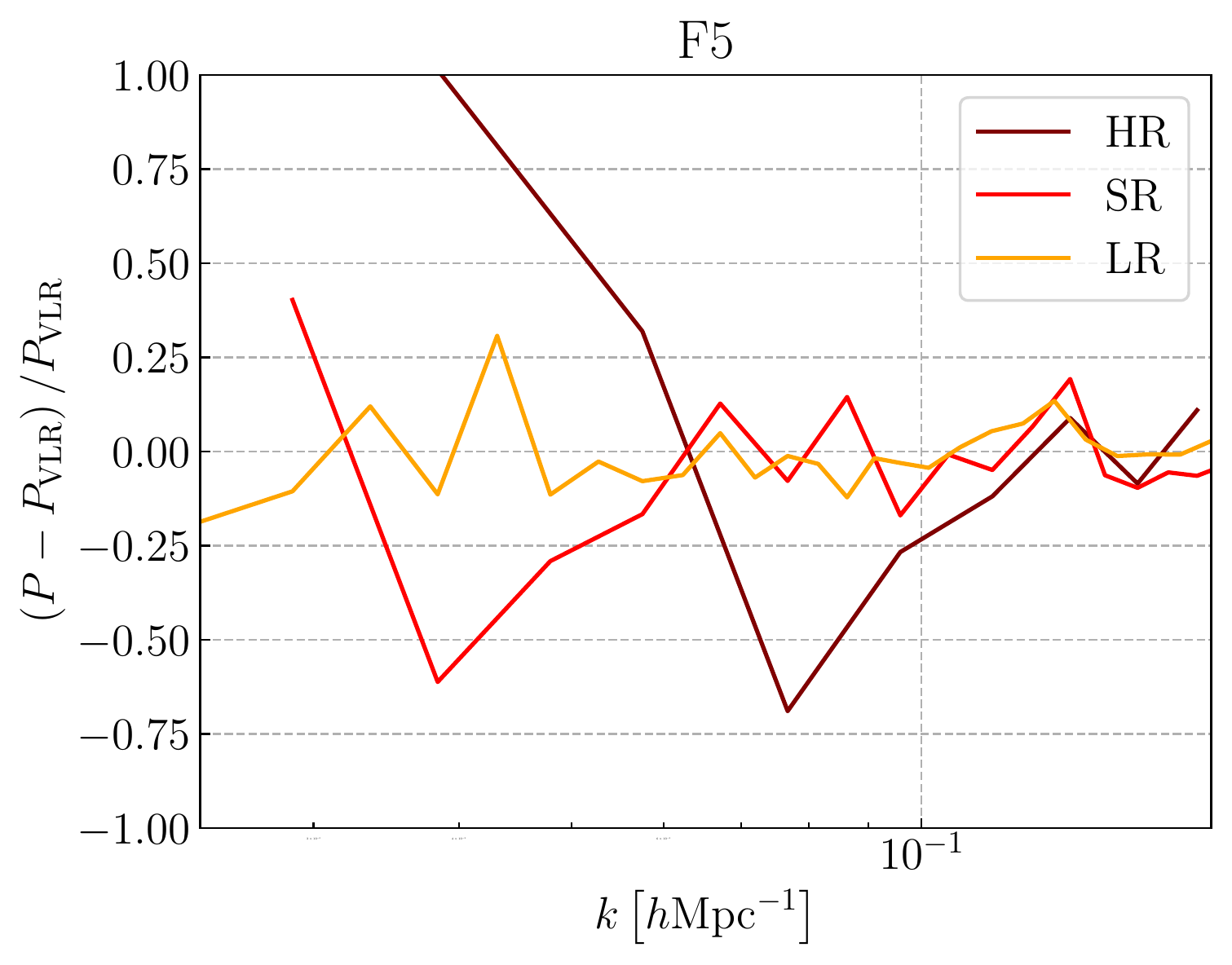}
  \end{subfigure}
  \begin{subfigure}{0.48\linewidth}
	\includegraphics[width=\linewidth]{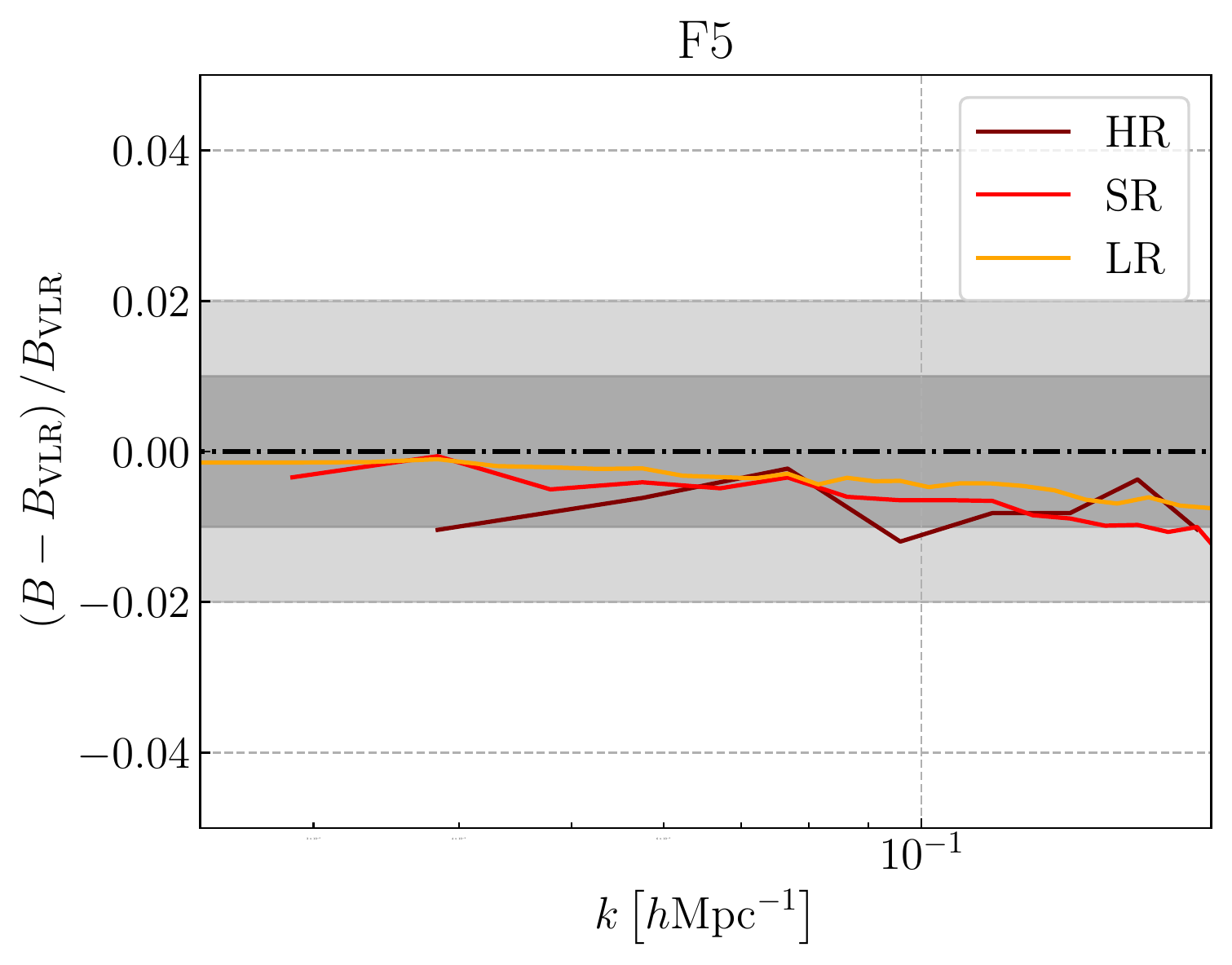}
  \end{subfigure}
  \caption{\label{fig:boxsize_large_scales_z0} Results of the large scale convergence test at $z=0$ for the F5 model.
    The VLR box is used as a reference and the measurements from the HR, SR and LR boxes are compared to it.
    The left panel shows the relative difference in the total matter power spectrum.
    The error of the SR box gets as high as $50\%$ at the largest scales covered by the emulator.
    The right panel shows the same test for the matter power spectrum boost.
    In this case, the large scale error in the SR box is smaller than $1\%$ even for the first $k$ bin.}
\end{figure*}

We now perform the same kind of analysis but focusing on large scales.
This time the reference box-size is the VLR and we compute for the LR, SR and HR simulation boxes the quantity
\begin{equation}
  \Delta_{\mathrm{vol}}(z,k) = \frac{Y-Y_{\mathrm{VLR}}}{Y_{\mathrm{VLR}}},
\end{equation}
where, as in the previous section, $Y$ stands for the power spectrum $P$ or its boost $B$.

The results for the model F$5$ at $z=0$ are shown in Figure~\ref{fig:boxsize_large_scales_z0}.
At the largest scales available for the SR and HR resolutions, the relative difference on the power spectrum is of order $50\%$ and $100\%$ respectively.
However, when considering the boost, the large scale errors remain under the $1\%$ level even for the fundamental mode of the HR box.
We have checked that this results hold for the three models F$4$, F$5$ and F$6$ and all redshifts $0<z<2$.

With this test, we are really probing a combination of cosmic variance and finite volume errors.
However, the purpose here is not to carry a detailed investigation of such effects, which would require a more dedicated study.
These results simply show that by taking the ratio of the matter power spectrum in $f(R)$ gravity with the corresponding $\Lambda$CDM cosmology, there is a very large cancellation of large scale errors (of order $\sim100\%$).
Such a large cancellation happens because the $f(R)$ and $\Lambda$CDM simulations are evolved from the same initial conditions.
Furthermore, the impact of $f(R)$ gravity on the matter distribution is weak at linear scales for the values of $\fR$ considered in this work.
As a consequence, the large scale random fluctuations remain almost the same in the $f(R)$ and $\Lambda$CDM simulations up to $z=0$.
This is another advantage of considering the boost instead of the full power spectrum.
However, we are still limited by the fundamental mode of our simulation box and cannot resolve features in the power spectrum that are smaller than $\Delta k \sim k_{\mathrm{F}}\simeq0.018 h\mathrm{Mpc}^{-1}$.

\subsubsection{Sample variance: impact of the number of realisations}
\label{subsubsec:real_number}

\begin{figure*}
  \centering
  \includegraphics[width=\linewidth]{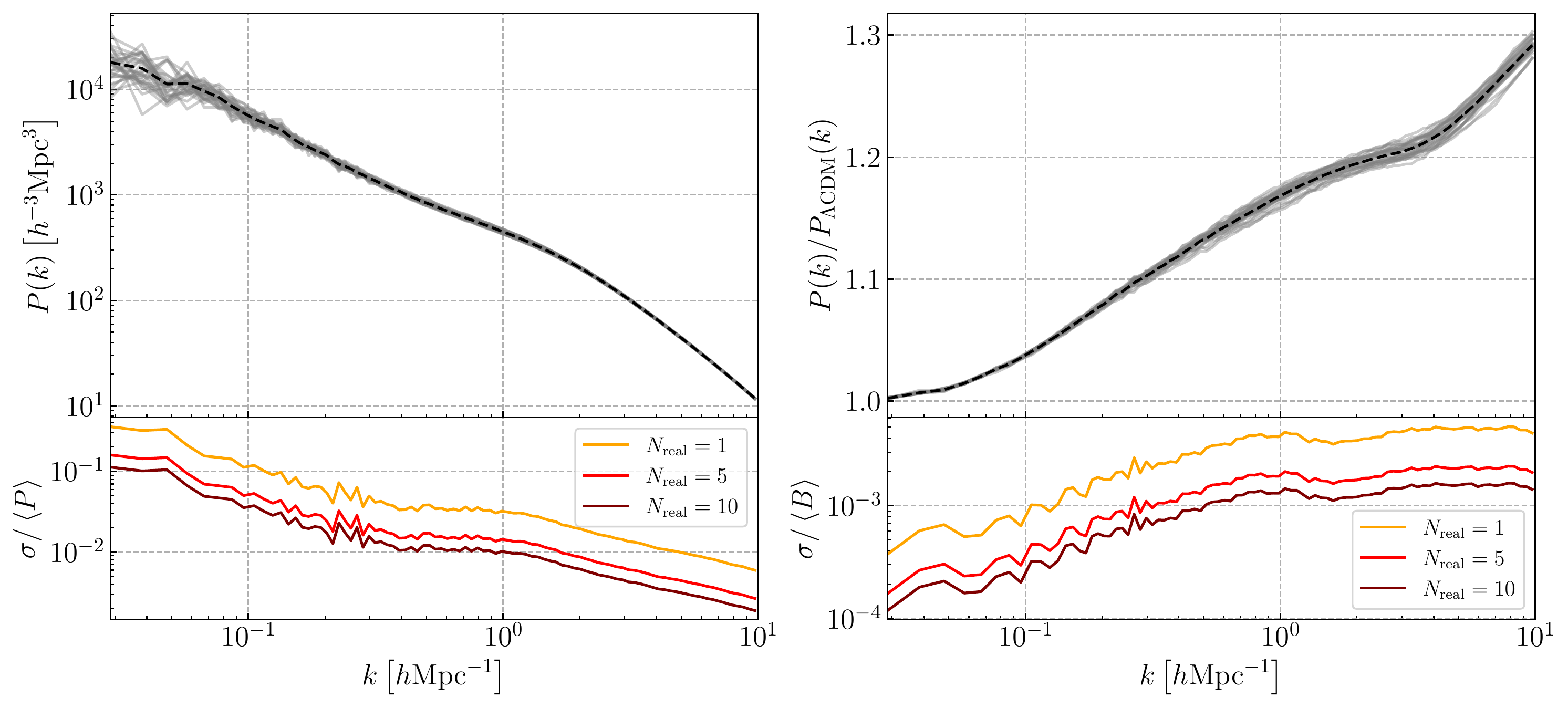}
  \caption{\label{fig:number_real_z0} \textit{Upper panels:} Matter power spectrum (\textit{left}) for our reference cosmological model F$5$ and the corresponding boost (\textit{right}) with respect to $\Lambda$CDM for $30$ independent realisations (gray lines).
  The mean over the different realisations is shown as a black dashed line.
  \textit{Lower panels:} Relative standard errors of the means computed over $N_{\mathrm{real}}$ independent realisations.
  Cosmic variance induces large scale errors of order $\mathcal{O}\left(10\%\right)$ in the full power spectrum, while reducing to $\mathcal{O}\left(0.01\%\right)$ in the case of the boost.
  This is true even with a single realisation.
}
\end{figure*}

As a complementary study to the one presented in the previous section, we assess the impact of the number of realisations on the matter power spectrum boost measurements, i.e. pure sample variance errors.
In order to do so, we use the set of $30$ independent realisations for our reference cosmological model F$5$ and the corresponding $\Lambda$CDM cosmology, run with the SR simulation box (see Section~\ref{subsec:simu_set_err}).
We start by measuring the matter power spectrum for each realisation of the F$5$ model as well as the corresponding boosts with respect to $\Lambda$CDM.
These quantities, along with their means computed over the $30$ realisations, are shown in the upper panels of Figure~\ref{fig:number_real_z0}.
The lower panels show the standard errors of the means computed over a number $N_{\mathrm{real}}=1,5,10$ of independent realisations.
They are divided by the mean over the full set of realisations in order to obtain a relative standard error.

We recover a similar result as in the last section.
There is a large cancellation of cosmic variance in the matter power spectrum boost with respect to the full power spectrum.
Indeed, for the boost, the relative standard error is of order $\mathcal{O}\left(0.01\%\right)$ on the largest scales available, while for the full power spectrum it is of order $\mathcal{O}\left(10\%\right)$.
This is true even with a single realisation.
The reasons behind this are the same as explained in the last section.

The cosmic variance cancellation is weaker at small scales.
In fact, Figure~\ref{fig:number_real_z0} shows that for the boost cosmic variance is an overall increasing function of $k$.
For our emulator, we have decided to perform $N_{\mathrm{real}}=5$ realisations for each training model.
We can see that this is enough to keep statistical errors below $1\%$ at the $3\sigma$ level at all scales.
In Figure~\ref{fig:number_real_z0}, we only show the results at $z=0$.
However, we have checked that the conclusions from this section hold at all redshifts relevant for our emulator, i.e. for $0<z<2$.

\subsection{Emulation errors}%
\label{subsec:emu_err}

\begin{figure}
  \centering
    \includegraphics[width=\linewidth]{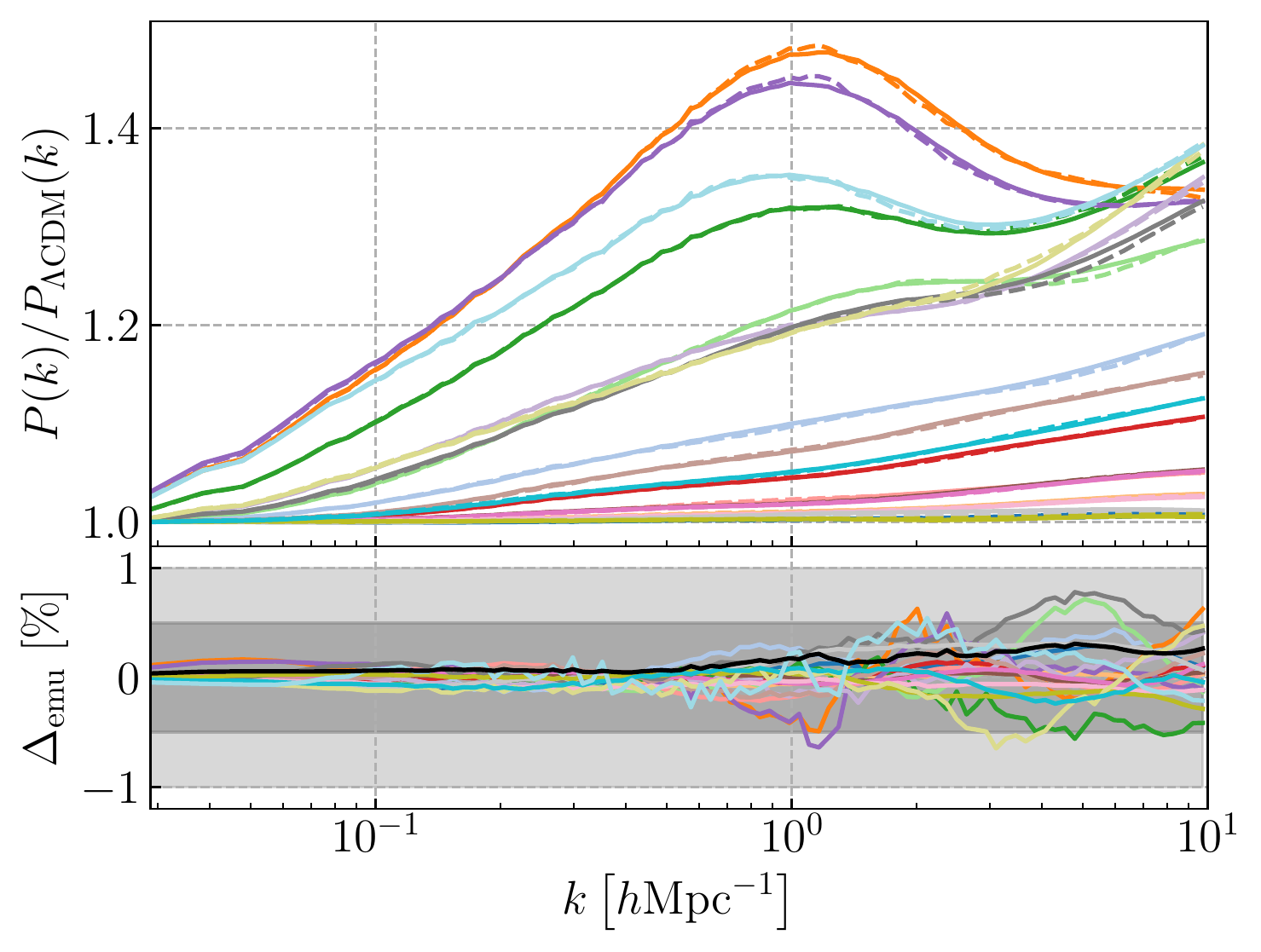}
  \caption{Results of the emulation accuracy test at $z=0$.
    The emulator is trained using all available models except one slice from the primary SLHD, which is used as a validation sample.
    The top panel shows the matter power spectrum boost predicted by the emulator for each test model (solid lines) and the measurements from the simulations (dashed lines).
    The bottom panel shows the relative differences between the predictions of the emulator and the simulations.
    The solid black line gives the corresponding root-mean-square-relative-error (RMSRE), as defined by Eq.~\eqref{eq:RMSRE}.
    The dark and light grey bands mark the $0.5\%$ and $1\%$ accuracy limits respectively.
    All the validation models have an emulation error smaller than $1\%$, even for the most extreme ones.
    The RMSRE is smaller than $0.5\%$ level at all the scales covered by the emulator.
  }
  \label{fig:emulation_errors}
\end{figure}

In this section, we assess the emulation errors that are purely due to the GP interpolation between training nodes.
For that we make use of the sliced structure of our experimental design by following the same procedure as in~\citet{Nishimichi_2019}.
We train the emulator using only $3$ slices from the primary SLHD and the $30$ refinement models.
The remaining slice (i.e. $20$ cosmological models) is used as a validation sample.
The advantage of sampling the parameter space with a maximin-distance SLHD is that the testing set of cosmologies covers efficiently the whole parameter space while minimising the overlap with the training points.
This way the comparison of the predictions from the emulator against the measured boosts from the validation sample gives a representative estimation of the emulation accuracy across the whole parameter space.
The results from this test are presented in Figure~\ref{fig:emulation_errors}.
It shows the comparison of the predictions from the emulator and the simulation measurements for the $20$ cosmological models of the validation sample at $z=0$.
The majority of models have relative errors that stay under the $0.5\%$ limit at all scales.
In the worst cases the emulation errors are still smaller than $1\%$.
We have also computed the root-mean-squared-relative-error (RMSRE)
\begin{equation}\label{eq:RMSRE}
  \mathrm{RMSRE}(k) = \sqrt{\frac{1}{N_{\mathrm{models}}} \sum_{i=0}^{N_{\mathrm{models}}-1}\left(\frac{B_{\mathrm{emu}}(k;\theta_{i})}{B_{\mathrm{sim}}(k;\theta_{i})}-1\right)^{2}},
\end{equation}
where $N_{\mathrm{models}}$ is the number of validation models, $\theta_{i}$ is the cosmological parameter vector of the validation model $i$, $B_{\mathrm{emu}}$ is the boost predicted by the emulator and $B_{\mathrm{sim}}$ is the boost measured from the simulation.
The RMSRE, given by the black line in the lower panel of Figure~\ref{fig:emulation_errors}, is smaller than $0.5\%$ for all values of $k$.
We have verified that these results remain valid for the $19$ redshift nodes of our training set.
Appendix~\ref{app:emu_err_z} presents the same accuracy test at $z=1$ and $z=2$.
The final version of our emulator is trained using the full set of $110$ models, and therefore, we expect the real emulation errors to be smaller than this estimation.

This type of comparison is an assessment of the emulation errors exclusively.
To estimate the total error budget of \textsc{e-mantis}, it is important to also take into account the accuracy of the training data.
In Section~\ref{subsubsec:res_effects}, we have estimated a maximum mass resolution error of $3\%$ in the range $0<z<2$ and $k\lesssim 7$ $h\mathrm{Mpc}^{-1}$.
The pure emulation errors are therefore negligible with respect to the systematic errors in the training data.
This is made possible by the large number of models in our experimental design, and, in particular, by the refinement strategy.
More quantitative details are given in Appendix~\ref{app:emu_err_nreal_refinements}.
The large scale errors estimated in Sections~\ref{subsubsec:vol_effects} and~\ref{subsubsec:real_number} are smaller than $1\%$ and, as a consequence, they can be neglected with respect to the mass resolution systematic effects.
Appendix~\ref{app:emu_err_z} shows that the redshift interpolation errors are also negligible.
The final accuracy of \textsc{e-mantis} is exclusively driven by the mass resolution errors in the training data.
As shown in Section~\ref{subsubsec:res_effects}, such errors are always smaller than $3\%$ for all cosmological models considered and for $0<z<2$ and $\lesssim7$ $h\mathrm{Mpc}^{-1}$.
This is a conservative bound and, in most cases, the resolution errors remain smaller than $1\%$.

\section{Results}%
\label{sec:results}

\subsection{Comparison with other predictions}%
\label{subsec:comparison_emu}

\begin{figure}
  \centering
  \includegraphics[width=\linewidth]{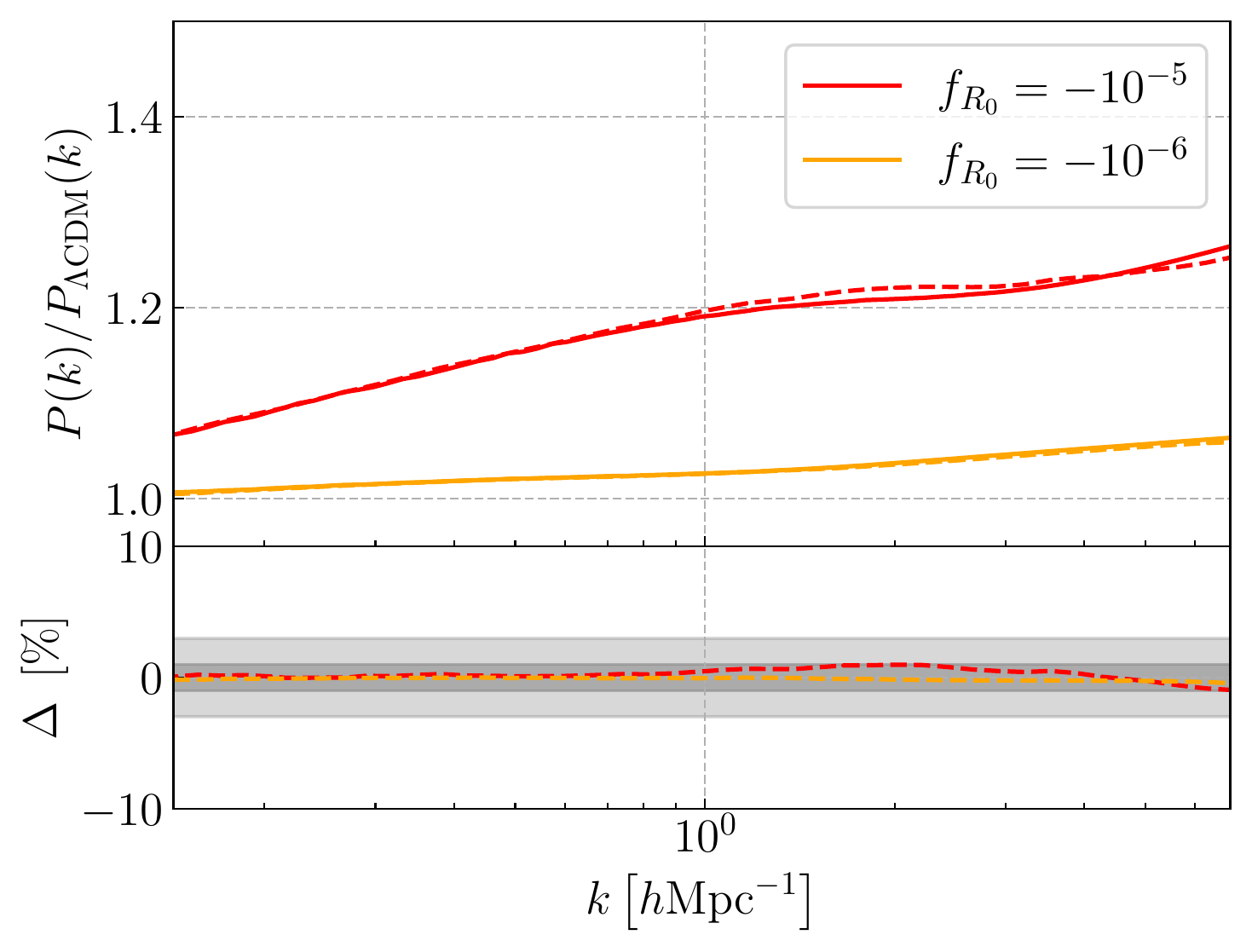}
  \caption{\label{fig:emu_vs_WCP}
    Comparison of the predicted boost by \textsc{e-mantis} (solid lines) with the \textsc{ecosmog} simulation data from~\citet{WCP} (dashed lines) at $z=0$.
    The top panel gives the power spectrum boosts and the lower panel their relative difference with respect to \textsc{e-mantis}.
    The light and dark grey bands mark the $1\%$ and $3\%$ levels respectively.
    The agreement is better than $1\%$ at all scales available ($k \lesssim 7 h\mathrm{Mpc}^{-1}$) and for both models F$5$ and F$6$.
    This comparison serves as a validation of \textsc{e-mantis} against an external simulation run with the same N-body code, which the authors of~\citet{WCP} have shown to be in good agreement with other simulations codes such as \textsc{mg-gadget} and \textsc{isis}.
  }
\end{figure}

In this section, we compare the output of \textsc{e-mantis} to existing predictions in the literature for the matter power spectrum boost in $f(R)$ gravity.
In~\citet{WCP}, the authors perform a comparison of different modified gravity simulation codes.
They show that the predicted matter power spectrum boosts in $f(R)$ gravity agrees within $1\%$, at least up to $k\lesssim7h\mathrm{Mpc}^{-1}$, for three different N-body simulation codes: \textsc{ecosmog}~\citep{Li2012}, \textsc{mg-gadget}~\citep{mg-gadget} and \textsc{isis}~\citep{isis}.
Figure~\ref{fig:emu_vs_WCP} compares the power spectrum boosts they obtain with \textsc{ecosmog} with the predictions from \textsc{e-mantis}.
For this comparison, we set $\Omega_{m}=0.269$ and $\sigma_{8}=0.8$, which are the cosmological parameters used in~\citet{WCP}.
Their simulations have a slightly better resolution than ours.
Indeed, they evolve the same number of DM particles ($512^{3}$) in a smaller simulation box ($L_{\mathrm{box}}=250$ vs $328.125 h^{-1}\mathrm{Mpc}$), giving a mass resolution of $m_{\mathrm{part}}=8.71\times10^{9}h^{-1}M_{\odot}$ (vs an average of $m_{\mathrm{part}}\sim2.3\times10^{10}h^{-1}M_{\odot}$ for our emulation suite).
They use a smaller refinement particle criterion ($m_{\mathrm{ref}}=8$ vs $14$) and also an additional refinement level ($l_{\mathrm{max}}=16$ vs $15$) for F$5$.
In spite of these differences, the predictions from \textsc{e-mantis} agree within $1\%$ for both models F$5$ and F$6$ and at all available scales ($k\lesssim7$ $h\mathrm{Mpc}^{-1}$).

\begin{figure}
  \centering
  \includegraphics[width=\linewidth]{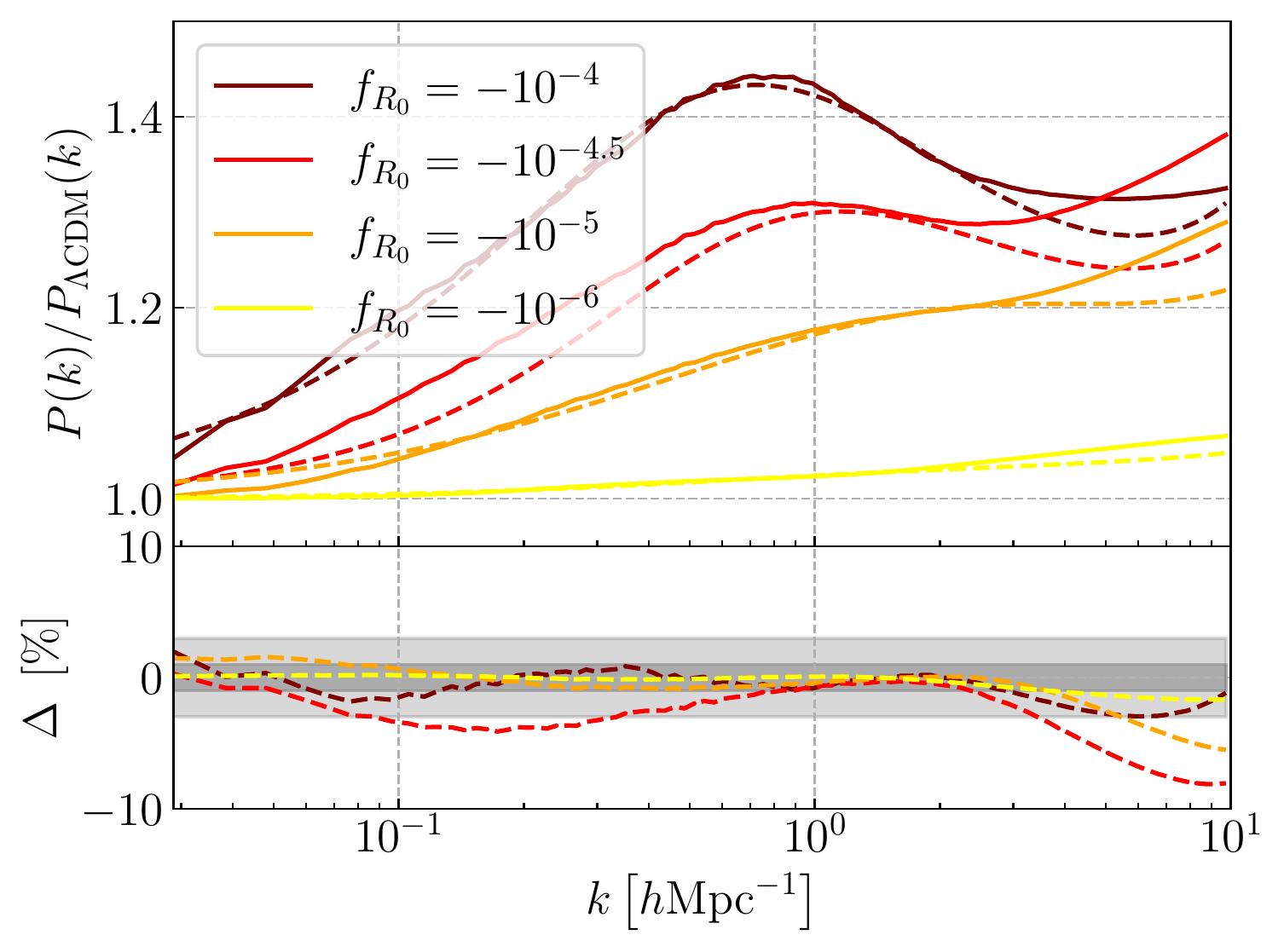}
  \caption{\label{fig:emu_vs_winther_fit}
    Comparison with the fitting formula from~\citet{winther_emulator} at $z=0$, which, like \textsc{e-mantis}, is based on N-body simulations run with \textsc{ecosmog}.
    The top panel shows the predicted power spectrum boosts by \textsc{e-mantis} (solid lines) and the fitting formula (dashed lines) for different values of $\fR$.
    The bottom panel gives the relative difference of the fitting formula prediction with respect to \textsc{e-mantis}.
    The light and dark grey bands mark the $1\%$ and $3\%$ levels respectively.
    For the models F$4$ and F$6$ the agreement is better than $3\%$ at all scales considered.
    In the case F$5$ there is a $3\%$ agreement for $k\lesssim5-6$ $h\mathrm{Mpc}^{-1}$ and then the difference increases up to $\sim5\%$.
    For F$4.5$, which is not one of the fitting nodes of~\citet{winther_emulator}, the difference is more important.
    The fitting formula interpolates the boost between three values of $f_{R_0}$, while \textsc{e-mantis} is trained on $110$ different cosmological models on a $3$D latin hypercube.
  }
\end{figure}

Another available prediction for the matter power spectrum boost in $f(R)$ gravity is the fitting formula introduced by~\citet{winther_emulator}.
The authors perform a fit of the boost using the \textsc{elephant} simulations~\citep{elephant_sim_suite}.
These simulations have been done with \textsc{ecosmog} and evolve $1024^{3}$ DM particles in a cubic box of side $L_{\mathrm{box}}=1024$ $h\mathrm{Mpc}^{-1}$ with a mass resolution of $7.78\times10^{10}h^{-1}M_{\odot}$.
The fitting formula is calibrated from three different $f(R)$ models: $\left|\fR\right| = \left\{10^{-4}, 10^{-5}, 10^{-6}\right\}$.
The other cosmological parameters are kept fixed.
We compare the predicted boost from \textsc{e-mantis} to the fitting formula for the models F$4$, F$5$ and F$6$.
The accuracy of the fitting formula should be the best for these three models, since they are the fitting nodes.
We also carry a comparison for F$4.5$ in order to test the interpolation of the fitting formula.
We set $\Omega_{m}= 0.281$ and $\sigma_{8}=0.82$, which are the values used in~\citet{winther_emulator}.
Figure~\ref{fig:emu_vs_winther_fit} gives the resulting comparison.
The agreement between both predictions is at the $3\%$ level at all scales for F$4$ and F$6$.
For F$5$ the $3\%$ agreement holds for $k\lesssim5-6$ $h\mathrm{Mpc}^{-1}$.
At smaller scales, the prediction of the fitting formula drops below that of \textsc{e-mantis}, up to a difference of $\sim5\%$.
One possible explanation for such a difference could be that our training simulations have a higher mass resolution than the \textsc{elephant} simulations.
In any case, such differences are well within the estimated $3\%$ accuracy of \textsc{e-mantis} combined with that of the fitting formula.
Indeed, in~\citet{winther_emulator}, the authors cite an accuracy of $1\%$ for $k\lesssim1h\mathrm{Mpc}^{-1}$ and of $5\%$ for $1<k<10h\mathrm{Mpc}^{-1}$.
In the case of F$4.5$, the difference between both predictions is slightly larger.
There is a $\sim3\%$ difference at intermediate scales.
Additionally, the prediction from the fitting formula drops by $\sim8\%$ at $k=10h\mathrm{Mpc}^{-1}$.
Both predictions are based on simulations run with the same N-body code and using a similar numerical resolution.
However, we expect the emulation of the boost by \textsc{e-mantis} to be more accurate than that of the fitting formula.
Indeed, \textsc{e-mantis} is trained on $110$ different cosmological models on a $3$D latin hypercube, while in~\citet{winther_emulator} they simply interpolate between three values of $f_{R_0}$.

\begin{figure}
  \centering
  \includegraphics[width=\linewidth]{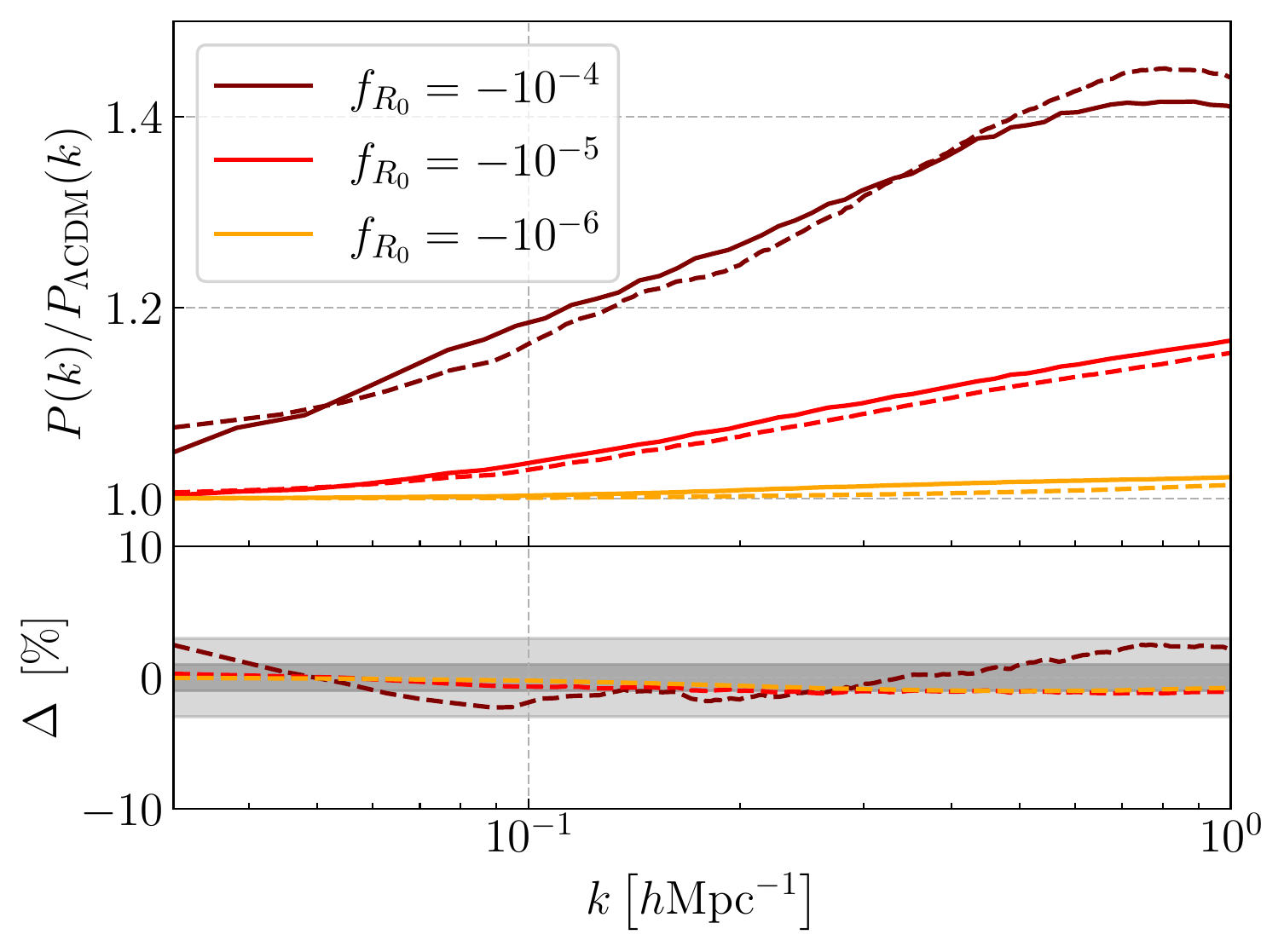}
  \caption{\label{fig:emu_vs_COLA}
    Comparison with the \textsc{mgemu} emulator~\citep{ramachandra_emulator}, based on COLA simulations, at $z=0$.
    The top panel shows the predicted power spectrum boosts by \textsc{e-mantis} (solid lines) and \textsc{mgemu} (dashed lines) for different values of $\fR$.
    The bottom panel gives the relative difference of the \textsc{mgemu} predictions with respect to \textsc{e-mantis}.
    The light and dark grey bands mark the $1\%$ and $3\%$ levels respectively.
    The models F$6$ and F$5$ agree to the $1\%$ level at all scales considered.
    In the case of F$4$, the agreement is at the $3\%$ level.
    We restrict the comparison to scales $k\leq 1h\mathrm{Mpc}^{-1}$, which is the limit of validity of the COLA predictions.
  }
\end{figure}

\begin{figure}
  \centering
  \includegraphics[width=\linewidth]{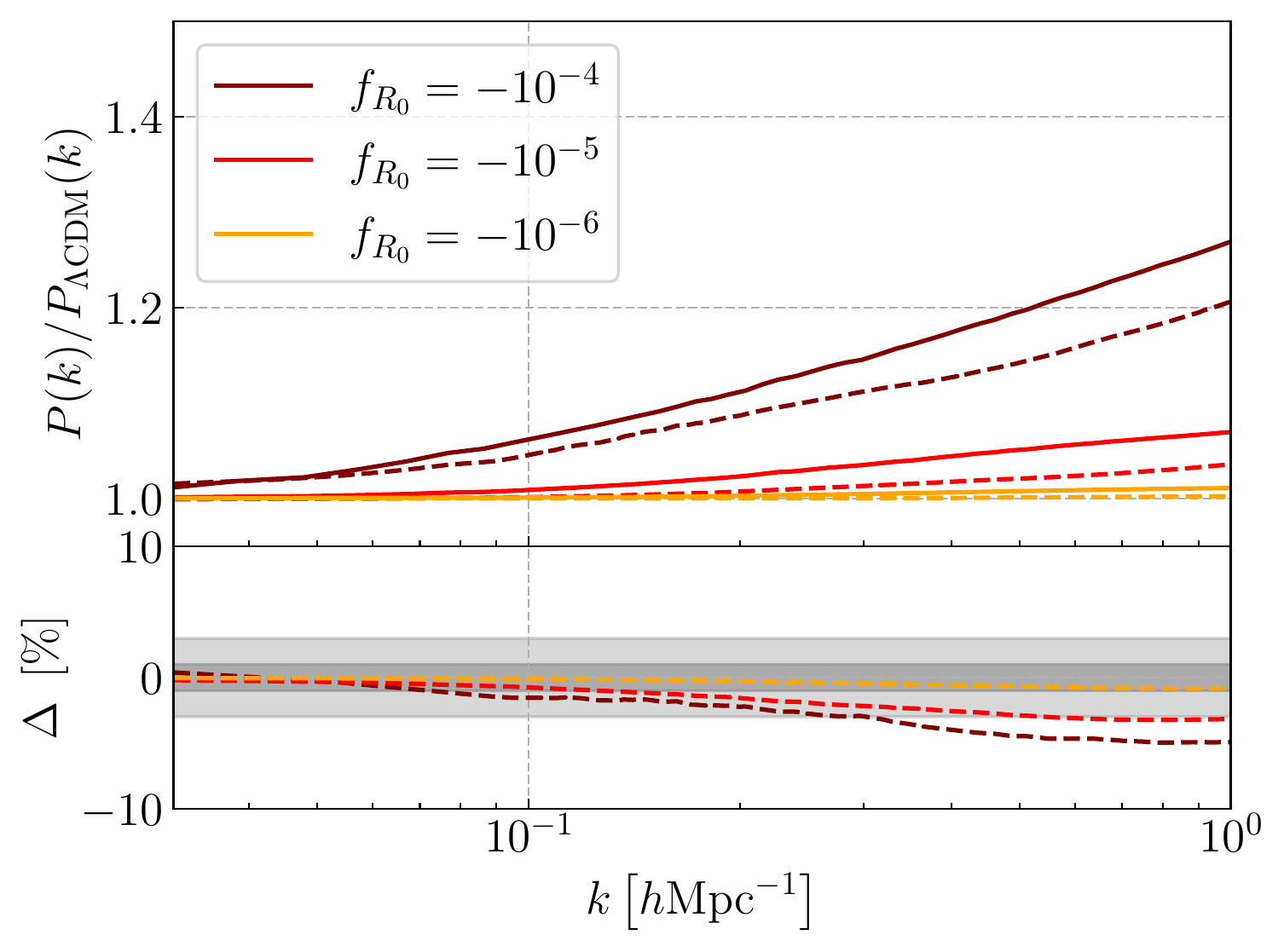}
  \caption{\label{fig:emu_vs_COLA_z1}
  Same as Figure~\ref{fig:emu_vs_COLA}, but at $z=1$.
  We notice that \textsc{mgemu} underestimates the power spectrum boost with respect to \textsc{e-mantis}, by up to $\sim5\%$ for F$4$ and $\sim3\%$ for F$5$.
  As pointed out in~\citet{ramachandra_emulator}, the COLA simulations implement an approximate solver for the MG screening, which produces this effect.
  As in Figure~\ref{fig:emu_vs_COLA}, we limit the comparison to the scales covered by \textsc{mgemu}, i.e. $k\leq1h\mathrm{Mpc}^{-1}$.
  }
\end{figure}

We perform a third comparison with the \textsc{mgemu} emulator introduced in~\citet{ramachandra_emulator}.
In this work, the authors build an emulator for the matter power spectrum boost in $f(R)$ gravity using COmoving Lagrangian Acceleration (COLA) simulations.
We refer the reader to the aforementioned paper for more specific details.
The comparison at $z=0$ between both emulators is given in Figure~\ref{fig:emu_vs_COLA}.
We set set $n_{s} = 0.9649$ and $n=1$ in the COLA emulator to match the values used in our training dataset.
We also use $\Omega_{m}=0.3153$ and $\sigma_{8}=0.8111$, which are the best fit values from~\citet{Planck18}.
We only consider scales $k\leq1h\mathrm{Mpc}^{-1}$ since the usage of the COLA emulator for larger values of $k$ is not recommended by its authors.
Indeed, the COLA simulations are less accurate than N-body ones at small scales.
Both emulators agree at the $1\%$ level for the models F$5$ and F$6$.
For F$4$, the agreement is at the $3\%$ level.
It is important to keep in mind that the model F$4$ is at the edge of the emulated parameter space of both emulators, which might lead to larger emulation errors in both predictions.
The COLA emulator has a reported accuracy of $5\%$ and, therefore, the differences with \textsc{e-mantis} are well within the expected error bars.
In Figure~\ref{fig:emu_vs_COLA_z1}, we show the same comparison, but at a redshift of $z=1$.
It can be seen that \textsc{mgemu} tends to underestimate the power spectrum boost with respect to \textsc{emantis}.
The difference reaches the $\sim3\%$ level for F$5$ and $\sim5\%$ for F$4$.
In~\citet{ramachandra_emulator}, the authors explain that the COLA simulations used to train \textsc{mgemu} implement an approximate method to solve for the screening of the $f(R)$ extra field.
This produces an underestimation of the boost at high redshifts, with respect to N-body simulations, which solve for the full MG equation of motion.

\begin{figure}
  \centering
  \includegraphics[width=\linewidth]{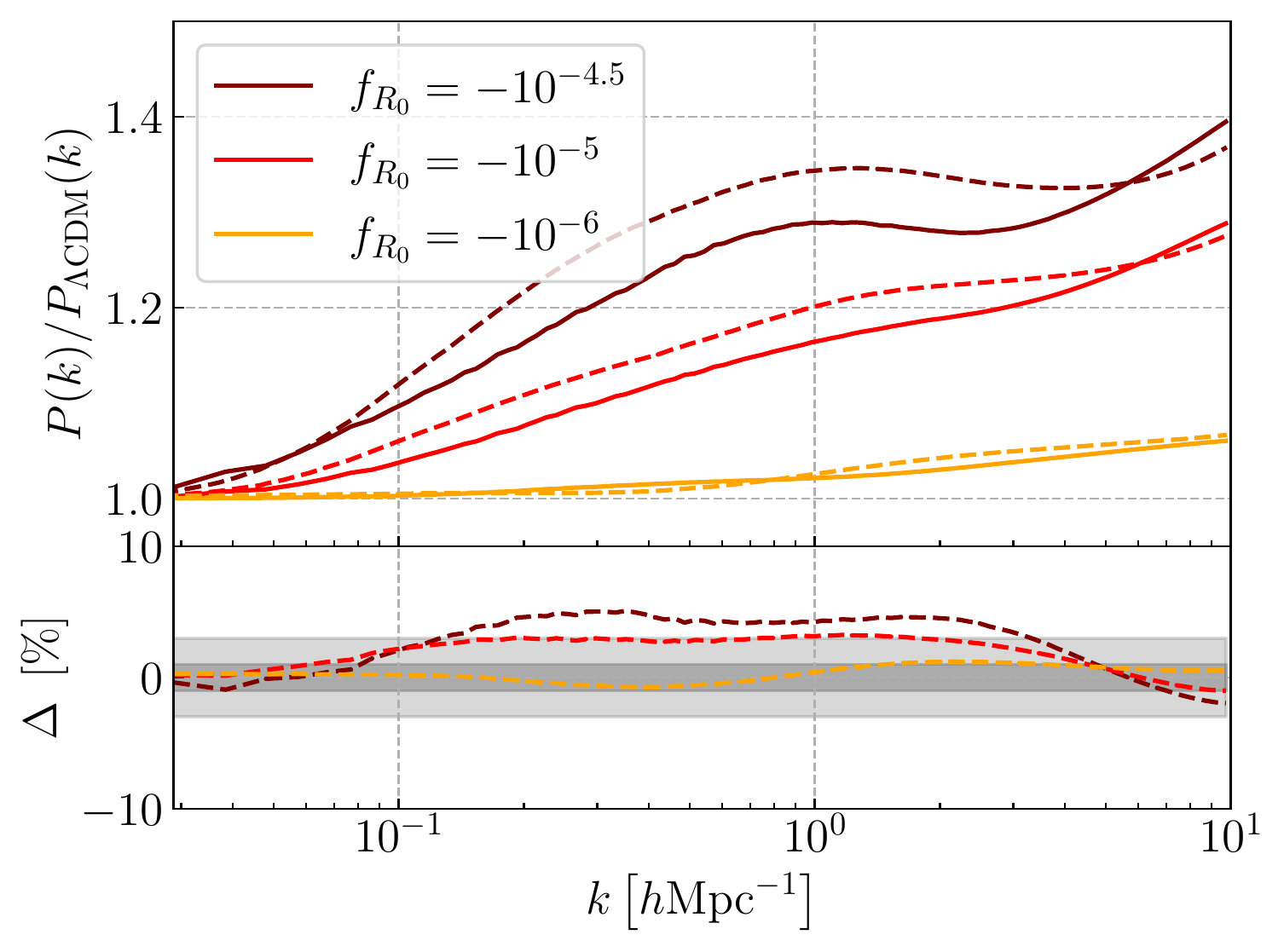}
  \caption{\label{fig:emu_vs_forge}
    Comparison with the \textsc{forge} emulator~\citep{Arnold2021} at $z=0$.
    The top panel shows the power spectrum boost for different $f(R)$ as predicted by \textsc{forge} (dashed lines) and \textsc{e-mantis} (solid lines).
    The bottom plot gives the relative difference of the \textsc{forge} prediction with respect to \textsc{e-mantis}.
    The light and dark grey bands mark the $1\%$ and $3\%$ levels respectively.
    For F$6$, there is a $1\%$ level agreement at all scales considered.
    In the case of F$5$ and F$4.5$, there is a $3\%$ and $5\%$ difference respectively at intermediate to small scales.
    The \textsc{forge} predictions have been corrected to compensate for the effect of \textsc{halofit}, since \textsc{forge} predicts $P_{f(R)}/P_\mathrm{\textsc{halofit}}$ instead of the boost $P_{f(R)}/P_{\Lambda\mathrm{CDM}}$, which is the robust quantity of interest in this article (see Section \ref{subsubsec:boost}).
  }
\end{figure}

We perform a last comparison with the \textsc{forge} emulator introduced by~\citet{Arnold2021}.
This emulator is built using cosmological N-body simulations run with the modified gravity \textsc{arepo} code~\citep{arnold_arepo}.
\textsc{forge} emulates the power spectrum boost in $f(R)$ gravity with respect to \textsc{halofit}~\citep{halofit} instead of a N-body $\Lambda$CDM prediction.
It would be incorrect to directly compare the outputs of both emulators, since they do not emulate the same base quantity.
The first cosmological model of \textsc{forge} (node-0 from Table 1 of~\citet{Arnold2021}) is a $\Lambda$CDM model, which we can use to get $P_{\Lambda\mathrm{CDM}}/P_{\textsc{halofit}}$.
We use this quantity to rescale the boost predicted by \textsc{forge} and compare it to the output of \textsc{e-mantis}.
In order to minimise the emulation errors when removing the \textsc{halofit} component from the \textsc{forge} prediction, we set the $\Lambda$CDM parameters to the cosmology of the \textsc{forge} node-0: $\Omega_{m}=0.31315$, $\sigma_{8}=0.82172$ and $h=0.6737$.
Figure~\ref{fig:emu_vs_forge} shows the comparison between \textsc{forge} and \textsc{e-mantis} for different $f(R)$ models at $z=0$.
For the model F$6$, there is a $1\%$ agreement at all the emulated scales.
However, for models F$5$ and F$4.5$, the difference is larger, reaching $3\%$ and $5\%$, respectively at intermediate to small scales.
The reported accuracy of \textsc{forge} is of order $2\%$ for most of the models covered by it and in particular the ones used in this comparison.
Combined with the $3\%$ estimated accuracy of \textsc{e-mantis}, the total error budget in this test is of $3.6\%$.
In the case of F$4.5$, the difference between both emulators is larger.
The fact that we have used the \textsc{forge} node-$0$ to remove the \textsc{halofit} component from its prediction might introduce its own additional source of errors.
However, this is the best comparison we can perform given the output quantities of both emulators.
Moreover the FORGE emulator does not benefit from the cancellation of numerical systematics that is obtained by considering the boost instead of the power spectrum.
Additionally, the F4.5 models is at the edge of the \textsc{forge} emulated space, which might lead to increased emulation errors. 
Furthermore, there might be some systematic difference between the N-body predictions of \textsc{ecosmog} and \textsc{arepo} that are not accounted for in the estimated accuracy of the emulators.
To conclude, the two emulators are complementary as they are  sensitive to different kind of systematics.
A more in depth study would require a detailed comparison of both simulation codes, which is out of the scope of this work.

Overall there is a good agreement between the predictions of \textsc{e-mantis} and other existing predictions from the literature.

\subsection{Example of application}%
\label{subsec:MCMC}

\begin{figure*}
    \centering
    \includegraphics[width=\linewidth]{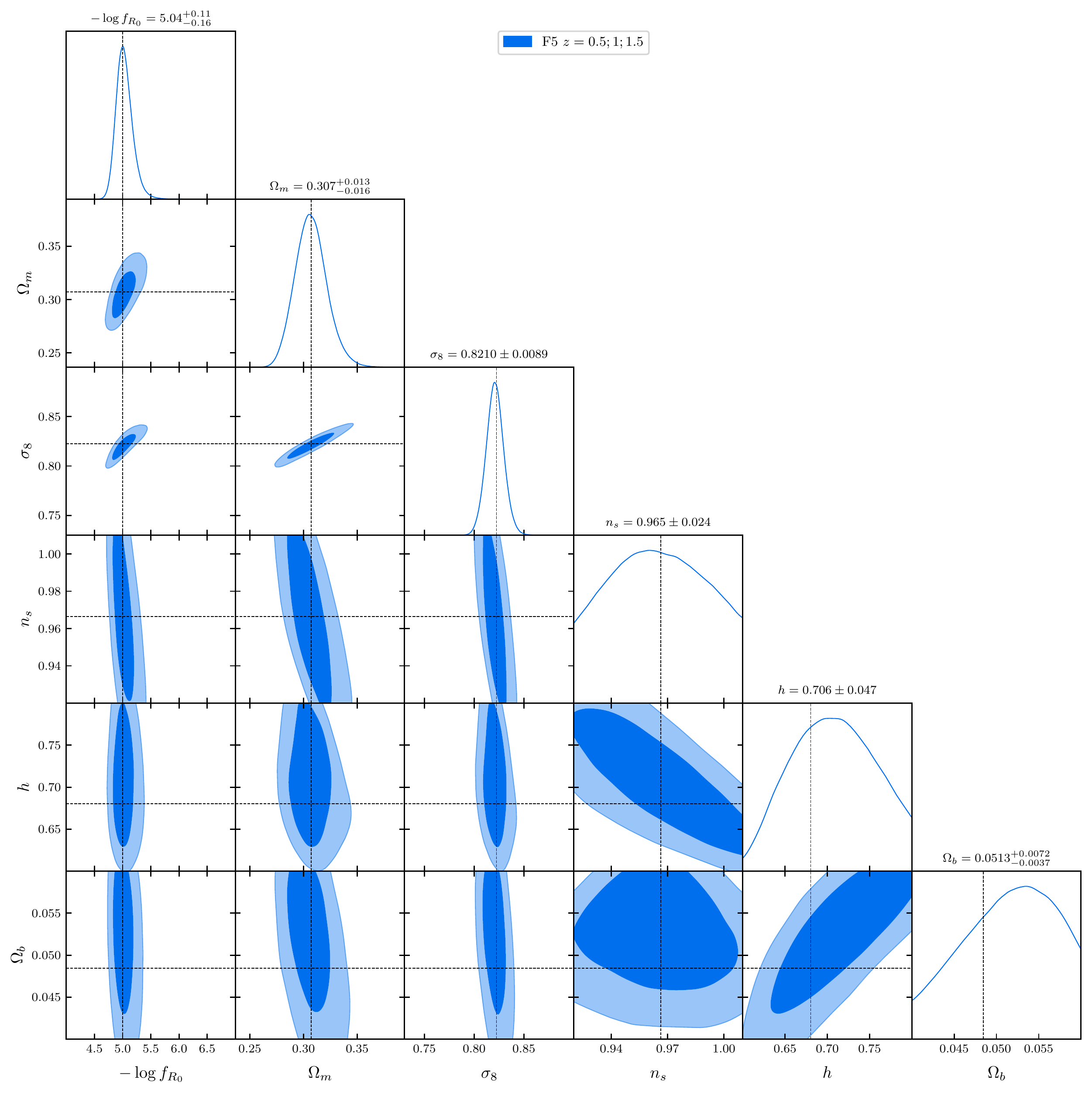}
    \caption{
    Marginalised $1$D and $2$D posterior distributions obtained from the MCMC described in Section~\ref{subsec:MCMC} for the 6 cosmological parameters $\left\{-\log{\left|\fR\right|}, \OmegaM, \sigma_8, n_s, h, \Omega_b\right\}$.
    The dark and light blue contours give the 1-$\sigma$ and 2-$\sigma$ confidence levels respectively.
    The fiducial values for each parameters, marked by the black dotted lines, always lay within the 1-$\sigma$ contours.
    The matter power spectrum $P(k)$ is obtained from simulation snapshots at $z=0.5,1,1.5$, while we use \textsc{e-mantis} for the predictions of the boost combined with \textsc{bacco} for the $\Lambda$CDM predictions.
    }
    \label{fig:MCMC}
\end{figure*}

In this section, we present a toy example of how to use the \textsc{e-mantis} emulator.
We aim  at recovering the cosmological parameters from a $f(R)$CDM simulation by running a Markov Chain Monte Carlo (MCMC) on the measured matter power spectrum.

The emulator presented in this work is able to compute the boost of the matter power spectrum $B(k)=P_{f(R)}/P_\mathrm{\Lambda CDM}$.
In order to get the full power spectrum in $f(R)$ gravity, such quantity needs to be combined with a non-linear $\Lambda$CDM prediction.
For the purpose of this section, we use the \textsc{bacco} emulator~\citep{bacco_emulator}.
Other $\Lambda$CDM emulators such as \textsc{CosmicEmu}~\citep{CosmicEmu_IV} or \textsc{EuclidEmulator2}~\citep{euclid_emulator_2} would also be valid alternatives.

In order to run the MCMC, we need simulation data to be used as the data vector.
The volume and mass resolution of the simulations presented throughout this work have been designed to predict the matter power spectrum boost with a few per cent errors.
However, they are not able to estimate the full power spectrum with the same level of accuracy.
Figure~\ref{fig:number_real_z0} shows that reducing statistical errors on the full power spectrum under the $1\%$ level at scales $0.1 h\mathrm{Mpc}^{-1} < k < 1 h\mathrm{Mpc}^{-1}$ would require performing at least $\mathcal{O}(100)$ $f(R)$ realisations with the same box size.
Alternatively, an accurate modeling of the non-linear covariance at those scales would avoid getting biased constraints even in the presence of significant errors in the data vector.
However, computing a fully non-linear covariance requires an even higher number of simulations, at least of order $\mathcal{O}\left(10^3-10^4\right)$\citep{Blot2015}.
Such a numerical effort goes beyond the scope of this section.
We resort to building a composite power spectrum in $f(R)$ gravity using the same boost strategy as in the construction of our emulator.
We compute the full power spectrum in $\Lambda$CDM using a series of $384$ independent realisations, each evolving $512^{3}$ dark matter particles in a simulation volume of $\left(328.125\, h^{-1}\mathrm{Mpc}\right)^{3}$.
This set of simulations is part of an ongoing simulation series which will be presented in a future work.
The values of the cosmological parameters used for these simulations are given in Table~\ref{tab:MCMC_priors}.
These simulations are run with the same numerical parameters previously described (see Table~\ref{tab:emulation_suite}).
Additionally, we perform $5$ realisations with the same volume and mass resolution for the F$5$ model in order to compute the power spectrum boost.
Figure~\ref{fig:number_real_z0} shows that the statistical errors on the power spectrum boost with $5$ realisations is much smaller than $1\%$.
We build a full power spectrum for F$5$ by multiplying the $\Lambda$CDM power spectrum by the F$5$ boost.
For simplicity, we use the same binning in $k$ as in \textsc{e-mantis} and described in Section~\ref{subsubsec:smoothing_boost}.
Such procedure gives us a full power spectrum in $f(R)$CDM cosmology with per cent level accuracy, while minimising the numerical effort.

We assume a linear Gaussian covariance for the power spectrum data.
Additionally, it is important to also take into account the errors from the emulator predictions, which at scales small enough dominate over the Gaussian covariance.
The authors of~\citet{bacco_emulator} estimate the \textsc{bacco} emulator to be accurate at the $2\%$ level in the case of $\Lambda$CDM.
According to the convergence studies presented in section~\ref{sec:conv_stud}, we consider a $3\%$ error on the power spectrum boost predicted by \textsc{e-mantis}.
The total relative systematic error introduced by the emulators has therefore a value of $\sigma_\mathrm{sys} = 0.036$, which we add quadratically to the diagonal Gaussian covariance.
The final covariance between two modes $k_1$ and $k_2$ takes the form
\begin{equation}\label{eq:covariance}
\mathrm{cov}(k_1,k_2) = \left[\frac{2}{N_{k_1}} + \sigma_\mathrm{sys}^2\right]P^2(k_1)\delta_{k_1k_2},
\end{equation}
where $N_{k_i} = k_{i}^2\Delta k_{i} V / (2\pi^2)$ is the number of independent Gaussian variables in the bin centred on $k_i$ and of width $\Delta k_i$, $V$ is the simulation volume and $\delta_{k_1k_2}$ is Kronecker delta.
The first term inside the brackets corresponds to the linear Gaussian contribution (see for example~\citet{Scoccimarro_1999}).

\begin{table}
  \centering
  \caption{
  Table summarising the truth values and the flat prior ranges used in the MCMC of Section~\ref{subsec:MCMC} for each parameter.
  The best fit values found by the MCMC, with their 1-$\sigma$ confidence levels, are also given.
  }
  \label{tab:MCMC_priors}
  \begin{tabular}{|c|c|c|c|}
    \hline\hline
    Parameter  & True value & Prior range & Best fit \\
    \hline\hline
    $-\log{\left|\fR\right|}$ & 5 & $\left[4,7\right]$ & $5.04_{-0.17}^{+0.11}$ \\
    \hline
    $\OmegaM$  & 0.3071 & $\left[0.2365,0.3941\right]$ & $0.307_{-0.016}^{+0.013}$\\
    \hline
    $\sigma_8$  & 0.8224 & $\left[0.73,0.9\right]$ & $0.8210\pm 0.0089$\\
    \hline
    $n_s$ & 0.96641 & $\left[0.92,1.01\right]$ & $0.965\pm0.024$\\
    \hline
    $h$ & 0.6803 & $\left[0.6,0.8\right]$ & $0.706\pm0.047$\\
    \hline
    $\Omega_b$ & 0.048446 & $\left[0.04,0.06\right]$ & $0.0513_{-0.0038}^{+0.0072}$\\
    \hline
  \end{tabular}
\end{table}

We combine the binned matter power spectrum data described earlier, the predictions from the emulators and the covariance from Eq.~\eqref{eq:covariance} into a Gaussian likelihood function.
We assume flat priors for the parameters $\left\{-\log{\left|\fR\right|}, \OmegaM, \sigma_8, n_s, h, \Omega_b\right\}$.
The variations we explore for each parameter correspond to the widest range allowed by the intersection of \textsc{e-mantis} and \textsc{bacco} and are given in Table~\ref{tab:MCMC_priors}.
We use the Python package \textsc{emcee}~\citep{emcee} to run the MCMC.
We employ the  “stretch move” sampling method from~\citet{Goodman_Weare}, which is the default algorithm used by \textsc{emcee}.

Our initial tests show that when using a single redshift the posterior distribution exhibits small multi-modal features.
These are due to the degeneracy between the parameters $\fR$, $\OmegaM$ and $\sigma_8$.
In order to break such degeneracies, we combine data from multiple redshifts by assuming that the datasets at different redshifts are independent from each other.
\textsc{e-mantis} is able to make predictions up to $z=2$.
However, \textsc{bacco} is limited to $z<1.5$.
In practice, we combine the matter power spectra from $z=0.5, 1$ and $1.5$ into a single data vector and repeat the covariance from Eq.~\eqref{eq:covariance} for each redshift.

We run an MCMC with $256$ walkers or independent chains and around $\sim100\ 000$ steps for each one of them, which took $\sim10$ hours to complete on a single thread of a laptop 11th Gen Intel Core i7-1165G7 @ 2.80Ghz cpu.
The resulting posterior distribution of the cosmological parameters is shown in Figure~\ref{fig:MCMC}.
The true values for each parameter, marked with the dotted black lines, always lay within the 1-$\sigma$ confidence contours (see also Table~\ref{tab:MCMC_priors}).
This simple toy use case illustrates how to use the boost predicted by \textsc{e-mantis} in combination with an independent $\Lambda$CDM prediction.
It serves as an additional cross-validation of the newly built emulator.

\section{Conclusions}

In this work, we present a new emulator, named \textsc{e-mantis}, that predicts the matter power spectrum boost in $f(R)$ gravity with respect to $\Lambda$CDM, $B(k)=P_{f(R)}/P_{\Lambda\mathrm{CDM}}$.
We run N-body cosmological simulations with the modified gravity code \textsc{ecosmog} in order to predict such quantity for different cosmological models.
Building an emulator for the boost instead of the raw power spectrum presents several advantages, which we summarise here.

Section~\ref{subsubsec:parameter_range} shows that the power spectrum boost mainly depends on only three cosmological parameters: $\fR$, $\Omega_{m}$ and $\sigma_{8}$.
Indeed, the variation of $B(k)$ with $h$, $n_{s}$ and $\Omega_{b}$ is smaller than $1\%$ up to $k=10h\mathrm{Mpc}^{-1}$, even for variations of these parameters of $\pm25\%$ around their best fit values from~\citet{Planck18}.
In appendix~\ref{app:cosmo_cross_dependence}, we have checked that simultaneous variations of those parameters have an effect which remains smaller than $2-3\%$ in the worst cases, mostly around the edges of the parameter space.
This error is less than the maximum systematic error in our training data.
Therefore, we can neglect the influence of $n_s$, $h$ and $\Omega_b$ on the boost when building the emulator with a few percent accuracy.
We then only need to sample a $3$-dimensional parameter space with N-body simulations instead of a $6$-dimensional one.
As a consequence, the required number of simulations to reach a given emulation accuracy is reduced.

Section~\ref{subsec:sys_sym_err} shows that the boost is less affected by statistical and systematic errors than the raw power spectrum.
Indeed, there is a very large cancellation of cosmic variance and large scale errors.
This is due to the fact that the $f(R)$ and $\Lambda$CDM simulations are evolved from the same initial conditions and that $f(R)$ gravity has a weak effect at large scales.
The same kind of cancellation happens with the systematic errors at small scales due to the finite mass resolution of the simulations.
In the end, we require less accurate simulations to resolve the boost to a given precision that what would be needed for the raw power spectrum.
We find that with our simulations, which cover an effective volume of $\left(560h^{-1}\mathrm{Mpc}\right)^{3}$ with a particle mass resolution of $m_{\mathrm{part}}\sim2\times10^{10}h^{-1}M_{\odot}$, we are able to compute the power spectrum boost with an accuracy better than $3\%$ for $0.03 \ h\mathrm{Mpc}^{-1} < k < 7 \ h\mathrm{Mpc}^{-1}$ and $0 < z < 2$.
In practice, as shown in Figure~\ref{fig:boxsize_small_scales_contour}, the systematic error on the power spectrum boost measurement depends on the value of $\fR$, redshift and scale.
However, the $3\%$ estimation is a conservative value for most models, scales and redshifts.
Even though in most cases the actual accuracy is better than $1\%$.

We sample the $3$D cosmological parameter space $\left\{\fR, \Omega_{m}, \sigma_{8}\right\}$ with $80$ models selected from a Sliced Latin Hypercube Design (SLHD) using the method introduced by~\citet{LHS_sliced}.
We also sample an additional $30$ models in the regions of the parameter space where the emulation errors are the highest.
For each of the $110$ cosmological models we run $5$ independent realisations both in $f(R)$ gravity and the corresponding $\Lambda$CDM cosmology.
From this training data we build an emulator using a Gaussian Process regression.
Section~\ref{subsec:emu_err}, along with Appendix~\ref{app:emu_err_z}, shows that the average emulation errors are smaller than $0.5\%$ for all scales and redshifts.
The accuracy of \textsc{e-mantis} is therefore dominated by the systematic errors in the training data.

Overall, building an emulator for the boost instead of the raw power spectrum greatly reduces our computational needs.
This is particularly relevant for $f(R)$ simulations, since they are more time consuming than $\Lambda$CDM ones.
Indeed, we find that with our specifications a $f(R)$ simulation is between two to ten times slower than a $\Lambda$CDM one, depending on the value of $\fR$.

In Section~\ref{subsec:comparison_emu}, we have compared the predictions of \textsc{e-mantis} to other existing predictions for the matter power spectrum boost in the literature.
First, we have considered the N-body simulations results from~\citet{WCP}, where they compare the boost predicted by different state-of-the-art modified gravity N-body codes.
This comparison serves as a validation, and shows that the predictions of \textsc{e-mantis} are in good agreement with the outputs of \textsc{ecosmog}, \textsc{isis} and \textsc{mg-gagdet}.
Then, we have compared \textsc{e-mantis} to the fitting formula from~\citet{winther_emulator}, which is calibrated on N-body simulations run with the same code and using a similar resolution.
We have seen that there is a good agreement for F$4$, F$5$ and F$6$, which are the models the fitting formula interpolates from.
The \textsc{e-mantis} emulator improves upon the predictions of this fitting formula by expanding the interpolation to the parameters $\OmegaM$ and $\sigma_8$ and by using 110 models (instead of 3) to emulate the boost.

We have also compared our predictions to the \textsc{mgemu} emulator introduced in~\citet{ramachandra_emulator}, which is based on COLA simulations.
\textsc{mgemu} is able to give predictions taking into account more cosmological parameters, such as $n_s$ and the modified gravity parameter $n$.
However, the range of validity of their predictions is restricted to scales $k\leq1h\mathrm{Mpc}^{-1}$, since COLA simulations are less accurate than N-body.
The overall agreement is good at $z=0$ and for those scales.
Additionally, we have seen in our comparison that \textsc{mgemu} tends to underestimate the boost at higher redshifts.
The authors of~\citet{ramachandra_emulator} explain that this is due to the approximate implementation of the screening mechanism in the COLA simulations.
This effect can be larger than the error induced by neglecting $n_s$ alone.
Therefore, \textsc{e-mantis} can give more accurate predictions in the case of $n=1$ even though the effect of $n_s$ is not taken into account.

Finally, we have performed a comparison with the \textsc{forge} emulator presented in~\citet{Arnold2021}.
To date, this is the only emulator for the matter power spectrum boost in $f(R)$ gravity based on N-body simulations.
However, they use a different N-body code, the modified gravity version of \textsc{arepo}~\citep{arnold_arepo}.
They consider a slightly larger parameter space than \textsc{e-mantis}, since they also include the parameter $h$.
However, we have shown in Section~\ref{subsubsec:parameter_range}, that the largest error comes from neglecting $n_s$.
Another difference, is that \textsc{e-mantis} emulates the boost with respect to a $\Lambda$CDM simulation, which we have shown in Section~\ref{subsec:sys_sym_err} allows for a large cancellation of systematic and statistical errors.
On the other side, \textsc{forge} emulates the boost with respect to \textsc{halofit} and  therefore does not benefit from this effect.
Our comparison shows a good agreement between both emulators for F$6$.
However, there seems to be a slight disagreement for larger values of the $\fR$ parameter, which is not expected given the emulation errors of both emulators.
Such difference could come from the usage of Halofit as a reference which makes the comparison difficult, an underestimate of the error bars in one of the emulators, or a systematic difference between \textsc{ecosmog} and \textsc{arepo} (which was not included in the comparison of~\citet{WCP}).
A more detailed investigation would be required to solve this issue.
Future generation surveys will likely require running their analyses using complementary theoretical predictions, in order to avoid getting biased cosmological constraints.
For this reason, it is important that several emulators based on different numerical codes, such as \textsc{forge} and \textsc{e-mantis}, are available to the community.

In Section~\ref{subsec:MCMC}, we have used \textsc{e-mantis} to recover the cosmological parameters of a numerical simulation by running an MCMC on its matter power spectrum.
This simple toy example illustrates how to combine the boost predicted by \textsc{e-mantis} with an independent $\Lambda$CDM prediction in order to get the full matter power spectrum prediction in $f(R)$CDM cosmology.

In this work we have ignored the effect of baryonic physics on the matter distribution.
Its impact on the matter power spectrum boost in $f(R)$ gravity is non-negligible for scales $k\gtrsim 2\ h\mathrm{Mpc}^{-1}$~\citep{arnold2019_hydro}.
Because of this, the \textsc{e-mantis} predictions should only be used on their own for larger scales.
In order to get accurate predictions for smaller scales, some correction taking into account baryonic feedback is required.
Currently, there are no available emulators for this type of correction in $f(R)$ gravity.
However, \citet{arnold2019_hydro} has shown that the full physics $f(R)$ power spectrum boost can be recovered by applying a GR baryonic correction to the dark matter only prediction.
Therefore, the \textsc{e-mantis} boost can be combined with a $\Lambda$CDM baryonic emulator, such as \textsc{BCemu}~\citep{BCemu}, in order to get accurate predictions at scales smaller than $k\gtrsim2\ h\mathrm{Mpc}^{-1}$. 
Alternatively, if a baryonic correction emulator in $f(R)$ gravity is developed in the near future, the non-linear dark matter only boost predicted by \textsc{e-mantis} (or another equivalent tool) will still be required in order to get the full matter power spectrum.
For this reasons, we believe that the emulator presented in this work could be a useful tool in order to constrain $f(R)$ gravity with the next generation of weak lensing surveys, such as LSST and \textit{Euclid}.

In the future, we plan to extend \textsc{e-mantis} to other observables, such as dark matter halo statistics and profiles.
We also plan to extend the current parameter space and cover alternative dark energy models.

\section*{Acknowledgments}

This work was granted access to HPC resources of TGCC through allocations made by GENCI (Grand Équipement National de Calcul Intensif) under the allocations 2020-A0090402287 and 2021-A0110402287. BL is supported by the European Research Council through a starting Grant (ERC-StG-716532 PUNCA), and by the UK Science and Technology Funding Council (STFC) Consolidated Grant No. ST/I00162X/1 and ST/P000541/1.
We thank the Euclid Consortium for the "Sponsor PhD Grant" of ISC.
The plots presented in this paper were produced with the \textsc{matplotlib}~\citep{matplotlib} package.
A significant amount of the numerical computations were carried out using the \textsc{SciPy}~\citep{scipy}, \textsc{NumPy}~\citep{numpy} and \textsc{scikit-learn}~\citep{scikit-learn} packages. 

\section*{Data Availability}

The emulator presented in this work, as well as the training power spectrum boosts, are publicly available at the following address: \url{https://doi.org/10.5281/zenodo.7738362}.
The rest of the data underlying this article will be shared on reasonable request to the corresponding author.



\bibliographystyle{mnras}
\bibliography{biblio} 




\appendix

\section{Impact of varying multiple cosmological parameters on the matter power spectrum boost}
\label{app:cosmo_cross_dependence}

In Section~\ref{subsubsec:parameter_range}, we have shown that the individual effect of the parameters $n_s$, $h$ and $\Omega_b$ on the matter power spectrum boost, around the reference model F$5$, is smaller than $1\%$.
Because of this, we have decided to neglect the effect of those parameters when building our emulator.
We then only need to interpolate in a $3$ dimensional space instead of a $6$ dimensional one, which reduces the number of training simulations required to reach a certain emulation accuracy level.
However, simultaneous variations of multiple of those parameters might produce a non-negligible effect.
Additionally, the effect of those parameters in other regions of the \textsc{e-mantis} parameter space might also be stronger.
Indeed, in our study of Section~\ref{subsubsec:parameter_range}, we have only considered variations around the reference model F$5$, which is placed towards the center of the emulation parameter space.

Probing all possible combinations of cosmological parameters would effectively require  building an emulator in a $6$ dimensional parameter space.
This would be computationally expensive, specially given the cost of running $f(R)$ gravity simulations.
Instead, we extend the \textit{cosmo} simulation suite with a few selected models in order to get an estimation of the error that \textsc{e-mantis} makes by neglecting the effect of $n_s$, $h$ and $\Omega_b$.
All the additional simulations discussed in this appendix are run with the same characteristics as in the \textit{cosmo} suite, except that we use one single random realisation instead of five.

First, we study the impact of varying two of the neglected parameters simultaneously around the reference cosmological model F$5$.
According to Figure~\ref{fig:cosmo_dependence}, the two neglected parameters with the largest impact are $n_s$ and $h$.
We run two simulations, which vary $n_s$ and $h$ by $+25\%$ and $-25\%$ respectively.
Figure~\ref{fig:cosmo_cross_h_ns} shows that the effects of both parameters mostly add up linearly (see Figure~\ref{fig:cosmo_dependence}).
The error made by neglecting both parameters is smaller than $1\%$ for $k\leq 2 h\mathrm{Mpc}^{-1}$ and reaches $\sim 2\%$ at $k=7 h\mathrm{Mpc}^{-1}$.
We stress that this is a worst case scenario, since we have varied both parameters by extreme values.
Figure~\ref{fig:cosmo_dependence} shows that the impact of $\Omega_b$ is much smaller than $1\%$ at all scales considered.
Therefore, we do not expect these results to significantly change when $\Omega_b$ takes different values.

We now assess the error made by individually neglecting the parameters $n_s$ and $h$ around other values of the main \textsc{e-mantis} parameters than the reference model F$5$.
We are interested in some worst case scenarios.
For instance we want to estimate the error made by \textsc{e-mantis}, when neglecting the individual impact of $n_s$ and $h$ around the edges of the parameter space in $\sigma_8$ and $\OmegaM$.
In order to do so, we run simulations by varying one of the main parameters as well as one of the neglected ones.
Figure~\ref{fig:cosmo_double_param_rel_err} shows the relative error made by neglecting $n_s$ or $h$ for some extreme values of $\sigma_8$ and $\OmegaM$.
We also perform the same study around F$6$, since it is a model of observational interest for future surveys.
One can see that the relative error is smaller than $1\%$ in most cases and reaches the $\sim1.5\%$ level in the worst considered case.
This should be compared to Figure \ref{fig:cosmo_dependence}: by changing the reference model the maximum errors have slightly increased from 1\% to 1.5\%.

By considering both results from Figures~\ref{fig:cosmo_cross_h_ns} and~\ref{fig:cosmo_double_param_rel_err}, we find that the maximum error on the boost made by neglecting the parameters $n_s$, $h$ and $\Omega_b$ is $2\%$.
An extrapolation to even more complex combinations suggests a possible upper limit of around $3\%$ (if some of the deviations add up).
To conclude, the maximum errors are expected at the 2-3\% level.
They mostly happen around the edges of the parameter space.
In most of the cases, the error is however smaller than $1\%$.

\begin{figure}
    \centering
    \includegraphics[width=\linewidth]{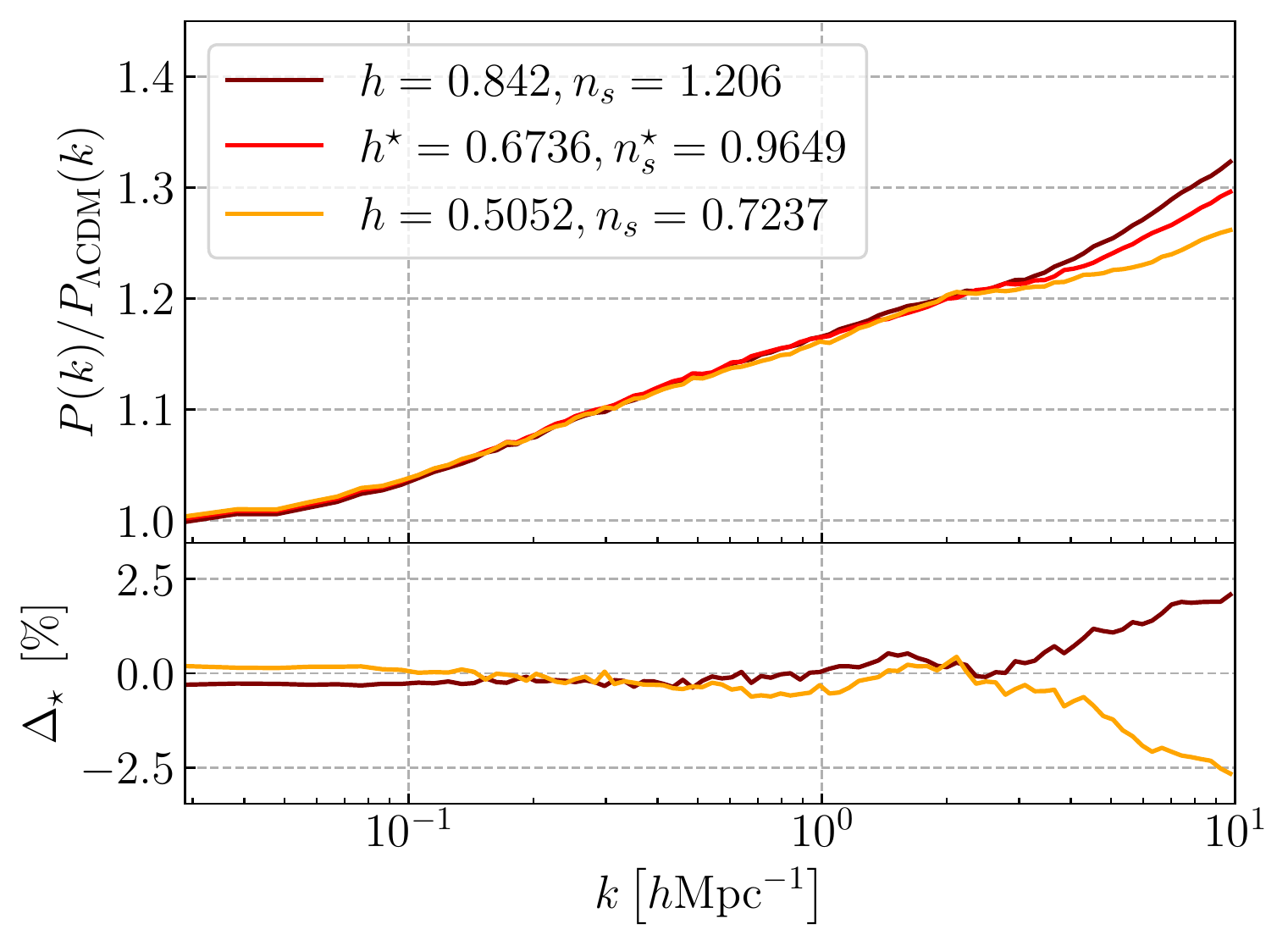}
    \caption{
    Effect of a simultaneous variation of the parameters $n_s$ and $h$ on the matter power spectrum boost. Both parameters are varied by $+25\%$ and $-25\%$ at the same time around the reference model F$5$. The top panel shows the boosts and the bottom panel the relative difference with respect to F$5$.
    The effect on the boost is smaller than $1\%$ up to $k\leq~2h\mathrm{Mpc}^{
    -1}$, and reaches $\sim2\%$ at $k=7~h\mathrm{Mpc}^{-1}$.
    We see that with respect to the results shown in Figure~\ref{fig:cosmo_dependence}, the individual effects of $n_s$ and $h$ mostly add up linearly.
    }
    \label{fig:cosmo_cross_h_ns}
\end{figure}

\begin{figure}
    \centering
    \includegraphics[width=\linewidth]{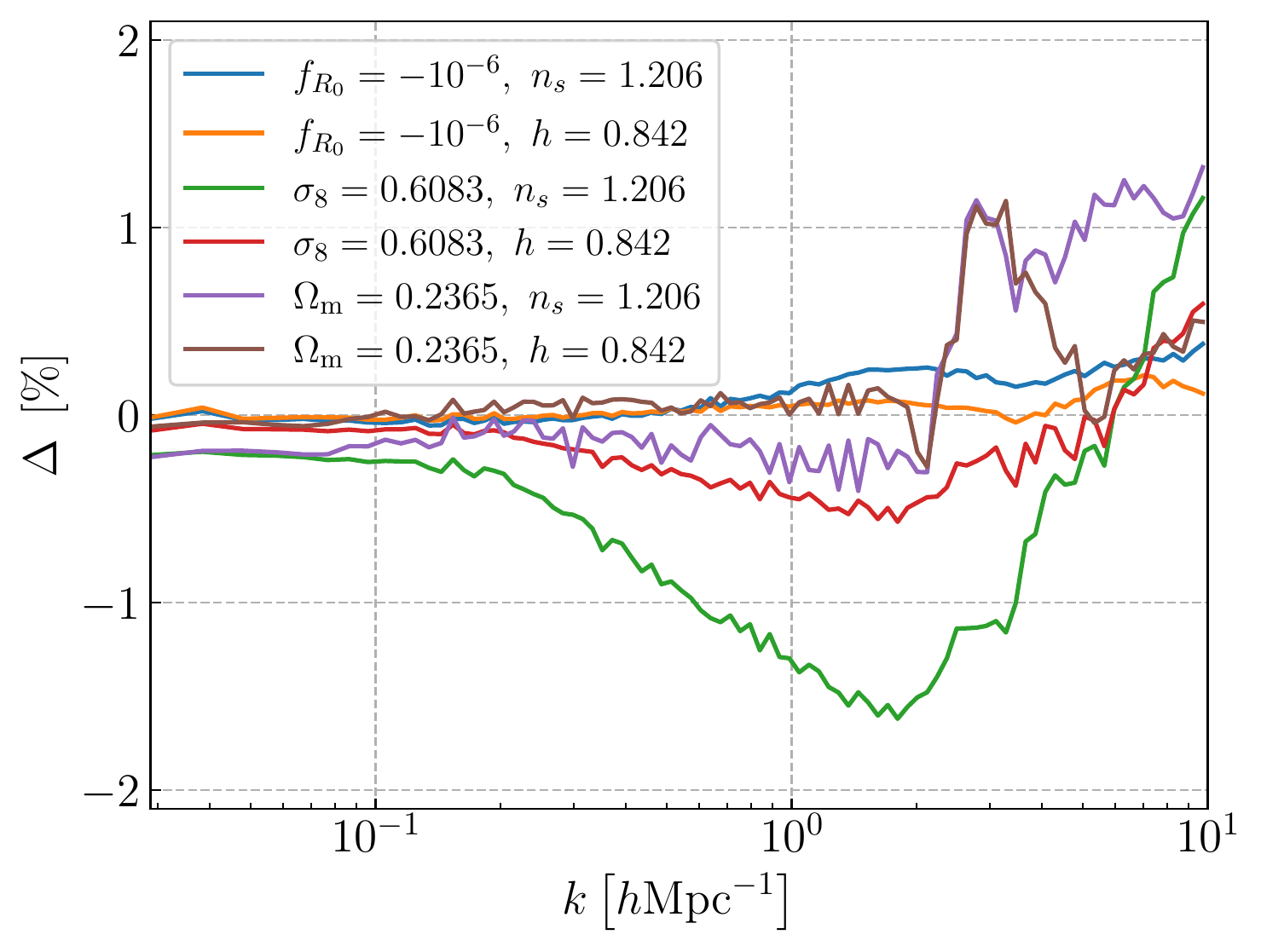}
    \caption{
    Error made by neglecting the effect of $n_s$ and $h$ for some selected values of the main \textsc{e-mantis} parameters different than the reference F$5$.
    The maximum error is around $\sim1.5\%$, while in most cases it stays smaller than $1\%$.
    }
    \label{fig:cosmo_double_param_rel_err}
\end{figure}

\section{Influence of the number of realisations and the refinement models on the emulation accuracy}
\label{app:emu_err_nreal_refinements}

\begin{figure*}
  \centering
  \begin{subfigure}{0.49\linewidth}
    \includegraphics[width=\linewidth]{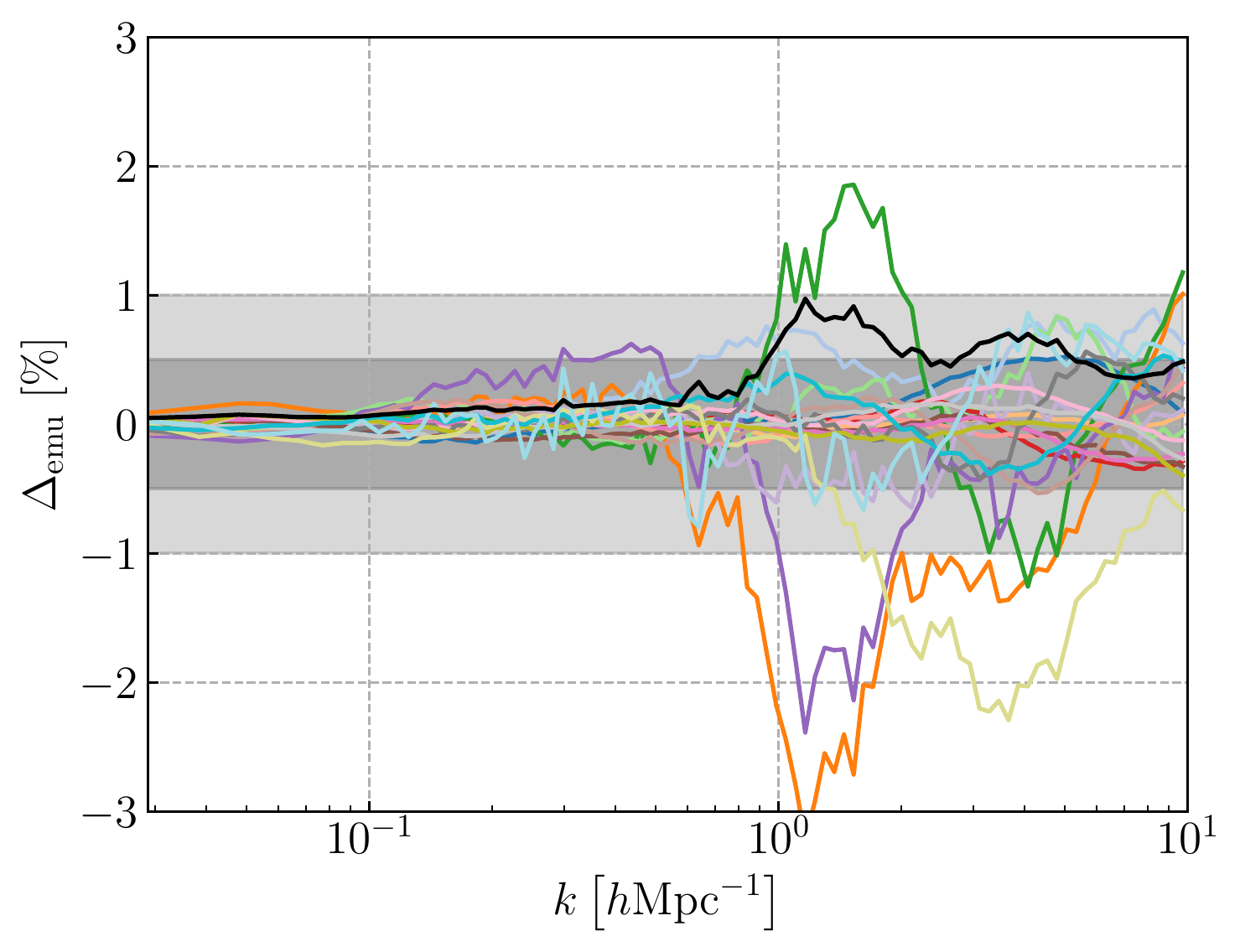}
  \end{subfigure}
  \begin{subfigure}{0.49\linewidth}
    \includegraphics[width=\linewidth]{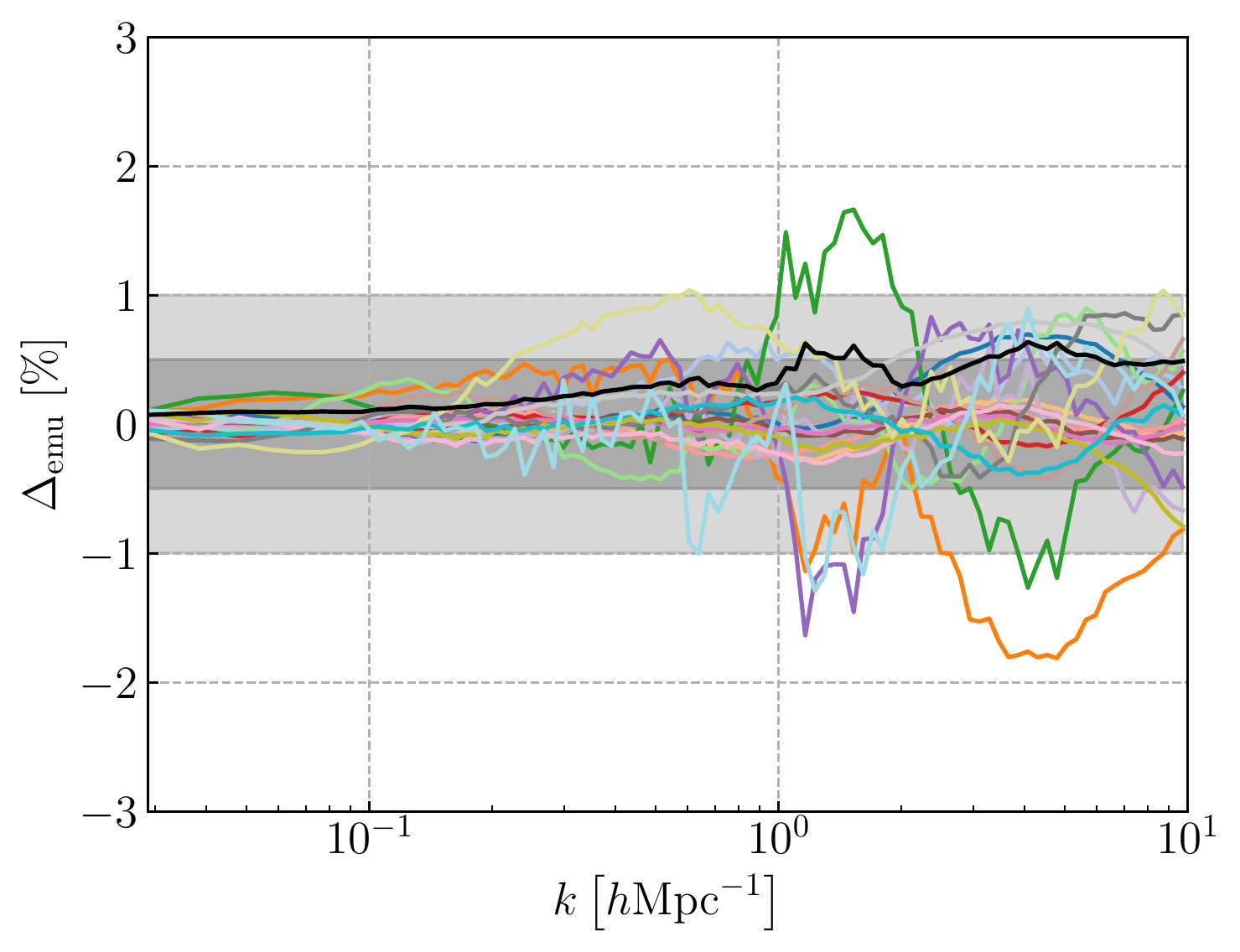}
  \end{subfigure}
  \begin{subfigure}{0.49\linewidth}
    \includegraphics[width=\linewidth]{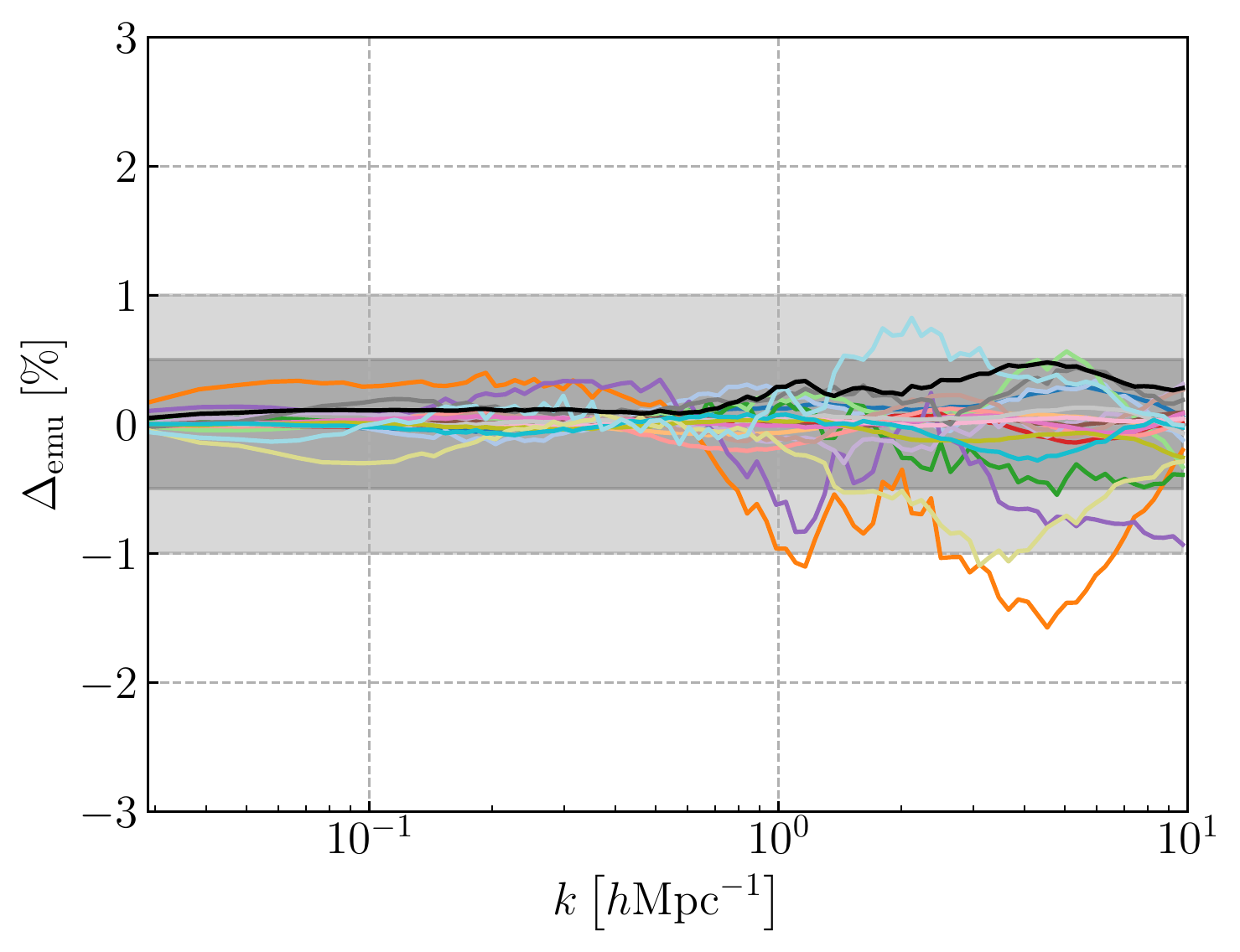}
  \end{subfigure}
  \begin{subfigure}{0.49\linewidth}
    \includegraphics[width=\linewidth]{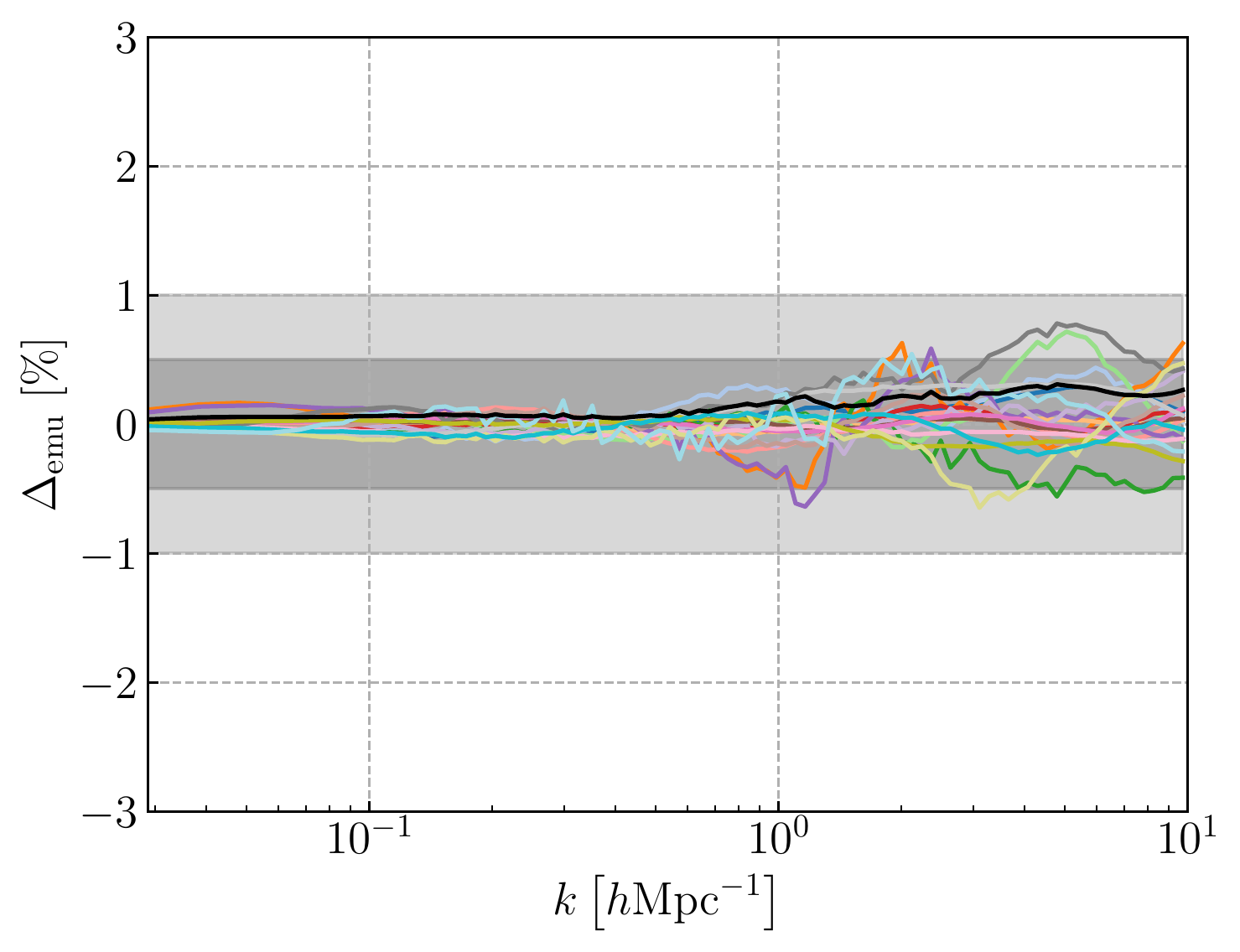}
  \end{subfigure}
  \caption{
    Emulation errors at $z=0$ for different training dataset configurations.
    \textit{Top left:} the emulator is trained on $3$ slices of the primary SLHD ($60$ models) and using a single realisation per cosmological model.
    \textit{Top right:} the $30$ refinement models are added to the training dataset.
    \textit{Botton left:} the emulator is trained without refinement models but using $5$ independent realisations per cosmological model.
    \textit{Bottom right:} the $30$ refinement models are once again added to the training dataset.
    In all the cases, the remaining slice from the primary SLHD is used as a validation sample.
    Each coloured line shows the emulation error for one of the $20$ validation models.
    The black line give the RMSRE, as defined by Eq~\eqref{eq:RMSRE}.
    The dark and light grey bands mark the $0.5$ and $1$ per cent accuracy limits.
    The $30$ refinement models as well as the $5$ realisations per cosmological model are needed to get emulation errors smaller than $1\%$ for all validation models and scales.
    In such a case, the RMSRE is well below the $0.5$ per cent level at all scales.
  }
  \label{fig:emulation_errors_nreal}
\end{figure*}

In this appendix, we show how both the number of realisations and the refinement models impact the final accuracy of the emulator.
We perform the same kind of test as in Section~\ref{subsec:emu_err}.
One slice of the primary SLHD is left out of the training models to be used as a validation sample.
The emulator is built with different training dataset configurations.
Then, for each configuration, the predictions of the emulator are compared to the simulations measurements from the validation models.
We consider four different configurations, where the emulator is trained using:
\begin{itemize}
  \item $60$ models from the primary SLHD and one realisation per cosmological model,
  \item $60$ models from the primary SLHD, the $30$ refinement models and one realisation per cosmological mode,
  \item $60$ models from the primary SLHD and five independent realisations per cosmological model,
  \item $60$ models from the primary SLHD, the $30$ refinement models and five independent realisations per cosmological model.
\end{itemize}
The results  of this test at $z=0$ are given in Figure~\ref{fig:emulation_errors_nreal}.
The refinement models as well as the five independent realisations per cosmological model make it possible to get an average emulation error smaller than $0.5\%$, and therefore negligible with respect to the systematic errors in the training data as estimated in Section~\ref{subsubsec:res_effects}.
Even for the worst validation models, the emulation errors remains well below $1\%$ at all relevant scales.
As expected, the signal is also smoother with $5$ realisations than with a single one.

\section{Emulation accuracy at different redshifts}
\label{app:emu_err_z}

\begin{figure*}
  \centering
  \begin{subfigure}{0.49\linewidth}
    \includegraphics[width=\linewidth]{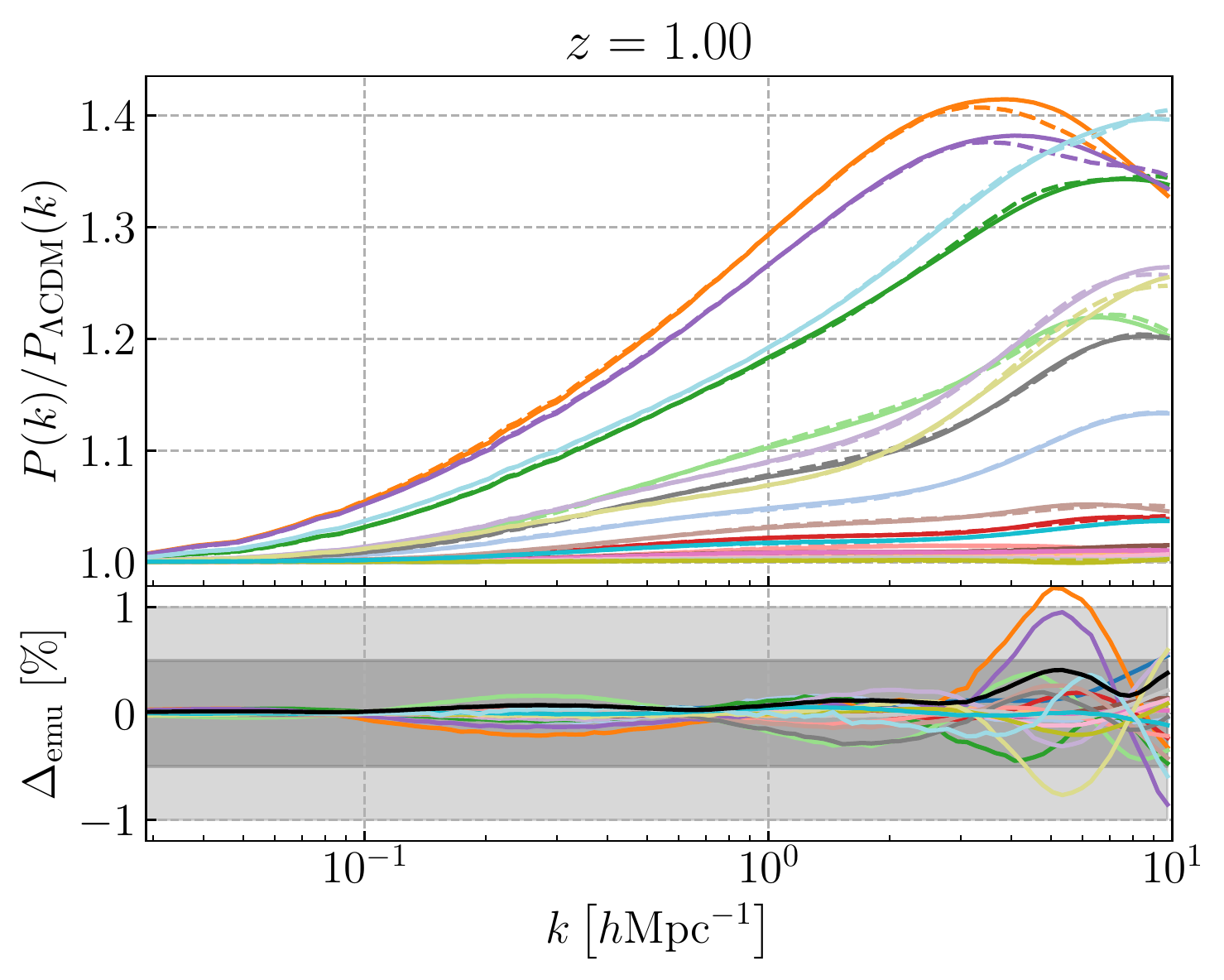}
  \end{subfigure}
  \begin{subfigure}{0.49\linewidth}
    \includegraphics[width=\linewidth]{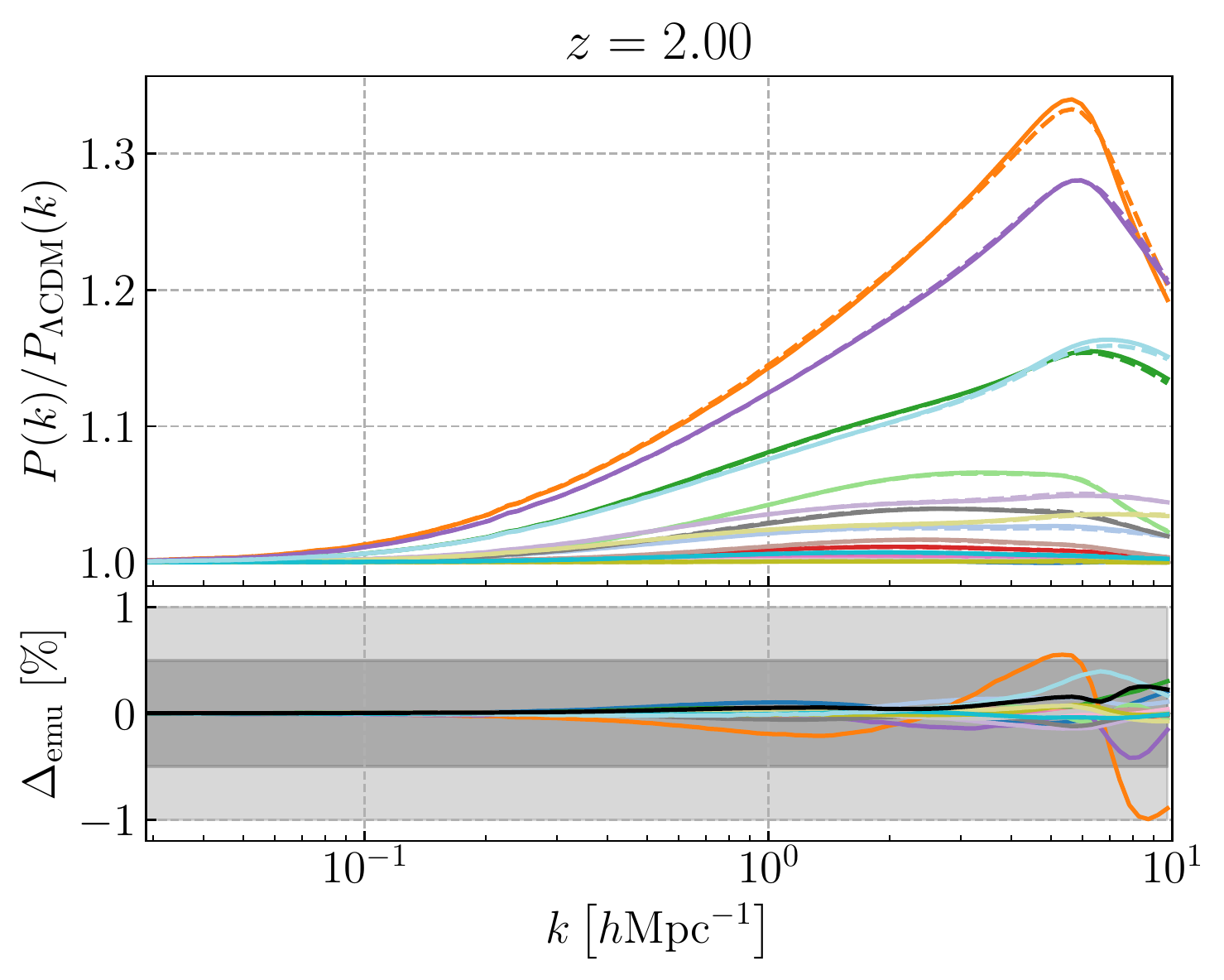}
  \end{subfigure}
  \caption{
    Results of the emulation accuracy test for the emulator at $z=1$ (left) and $z=2$ (right).
    The emulator is trained using the refinement models and $3$ slices from the primary SLHD.
    The remaining slice is used as a validation set.
    The top panels give the predicted boost by the emulator for each test model (solid lines).
    The dashed lines are measurements from the simulations.
    The bottom panels give the relative emulation errors for each validation model.
    The solid black line corresponds to the RMSRE as defined by Eq~\eqref{eq:RMSRE}.
    The dark and light grey bands give the $0.5\%$ and $1\%$ levels respectively.
    At both redshifts the emulation errors are smaller than $0.5\%$ for most models and scales.
    Only for a few extreme models the emulation errors reache the $1\%$ level at the smallest scales.
    The RMSRE is always smaller than $0.5\%$.
  }
  \label{fig:emulation_errors_z}
\end{figure*}

Our $f(R)$ matter power spectrum boost emulator is built using training data from $19$ redshift nodes (see Section~\ref{subsec:simu_set_emu}).
In Section~\ref{subsec:emu_err} we have estimated the emulation errors at $z=0$ by splitting the simulation data into a training and validation set.
Figure~\ref{fig:emulation_errors_z} gives the same results but at $z=1$ and $z=2$.
The conclusions drawn for $z=0$ are still valid at those redshifts.
More precisely, the emulation errors for the $20$ validation models only reach the $1\%$ level for the most extreme models and the RMSRE, defined in Eq~\eqref{eq:RMSRE}, remains under $0.5\%$ at all scales.
We have verified that these results hold for the $19$ redshift nodes of \textsc{e-mantis} between $0<z<2$.

\begin{figure}
  \centering
  \includegraphics[width=\linewidth]{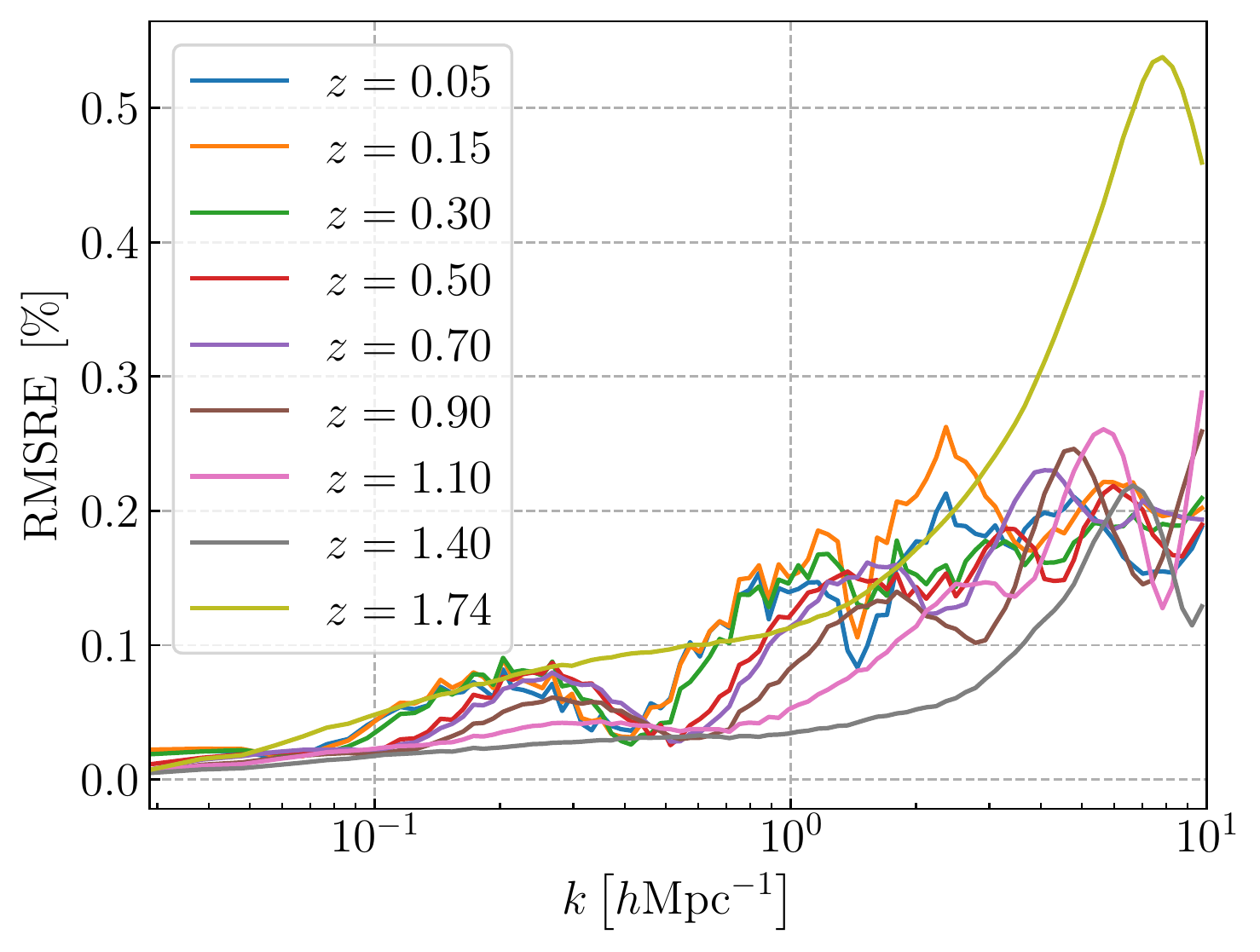}
  \caption{\label{fig:label}
    Redshift interpolation accuracy test.
    The emulator is trained using only $10$ redshift nodes.
    The remaining $9$ redshift nodes are used as a validation sample.
    For each validation redshift we compare the predictions from the emulator to the simulation measurements for the $80$ models of the primary SLHD.
    We give here the RMSRE, as defined by Eq.~\eqref{eq:RMSRE_interpz}, which for most redshifts is smaller than $0.3\%$ at all scales.
    The redshift interpolation errors in the final emulator, which uses twice as many redshift nodes as in this test, are expected to be smaller.
  }
\end{figure}

\textsc{e-mantis} is also able to perform predictions for redshift values outside the training nodes.
We proceed by building an independent emulator at each redshift node.
Then, for any arbitrary redshift, we perform a linear interpolation of the predicted boost between the two neighbouring redshift nodes.
In order to estimate the contribution to the emulation error budget of such a procedure, we train our emulator using the full set of available cosmological modes but only $10$ redshift nodes.
The remaining nodes are used as a validation sample.
On these $9$ validation nodes, we compare the prediction of the emulator to the simulation measurements for the $80$ models of the primary SLHD.
We exclude the refinement models from this comparison in order to have an homogeneous distribution of models across the cosmological parameter space.
We compute the corresponding root-mean-square-relative-error (RMSRE) at each validation redshift, defined as
\begin{equation}\label{eq:RMSRE_interpz}
  \mathrm{RMSRE}(k,z) = \sqrt{\frac{1}{N_{\mathrm{models}}} \sum_{i=0}^{N_{\mathrm{models}}-1}\left(\frac{B_{\mathrm{emu}}(k,z;\theta_{i})}{B_{\mathrm{sim}}(k,z;\theta_{i})}-1\right)^{2}},
\end{equation}
where $N_{\mathrm{models}}$ is the number of models used in the comparison, $\theta_{i}$ is the cosmological parameter vector of model $i$, $B_{\mathrm{emu}}$ is the power spectrum boost predicted by the emulator and $B_{\mathrm{sim}}$ is the boost measured from the simulations.
The obtained RMSREs are shown in Figure~\ref{fig:emulation_errors_z}.
For most redshifts the RMSRE is smaller than $0.3\%$ at all scales.
Only for $z=1.74$ the RMSRE reaches the $0.5\%$ level.
The final emulator has twice as many redshift nodes as in this test.
We therefore expect the real redshift interpolation errors to be smaller that the present estimation.
In any case, these errors are negligible with respect to the mass resolution errors, which can reach the $3\%$ level.


\bsp	
\label{lastpage}
\end{document}